\documentclass[12pt]{article}
\usepackage{mymacros}

\title{Towards complexity in de Sitter space from the double-scaled Sachdev–Ye–Kitaev model}
\author[a]{Sergio E. Aguilar-Gutierrez}
\affiliation[a]{Institute for Theoretical Physics, KU Leuven,\\ Celestijnenlaan 200D, B-3001 Leuven, Belgium}
\emailAdd{sergio.ernesto.aguilar@gmail.com}
\abstract{How can we define complexity in dS space from microscopic principles? Based on recent developments pointing towards a correspondence between a pair of double-scaled Sachdev–Ye–Kitaev (DSSYK) models/ 2D Liouville-de Sitter (LdS$_2$) field theory/ 3D Schwarzschild de Sitter (SdS$_3$) space in \cite{Narovlansky:2023lfz,Verlinde:2024znh,Verlinde:2024zrh}, we study concrete complexity proposals in the microscopic models and their dual descriptions. First, we examine the \textit{spread complexity} of the maximal entropy state of the doubled DSSYK model. We show that it counts the number of entangled chord states in its doubled Hilbert space. We interpret spread complexity in terms of a time difference between antipodal observers in SdS$_3$ space, and a boundary time difference of the dual LdS$_2$ CFTs. This provides a new connection between entanglement and geometry in dS space. Second, \textit{Krylov complexity}, which describes operator growth, is computed for physical operators on all sides of the correspondence. Their late time evolution behaves as expected for chaotic systems. Later, we define the \textit{query complexity} in the LdS$_2$ model as the number of steps in an algorithm computing n-point correlation functions of boundary operators of the corresponding antipodal points in SdS$_3$ space. We interpret query complexity as the number of matter operator chord insertions in a cylinder amplitude in the DSSYK, and the number of junctions of Wilson lines between antipodal static patch observers in SdS$_3$ space. Finally, we evaluate a specific proposal of \textit{Nielsen complexity} for the DSSYK model and comment on its possible dual manifestations.}

\begin{document}

\maketitle
\newpage
\section{Introduction}
The main results of our work are summarized in Table \ref{tab:results}. Below we provide some background; our motivation; and an outline of this manuscript.
\begin{table}[t!]   
\begin{center}    
\begin{tabular}  {|C{2.3cm}|C{3.3cm}|C{3cm}|C{3cm}|} \hline  
Proposal & Doubled DSSYK model & LdS$_2$ CFT & SdS$_3$ space\\ \hline
Spread complexity, $\mathcal{C}_{\rm S}$& Number of entangled chord states (\ref{eq:C S DDSSYK}) for the state (\ref{eq:ref energy state})& Boundary time difference (\ref{eq:CS LdS})& Static patch time difference (\ref{eq:relation y and spread})\\ \hline
Krylov complexity, $\mathcal{C}_{\rm K}$& Exponential growth (\ref{eq:late time CK}) of physical operators in (\ref{eq:O phys DDSSYK}) &Exponential growth (\ref{eq:late time CK}) of physical operators (\ref{eq:O phys LdS})& Exponential growth (\ref{eq:late time CK}) of physical operators (\ref{eq:O phys SdS})\\ \hline
Query complexity, $\mathcal{C}_{\rm Q}$& Number of matter chord insertions (\ref{eq:CQ DDSSYK}) & Number of fusions (\ref{eq:Query n fusions}) & Number of junctions of Wilson lines (\ref{eq:CAny})\\ \hline
Nielsen complexity, $\mathcal{C}_{\rm N}$& Distance (\ref{eq:CN after minimization}) between $\mathbb{1}$ and a unitary (\ref{eq:Nielsen intermediate}) in its group manifold & ??? & ???\\ \hline
\end{tabular}   
\caption{Different quantum complexity proposals (spread, Krylov, query, and Nielsen complexity) studied in this work and their interpretation for each side of the doubled DSSYK model/LdS$_2$ CFT/SdS$_3$ space correspondence. For comments about holographic duals to Nielsen complexity, see Sec \ref{sec:comments Nielsen}.}
\label{tab:results}
\end{center}
\end{table}

\subsection*{Static patch holography}
Since the early stages of the anti-de Sitter (AdS)/ conformal field theory (CFT) correspondence \cite{Maldacena:1997re,Gubser:1998bc,Witten:1998qj}, there has been a lot of interest in developing the holographic dictionary for de Sitter (dS) space \cite{Strominger:2001pn,Witten:2001kn,Maldacena:2002vr} to address some of the puzzling features of the cosmological horizons. For instance, there is a finite and constant entropy perceived by a worldline static patch observer due to the Hawking radiation coming from the cosmological horizon, which is given according to the Gibbons-Hawking formula \cite{Gibbons:1977mu}
\begin{equation}\label{eq:GH entropy}
    S_{\rm GH}=\frac{A(r_c)}{4G_N}~,
\end{equation}
where $A$ denotes the area of the cosmological horizon, with a radius $r_c$ according to the worldline observer. Given the finite value of this entropy, it was conjectured that the static patch of dS space can be described as a unitary quantum system carrying $\exp (S_{\rm GH})$ degrees of freedom \cite{Bousso:1999dw,Bousso:2000nf,Banks:2006rx,Anninos:2011af,Banks:2018ypk}, and this has been recently interpreted as a cosmological central dogma \cite{Shaghoulian:2021cef}, given its close similarities with the central dogma describing black holes as unitary systems with a finite number of degrees of freedom \cite{Almheiri:2020cfm}.

This realization naturally leads to the proposal of \textit{static patch holography} (see \cite{Spradlin:2001pw,Bousso:2002fq,Anninos:2012qw,Galante:2023uyf} for reviews, including other approaches to dS holography), which assumes there is a putative dual theory describing the static patch of dS space. There are two main approaches in this area, worldline holography \cite{Anninos:2011af,Leuven:2018ejp}, where, as the name suggests, the holographic dual is located on the worldline of the observer; and stretched horizon holography \cite{Susskind:2021esx}, where the dual is included in the so-called stretched horizon, a time-like surface within the static patch that is postulated to be very close to the cosmological horizon. This latter approach is motivated by recent studies where a double-scaled Sachdev–Ye–Kitaev\footnote{See \cite{Sachdev_1993,kitaevTalks} for starting work in the SYK model, and \cite{Rosenhaus:2018dtp,Sachdev:2024gas} for recent reviews.} (DSSYK) model is conjectured to reside within the stretched horizon of dS JT gravity \cite{Susskind:2022bia,Rahman:2022jsf}  (see also \cite{HVtalks}).\footnote{Alternatively, it can also be motivated by introducing $T\overline{T}+\Lambda_2$  deformations \cite{Gorbenko:2018oov,Lewkowycz:2019xse,Shyam:2021ciy,Coleman:2021nor,Batra:2024kjl} as they generate time-like Dirichlet boundaries within the static patch of dS$_3$ space, whose stability under thermal fluctuations has been examined in different works \cite{Svesko:2022txo,Banihashemi:2022jys,Banihashemi:2022htw,Anninos:2024wpy}.} We will \textbf{not assume} the existence of a stretched horizon in this work.

\subsection*{3D Schwarzschild-de Sitter space and its holograms}
Recently, there have been several exciting developments in dS holography based on the DSSYK model in the series of works \cite{Narovlansky:2023lfz,Verlinde:2024znh,Verlinde:2024zrh} (see also \cite{Milekhin:2023bjv,Xu:2024hoc}). It has been argued that a pair of DSSYK models can have a dual interpretation in terms of (1+2)-dimensional (non-rotating) Schwarzschild-de Sitter (SdS$_3$) space. {We will briefly review the different sides of the correspondence below.}

On the bulk side, SdS$_3$ space is a spherically symmetric solution to the equations of motion of the Einstein-Hilbert action with a positive cosmological constant:
\begin{equation}
I = \frac{1}{16 \pi G_N} \int \rmd^{3} x \, \sqrt{-g} \, \qty( \mathcal{R} - 2 \Lambda ) \, , \qquad
\Lambda = \ell_{\rm dS}^{-2} \, ,
\label{eq:action_EOM_gend}
\end{equation}
with $G_N$ the 3D Newton's constant, and $\ell_{\rm dS}$ the dS radius. The metric reads
\begin{equation}
\rmd s^2 = - f(r) \rmd t^2 + \frac{\rmd r^2}{f(r)} + r^2 \rmd \Phi^2 \, ,  \qquad
f (r) = 8G_NM - \frac{r^2}{\ell_{\rm dS}^2} \, ,
\label{eq:asympt_dS}
\end{equation}
where $M$ is the ADM energy with respect to $\mathcal{I}^+$ (e.g. see \cite{Gibbons:1977mu,Balasubramanian:2001nb,Ghezelbash:2001vs} for more details). Importantly, SdS$_3$ is locally isomorphic to dS$_3$ space; however, the term $M$ modifies the periodicity of $\Phi$ in (\ref{eq:asympt_dS}) by
\begin{equation}
    \Phi\sim\Phi+2\pi(1-\alpha)~,\quad\alpha\equiv1-\sqrt{1-8G_NM}~.
\end{equation}
\cite{Verlinde:2024znh} proposed to identify a holonomy variable measuring the conical deficit angle, $2\pi\alpha$, produced by matter sources along the poles of the sphere, with the Hamiltonian for SdS$_3$ space. They studied the canonical quantization of this proposal in the Chern-Simons (CS) formulation of SdS$_3$ space (see e.g. \cite{Witten:1988hc,Witten:1998qj,Castro:2011xb,Castro:2023dxp,Castro:2023bvo}) which turns out to take the same form of a pair of DSSYK models, subject to physical constraints.\footnote{Both models have the same quantum group symmetry, $SL_q(2)$. See \cite{Ohtsuki:2002ud} for a pedagogical introduction, and \cite{Berkooz:2022mfk,Blommaert:2023opb,Lin:2023trc,Blommaert:2023wad} for recently found connections with the DSSYK model, and holography \cite{Almheiri:2024ayc}.}

On the quantum mechanical side of the correspondence, each DSSYK model describes a strongly interacting system of $N$ Majorana fermions in $(0+1)$-dimensions with all to all $p$ body interactions governed by the Hamiltonian
\begin{equation}\label{eq:DDSSYK Hamiltonian}
    H^{\rm L/R}=\rmi^{p/2}\sum_{1\leq i_1<\dots<i_p\leq N}J_{i_1,\dots,~i_p}\psi^{\rm  L/R}_{i_1}\dots\psi^{\rm L/R}_{i_p}~,
\end{equation}
where L and R are labels to distinguish the different theories; $\psi^{\rm L/R}_{i_j}$ are Majorana fermions, obeying $\qty{\psi_{i},~\psi_{j}}=2\delta_{ij}$, with $i=1,\dots N$; and the coupling constants $J_{i_1\dots i_p}$ obey the following Gaussian distribution
\begin{equation}\label{eq:GEA J}
\expval{(J_{i_1\dots i_p})}=0~,\quad\expval{(J_{i_1\dots i_p})^2}=\frac{J^2}{\lambda\begin{pmatrix}
        N\\
        p
    \end{pmatrix}}~.
\end{equation}
The double scaling refers to
\begin{equation}\label{eq:double scaling}
    N,~ p \rightarrow \infty~,\quad \lambda = \frac{2p^2}{N}~ \text{fixed}~.
\end{equation}
This model has received much interest in the literature, see e.g. \cite{kitaevTalks,Maldacena:2016hyu,Cotler:2016fpe,Berkooz:2018qkz,Berkooz:2018jqr}. Intriguingly, the DSSYK model has a \emph{maximal entropy state} \cite{Lin:2022rbf}, which is one of the main characteristics of dS space associated with the entropy given by (\ref{eq:GH entropy}) \cite{Chandrasekaran:2022cip}\footnote{Different systems share this characteristic, they can be elegantly studied with the techniques of type II$_1$ von Neumann algebras \cite{Chandrasekaran:2022cip,Lin:2022rbf,Jensen:2023yxy,Kudler-Flam:2023qfl,Witten:2023qsv,Witten:2023xze,Seo:2022pqj,Gomez:2023wrq,Gomez:2023upk,Gomez:2023tkr,Gomez:2023jbg,Aguilar-Gutierrez:2023odp,Basteiro:2024cuh,Xu:2024hoc}. We will not enter into the details about this area. The reader is referred to \cite{Papadodimas:2013jku,Jefferson:2018ksk,Leutheusser:2021frk,Witten:2021unn} for early work on von Neumann algebras in quantum gravity, and \cite{Witten:2018zxz,Witten:2021jzq,Sorce:2023fdx,Casini:2022rlv} for recent reviews.} It was argued in \cite{Narovlansky:2023lfz} that the doubled DSSYK system (\ref{eq:DDSSYK Hamiltonian}) can describe the same correlators as dS$_3$ space once one imposes a Hamiltonian constraint on the physical states (i.e. gauge and diffeomorphism invariant states) of the system,
    \begin{equation}\label{eq:phy states}
        (H^{\rm L}-H^{\rm R})\ket{\psi_{\rm phys}}=0~,
    \end{equation}
which translates to the requirement that for the physical operators acting on the system,
    \begin{equation}\label{eq:phy ops}
        [H^{\rm L}-H^{\rm R},~\mathcal{O}_{\rm phys}]=0~.
    \end{equation}
Interestingly, the maximal entropy state corresponds to an energy eigenstate 
\begin{equation}\label{eq:ref energy state}
    \ket{\mathbb{E}_0}\equiv\ket{E_0^{\rm L},~E_0^{\rm R}}~.
\end{equation}
Under these considerations, it was found in \cite{Verlinde:2024znh} that one can develop a dictionary between the doubled DSSYK model and (2+1)-dimensional SdS$_3$ space, even away from the $G_N\rightarrow0$ regime previously employed in \cite{Narovlansky:2023lfz}. The holographic dictionary so far has succeeded in matching partition functions, correlators, and quasinormal modes of dS space. The state in (\ref{eq:ref energy state}) has been identified with the maximal entropy state of dS space, $\ket{\psi_{\rm dS}}$. {According to the interpretation in \cite{Verlinde:2024znh}, the microscopic theory dual might be located on the cosmological horizon, given that $E_0$ corresponds to the maximum of the spectral density $\rho(E)=\rme^{S(E)}$; or along the worldline of the observers.}

It might be surprising for the reader that there is a duality between 3D gravity (SdS$_3$) and a quantum mechanical theory (DSSYK), in contrast, for instance, to the holographic dictionary between (nearly)-AdS$_2$ space \cite{Maldacena:2016upp}, described by Jackiw–Teitelboim (JT) gravity \cite{JACKIW1985343,TEITELBOIM198341}, with the triple scaling limit of the SYK model (i.e. $\lambda\ll1$ and energies $E/J\ll1$) \cite{Sachdev:2010um,Maldacena:2016hyu}. In \cite{Verlinde:2024zrh}, it was shown that there is an alternative procedure in the CS quantization of SdS$_3$, depending on the order when the physical constraints are imposed. This results in a third member of the correspondence, a two-dimensional gravity theory that will be referred to as Liouville-de Sitter (LdS) in the remainder of the paper. This theory in Lorentzian-signature is defined in terms of two space-like Liouville-CFT$_2$,\footnote{See \cite{Nakayama:2004vk,Teschner:2001rv,Vargas:2017swx} for reviews, and \cite{Polyakov:1981rd} for initial work in this area.} as
\begin{equation}\label{eq:action LdS}
    \begin{aligned}
        I=&I[\phi_+]+I[\phi_-]~,\\
        I[\phi_\pm]=&\frac{1}{4\pi}\int_{\Sigma}\rmd^2\sigma~\sqrt{\abs{h}}\qty[h^{\mu\nu}\partial_\mu\phi_\pm{\partial_\nu}\phi_\pm+Q_\pm \mathcal{R}_h\phi_\pm+\mu_{\rm B}\rme^{2b_\pm\phi_\pm}]\\
        &+\frac{1}{2\pi}\int_{\partial\Sigma}\rmd\tau~ \abs{h}^{1/4}(Q_\pm k+\mu_B\rme^{b_\pm\phi_\pm})
    \end{aligned}
\end{equation}
where $\Sigma$ is the boundary manifold (such that $\partial\Sigma$ corresponds to the geodesic of S or N pole worldline observer in SdS$_3$ space \cite{Verlinde:2024zrh}); $\mu_B$ is called the boundary cosmological constant, which parametrizes the boundary conditions of the theory; $Q_\pm=b_\pm+b_\pm^{-1}$ is the background charge; $h_{ij}$ the boundary metric; $\mathcal{R}_h$ its scalar curvature; $k$ the boundary curvature; $\tau$ is a time-like coordinate along $\partial{\Sigma}$, and $b_\pm\in\mathbb{C}$ are constants which obey $b_+=(b_-)^*$ and $b_+^2\in\rmi\mathbb{R}$. Reflecting boundary conditions along $\partial\Sigma$ are imposed, corresponding to Fateev-Zamolodchikov-Zamolodchikov-Teschner (FZZT) branes \cite{Fateev:2000ik,Teschner:2000md} along the boundary, whose state is specified by $\mu_B$.

Upon quantization, the central charge of the $\pm$ sectors is complex, while the complete theory has a real central charge, given by
\begin{equation}
    c_\pm=1+Q_\pm^2~,\quad c_{+}+c_-=26~.
\end{equation}
It was found in \cite{Verlinde:2024zrh} that the correlation functions of physical operators (see Sec. \ref{sec:Krylov}) in this theory agree with those in the doubled DSSYK model, which together with the original description of SdS$_3$ space, provide compelling evidence for a holographic triality.

\subsection*{Main question: How to define complexity in dS space?}
An exciting possibility from this recent holographic framework in terms of the DSSYK model is to study notions of quantum information theory for SdS$_3$ space and the Liouville CFT side from the quantum mechanical dual description. Concretely, we ask:
\begin{quote}
\emph{Can the doubled DSSYK model provide first principles to properly define quantum information-theoretic notions of complexity in dS space?}
\end{quote}
There are different notions of complexity in quantum information theory. One of the most commonly used, computational complexity, can be defined in terms of states or operators (see \cite{Chapman:2021jbh} for a review). In the state definition, it is a measure of the difficulty of building a target state from a reference state by applying a given set of elementary operations. In terms of operators in quantum circuits, it is defined as the number of elementary gates, a discrete set of unitary operators, from a universal gate set, that is needed to model a particular unitary operator to a given precision \cite{nielsen_2010}. 

Complexity in quantum information theory plays a crucial role in establishing the advantages of quantum over classical computation; in classifying computational problems for algorithm optimization; as a measure of quantum chaos in many-body systems; among different uses in quantum mechanics and field theory \cite{Caputa:2017yrh,Jefferson:2017sdb,Chapman:2017rqy,Bhattacharyya:2018wym,Chapman:2018hou,Camargo:2018eof,Ge:2019mjt,Brown:2019whu,Balasubramanian:2019wgd,Chapman:2019clq,Caputa:2018kdj,Auzzi:2020idm,Caginalp:2020tzw,Flory:2020eot,Flory:2020dja,Chagnet:2021uvi,Basteiro:2021ene,Brown:2021uov,Balasubramanian:2021mxo,Brown:2022phc,Erdmenger:2022lov,Baiguera:2023bhm,Ali:2018fcz,Bhattacharyya:2018bbv,Bhattacharyya:2019kvj,Ali:2019zcj,Bhattacharyya:2019txx,Bhattacharyya:2020iic,Bhattacharyya:2022ren,Bhattacharyya:2023sjr,Bhattacharyya:2020art,Bhattacharyya:2020rpy,Bhattacharyya:2020kgu,Bhattacharyya:2024duw}. While computational complexity has several practical uses, it also suffers from several ambiguities in its definition due to the dependence on the details about reference and the type of elementary operations to reach the target state in the state definition; or related to the type of gate sets and the precision to approximate a given operator.

In the holographic context, several proposals have been motivated to match the state computational complexity of a dual state in a CFT. Importantly, they must capture the late time growth of the wormhole inside an eternal black hole \cite{Susskind:2014moa}. The pioneering proposals include the \emph{complexity equals volume} (CV) \cite{Susskind:2014rva,Stanford:2014jda}, \emph{complexity equals action} (CA) \cite{Brown:2015bva, Brown:2015lvg}, \emph{complexity equals spacetime-volume} (CV2.0) \cite{Couch:2016exn}. Recently, it has been observed that there exists an infinite number of gravitational observables that can all serve as holographic measures of complexity, referred to as \emph{complexity equals anything} (CAny) \cite{Belin:2021bga, Belin:2022xmt}, which are defined to reproduce the main features as computational complexity for a generic quantum circuit (although without accounting for the saturation of complexity due to finite system sizes), i.e.~a late time linear growth, and the switchback effect \cite{Stanford:2014jda}, which is a decrease in complexity growth due to perturbations.

In relation to dS space, there have been several studies about the behavior of the previous holographic complexity proposals (developed for AdS black holes) when applying them in SdS$_{d+1}$ space for observables that are anchored to the stretched horizon. Originally, in \cite{Jorstad:2022mls} (see related discussions in \cite{Susskind:2021esx,Chapman:2021eyy,Auzzi:2023qbm,Anegawa:2023wrk,Anegawa:2023dad,Baiguera:2023tpt}) it was discovered that certain proposals (including CV, CA, CV2.0) lead to hyperfast scrambling, which is defined as 
\begin{align}\label{eq:hyperfast}
\lim_{t\rightarrow t_c}\dv{\mathcal{C}}{t}\rightarrow\infty    
\end{align}
where $\mathcal{C}$ represents the holographic complexity observable computed for a given proposal, and $t_c$ is a critical (stretched horizon) time. However, there is a different set of proposals within the CAny framework for codimension-one extremal surfaces where instead there is an eternal late time growth in asymptotically dS spacetimes \cite{Aguilar-Gutierrez:2023zqm,Aguilar-Gutierrez:2023tic,Aguilar-Gutierrez:2023pnn,Aguilar-Gutierrez:2024rka,Aguilar-Gutierrez:2024rka}). Microscopically, (\ref{eq:hyperfast}) could be interpreted as a very fast scrambling of the degrees of freedom of the dual theory \cite{Susskind:2021esx}, faster than in maximally chaotic systems \cite{Maldacena:2015waa} (see Sec. \ref{sec:Krylov} for related comments).

Given that the previous studies have considered different gravitational observables without a clear holographic description in terms of complexity, our work aims to examine some microscopic notions of complexity and interpret their bulk description from the dS holographic dictionary based on the series of works in \cite{Narovlansky:2023lfz,Verlinde:2024zrh,Verlinde:2024znh}, and compare their evolution with the previous holographic complexity proposals.

\subsection*{Spread, Krylov, query and Nielsen complexity}
In this work, we will be particularly interested in concrete microscopic complexity proposals in connection with the doubled DSSYK model and its duals.

\emph{Spread complexity} \cite{Balasubramanian:2022tpr}, and \emph{Krylov complexity} \cite{Parker:2018yvk} are commonly used definitions of complexity that probe quantum chaos in generic quantum systems. Spread and Krylov complexity of a time-evolved state or operator respectively describe the average position along a 1D chain of ordered basis of states or operators. The spread complexity of a time-evolved pure state $\ket{\phi(t)}$ is defined as \cite{Balasubramanian:2022tpr}
\begin{equation}\label{eq:spread complexity}
    \mathcal{C}_{\rm S}(t)\equiv\sum_n n\abs{\bra{\phi(t)}\ket{K_n}}^2~,
\end{equation}
where $\ket{K_n}$ is the orthonormal, ordered Krylov basis. There is a similar definition for Krylov complexity of operators in terms of a Krylov basis, which we review in Sec. \ref{ssec:def Krylov} (a more complete review is found in \cite{Nandy:2024htc}). Importantly, it has been conjectured, based on different numerical and analytic results \cite{Parker:2018yvk} (see also \cite{Barbon:2019wsy,Caputa:2021sib}), that Krylov complexity can grow at most exponentially with time in maximally chaotic systems, where the exponent is proportional to the maximal Lyapunov exponent \cite{Maldacena:2015waa,Murthy:2019fgs} of out-of-ordered time correlators (OTOCs) \cite{Rozenbaum:2016mmv}. However, the exponential behavior of Krylov complexity can also appear in certain integrable systems in their early time regime \cite{Bhattacharjee:2022vlt}, and in free CFTs at late times \cite{Avdoshkin:2022xuw}. Nevertheless, it has been argued to be a commonly reliable probe of chaotic systems in their late time evolution \cite{Erdmenger:2023wjg}. A significant advantage of spread and Krylov complexity over other definitions is that they are unambiguously defined once the initial state or operator is specified, and they have already found numerous applications, e.g.~\cite{Bhattacharjee:2023dik,Chattopadhyay:2023fob,Camargo:2024deu,Aguilar-Gutierrez:2023nyk,Caputa:2022eye,Afrasiar:2022efk,Caputa:2022yju,Pal:2023yik,Barbon:2019wsy,Bhattacharjee:2022vlt,Dymarsky:2019elm,Dymarsky:2021bjq,Kundu:2023hbk,Bhattacharya:2022gbz,Mohan:2023btr,Yates:2021asz,Caputa:2021ori,Patramanis:2021lkx,Trigueros:2021rwj,Rabinovici:2020ryf,Rabinovici:2021qqt,Rabinovici:2022beu,Bhattacharjee:2022qjw,Chattopadhyay:2023fob,Bhattacharjee:2023dik,Bhattacharjee:2022ave,Takahashi:2023nkt,Camargo:2022rnt,Avdoshkin:2022xuw,Erdmenger:2023wjg,Hashimoto:2023swv,Camargo:2023eev,Iizuka:2023pov,Caputa:2023vyr,Fan:2023ohh,Vasli:2023syq,Gautam:2023bcm,Iizuka:2023fba,Huh:2023jxt,Anegawa:2024wov,Caputa:2024vrn,Chen:2024imd,Caputa:2024xkp,Chattopadhyay:2024pdj,Bhattacharya:2023yec,Bhattacharya:2024hto,Basu:2024tgg,Tang:2023ocr,Banerjee:2022ime,Nandy:2023brt,Bhattacharyya:2023grv,Bhattacharyya:2023dhp}. Recent discussions on the connections between these notions can be found in \cite{Alishahiha:2022anw,Caputa:2024vrn}. 

Importantly, in the AdS holographic context \cite{Rabinovici:2023yex} (see also \cite{Lin:2022rbf}) it was found that the spread complexity in the triple scaling limit of the SYK model, for a particular reference state (interpreted as the thermofield double state of the model), has a bulk interpretation in terms of a regularized geodesic distance between the asymptotic boundaries of a doubled sided black hole (i.e. wormhole length) in JT gravity.

On the other hand, {there are some first principle approaches to defining complexity with holographic CFTs, including \cite{Chen:2020nlj,Chen:2022fbg,Erdmenger:2022lov}.} In particular, the work \cite{Chen:2020nlj} has a natural proposal for state complexity denoted as ``\emph{query complexity}". It is defined as the number of steps taken in algorithm computing multipoint correlators through an iterative application of fusion rules in the CFT, which we express
\begin{equation}\label{eq:Query n fusions}
    \boxed{\mathcal{C}_{\rm Q}=\text{number~of~ fusions}~.}
\end{equation}
The algorithm can be translated into the language of CS theory and Wilson loops. This proposal was initially developed in the context of global AdS$_3$ space/vacuum CFT$_2$. The bulk interpretation of $\mathcal{C}_{\rm Q}$ can be expressed in terms of mean curvature and torsion, as we will discuss in Sec. \ref{ssec:query}.

In contrast, there is a more ambiguous notion of complexity, which provides upper and lower bounds to the computational complexity for quantum circuits \cite{Nielsen1}, known as Nielsen complexity (see also \cite{Nielsen2,Nielsen3,nielsen_2010}). In this geometric approach, circuit complexity is approximated by geodesics distances in a Lie group manifold that replaces a discrete set of gates approximating unitary operators, where the trajectories are generated by time-dependent Hamiltonians. Nielsen complexity corresponds to the minimal length of a geodesic curve connecting a target unitary operator and the identity operator, $\mathbb{1}$. A similar notion of Nielsen complexity can be introduced for states (see e.g. \cite{Chapman:2021jbh}). Although this method can be useful for studying the evolution of quantum circuits; in practice, it can be unfeasible to evaluate in many body systems, except for very simple cases. Several approximation methods have been proposed to obtain bounds on $\mathcal{C}_{\rm N}$, see e.g. \cite{Balasubramanian:2019wgd,Craps:2022ese,Craps:2023rur,Craps:2023ivc}. We will consider a concrete definition where explicit evaluations can be performed. Despite ambiguities, there are robust features about the scaling of circuit complexity with system size, which has motivated the definition of the CAny conjectures \cite{Belin:2021bga,Belin:2022xmt} for holographic complexity in asymptotically AdS spacetimes.

These different approaches to complexity and their connection with the AdS/CFT dictionary have motivated our study within the doubled DSSYK model and its duals, which we summarize below.

\subsection*{Outline of the paper}
The main purpose of our work is to study concrete notions of complexity in dS space based on the microscopic dual theories identified in \cite{Narovlansky:2023lfz,Verlinde:2024znh,Verlinde:2024zrh}, as well as to develop its geometrical interpretation in the bulk. Concretely, we study the definitions of spread, Krylov, and query complexity on all sides of the dS correspondence, and a particular proposal of Nielsen complexity in the DSSYK model.

First, using the doubled Hilbert space formalism of \cite{Okuyama:2024yya} we provide a natural interpretation of spread complexity in the doubled DSSYK model as a map that counts the number of entangled chord states, which are projected onto the maximal entropy state (\ref{eq:ref energy state}). In this formalism, there is a description of the DSSYK model reminiscent of a \textit{multi-scale entanglement renormalization ansatz} ({MERA}) network \cite{Vidal:2007hda,Vidal:2008zz}, as noticed by \cite{Okuyama:2024yya}. We use a known entry in the holographic dictionary relating chord number with a geodesic length\footnote{Note however, that this is not in contradiction with \cite{Aguilar-Gutierrez:2023nyk}, where it was found that spread and Krylov complexity cannot represent distances in metric spaces, as a geodesic length between two points is not the same as their distance.} measuring static patch time difference between the N and S poles in SdS$_3$ space \cite{Verlinde:2024znh}, to interpret boost symmetries in SdS$_3$ space in terms of spread complexity in the doubled DSSYK model. Since spread complexity counts entangled chord states in the doubled Hilbert space, our study provides a connection between entanglement and geometry similar to the ER=EPR conjecture \cite{Maldacena:2013xja}.

Secondly, we study the notion of Krylov complexity for physical operators. To our convenience, the correlation functions of physical operators in the maximal entropy state have been computed and matched by \cite{Narovlansky:2023lfz,Verlinde:2024znh,Verlinde:2024zrh}. Since Krylov complexity is completely determined in terms of the correlation functions of the chosen initial operator (see Sec. \ref{ssec:def Krylov}), our results describe the same physical operator growth for the doubled DSSYK model/LdS$_2$ CFT/SdS$_3$ space. We show that this complexity proposal displays an exponential growth behavior with respect to physical time, as expected for maximally chaotic systems. Its evolution is quite similar to previous studies on the Krylov complexity of the DSSYK model \cite{Parker:2018yvk,Bhattacharjee:2022ave}. Some details differ with respect to previous studies, given that we evaluate Krylov complexity for physical operators obeying the constraint (\ref{eq:phy ops}) instead of the $\psi_i$ fields themselves.

Later, we study the notion of query complexity for the LdS$_2$ CFT, originally proposed by \cite{Chen:2020nlj} for vacuum CFT$_2$ states dual to AdS$_3$ space. In this context, query complexity is the number of steps taken in an algorithm that reproduces multipoint correlators based on fusion rules in the CFT. However, our implementation of the type of correlation functions differs substantially from \cite{Chen:2020nlj} in that we consider pairwise contractions between matter operators inserted on the north and south poles of SdS$_3$ space (corresponding to the boundaries of the LdS$_2$ CFTs), instead of operators within a single AdS global time slice. Moreover, using the dictionary in \cite{Narovlansky:2023lfz,Verlinde:2024znh,Verlinde:2024zrh}, the protocol can be interpreted in terms of the doubled DSSYK model as counting the number of connections of matter operator chords in a cylinder amplitude, where the ends of the cylinder correspond to the pairs of DSSYK theories. Like in spread complexity, the correlation functions can be expressed in terms of MERA entangler and disentangler operators between the pairs of DSSYK theories. Lastly, we argue that query complexity in the bulk theory corresponds to the number of junctions of Wilson lines connecting the north and south poles of SdS$_3$ space in its CS formulation, which computes the correlation functions, and we describe how this approach relates to the holographic complexity proposals in asymptotically dS space \cite{Jorstad:2022mls,Auzzi:2023qbm,Anegawa:2023wrk,Anegawa:2023dad,Baiguera:2023tpt,Aguilar-Gutierrez:2023zqm,Aguilar-Gutierrez:2023tic,Aguilar-Gutierrez:2023pnn,Aguilar-Gutierrez:2024rka}.

Finally, motivated by the connections between computational complexity and holography, we study a particular notion of Nielsen operator complexity in the DSSYK model. We adopt a measure of complexity that is invariant under unitary transformations and time reversal, which allows for a tractable evaluation. We find a linear time growth, as expected for generic chaotic systems (see e.g. \cite{Chapman:2021jbh}.) Although, in contrast with the other proposals, we do not identify a holographic dual description, we notice that this type of evolution is compatible with certain holographic complexity proposals in asymptotically dS spacetimes \cite{Aguilar-Gutierrez:2023zqm,Aguilar-Gutierrez:2023tic,Aguilar-Gutierrez:2023pnn,Aguilar-Gutierrez:2024rka}. We then evaluate the low-energy limit of Nielsen complexity in the DSSYK, which is appropriate for describing JT gravity.

The structure of the manuscript is as follows. In Sec. \ref{sec:background} we provide some background material on the DSSYK model, its connection with MERA tensor networks  \cite{Okuyama:2024yya}, and we explain the definitions of spread, Krylov, query, and Nielsen complexity. The new results start in Sec. \ref{sec:Spread}, where we evaluate the spread complexity of the maximal entropy state (\ref{eq:ref energy state}) in the Hilbert space of the doubled DSSYK model and study its holographic manifestations. Sec. \ref{sec:Krylov} is dedicated to deriving the Krylov complexity for the doubled DSSYK model, LdS$_2$ CFT and SdS$_3$ space using the known correlation functions in all sides of the correspondence \cite{Verlinde:2024znh,Verlinde:2024zrh}, and analyzing its late time behavior, which follows that of a maximally chaotic system. In Sec. \ref{sec:Query}, we use the proposal for query complexity, originally developed in \cite{Chen:2020nlj}, to define complexity in the LdS$_2$ CFT, and study its manifestation in terms of chord diagrams in the DSSYK model, and bulk invariant quantities in SdS$_3$ space. Sec. \ref{sec:comments Nielsen} contains new results on Nielsen's geometric approach to operator complexity in the DSSYK model, and some comments about its relation with dS holography, and JT gravity. Finally, Sec. \ref{sec:conclusions} contains a summary of the findings and some outlook questions.

\section{Background material on chords and complexity}\label{sec:background}
This section provides the necessary background material on the DSSYK model, spread, Krylov, query, and Nielsen complexity to derive the new results in the following sections, and it also serves to introduce the notation.

\subsection{Review of the chord Hilbert space of the DSSYK model}
We will be interested in ensemble-averaged moments of the Hamiltonian, denoted as $\expval{\tr[H^k]}_J$. This consists of all pairwise Wick contractions between Hamiltonians where one performs an averaging over the Gaussian couplings, $J_{i_1\dots i_p}$, in (\ref{eq:GEA J}). It was discovered in \cite{Erd_s_2014,Cotler:2016fpe,Berkooz:2018qkz,Berkooz:2018jqr} that one can perform these evaluations using \textit{chord diagrams}, which are segments or circles with nodes that are connected in pairs by lines (chords). The rules in this expansion are deduced by appropriately commuting the Majorana fermions inside the trace of the moments (see \cite{Berkooz:2018jqr} for details) and considering the average over couplings, which reduces to a counting problem of the different contractions in the Hamiltonian moments. This can be expressed as:
\begin{equation}\label{eq:chord tech}
    \expval{\tr(H^k)}_J=\frac{J^{2k}}{\lambda^k}\sum_{\text{chord diagrams}}q^{\#(H\cap H)}~,
\end{equation}
where $q=\rme^{-\lambda}$, and \# is a shorthand for ``number of". This means that then there will be a relative weight $q^n$ when any given chord intercepts with $n$ other chords.

\begin{figure}[t!]
    \centering
    \includegraphics[width=0.35\textwidth]{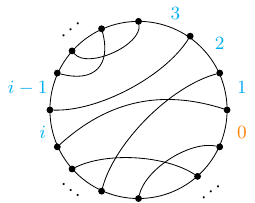}\hspace{0.5cm}\includegraphics[width=0.6\textwidth]{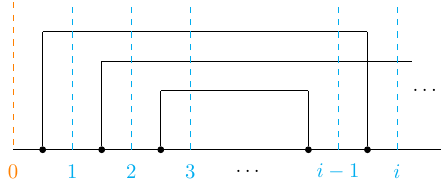}
    \caption{\textit{Left}: Example of disk chord diagram, where we label the different levels (cyan) before each vertex (black dot), where $0$ (orange) represents the level where we will cut the diagram. \textit{Right}: The chord diagram is sliced open (each level is represented with a dashed line). Each chord is a Wick contraction between the nodes (black dots) corresponding to the Hamiltonians in (\ref{eq:chord tech}), which can then end on the subsequent levels.}
    \label{fig:SlicedChordDiagram}
\end{figure}

We will give a brief overview (mostly based on \cite{Berkooz:2018jqr,Rabinovici:2023yex}) of how to use (\ref{eq:chord tech}) to evaluate amplitudes that only involve the Hamiltonian moments. First, consider slicing open the chord diagram at any chosen point, so that the total number of nodes (which depends on how many closed or open chords we consider) lie on a line rather than a circle, as shown in Fig. \ref{fig:SlicedChordDiagram}. This represents a transition from a state without chords, to one with $k$ chords, which transitions back to one without chords.

\begin{figure}[t!]
    \centering
    \includegraphics[width=0.35\textwidth]{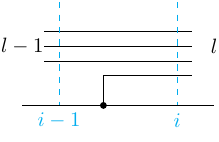}\hspace{0.75cm}\includegraphics[width=0.35\textwidth]{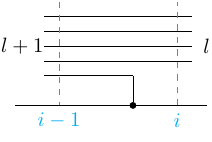}
    \caption{Two ways to end with $l$ open chords after vertex $i$. \textit{Left}: $l-1$ open chords before vertex $i$. \textit{Right}: $l+1$ open chords before vertex $i$.}
    \label{fig:OpenClosedChords}
\end{figure}

Let $v_{l}^{(i)}$ denote the sum of chord diagrams with $l$ open chords at the $i$-th vertex in the sliced amplitude, and starting at the zeroth vertex. For a generic vertex $i$ with $l$ open chords immediately after it, one has 2 possibilities (Fig. \ref{fig:OpenClosedChords}): (a) that $l-1$ of them were open at level $i-1$ and one chord opens just before the vertex $i$, and (b) that $l+1$ of them were open at level $i-1$ and one chord is closed at vertex $i$. In the latter case, the chord that closes off might cross with any of the other $l$ open chords. 
Considering the factor (\ref{eq:chord tech}), the recursion relation for the total number of \emph{involuntary interceptions} at a given vertex becomes
\begin{equation}\label{eq:relation v}
    v_{l}^{(i+1)}=\frac{J}{\sqrt{\lambda}}\qty(v_{l-1}^{(i)}+\sum_{j=1}^lq^jv_{l+1}^{(i)})~.
\end{equation}
This can be expressed in terms of the so-called ``\emph{transfer matrix}", $T$, defined by the relation above as $v_{l}^{(i+1)}=Tv_{l}^{(i)}$. We can then represent the sum of chord diagrams by acting with $T$ as
\begin{equation}\label{eq:solu T}
    v^{(i)}_l=T^i\ket{l}~,
\end{equation}
where we are considering $\mathcal{H}=\bigoplus_{l=0}^{\infty}\mathbb{C}\ket{l}$ as the auxiliary \emph{chord Hilbert space}. However, (\ref{eq:relation v}) and (\ref{eq:solu T}) imply that $T$ would not be symmetric on this basis. We will pick $\qty{\ket{n}}$ to be a orthonormalized chord basis (i.e. $\bra{n}\ket{m}=\delta_{nm}$), such that $T$ is symmetric in this basis
\begin{align}
    T\ket{n}&=\frac{J}{\sqrt{\lambda}}\qty(\sqrt{\frac{1-q^n}{1-q}}\ket{n-1}+\sqrt{\frac{1-q^{n+1}}{1-q}}\ket{n+1})~.\label{eq:Transfer matrix}
\end{align}
Moreover, we can define the following operators:
\begin{equation}\label{eq: Oscillators 1}
    A\ket{n}=\sqrt{\frac{1-q^n}{1-q}}\ket{n-1}~,\quad A^\dagger\ket{n}=\sqrt{\frac{1-q^{n+1}}{1-q}}\ket{n+1}~,
\end{equation}
which obey the q-deformed commutation relation:
\begin{equation}\label{eq: Oscillators 2}
    \qty[A,~A^\dagger]_q\equiv AA^\dagger-qA^\dagger A=1~.
\end{equation}
$T$ can be then described in terms of a q-deformed harmonic oscillator as:
\begin{equation}\label{eq:H DDSSYK momnetum}
\begin{aligned}
T&=\frac{J}{\sqrt{\lambda}}(A+A^\dagger)~.
\end{aligned}
\end{equation}
Thus, the chord Hilbert space can be seen as the Fock space of the q-deformed oscillator. For our later discussion, it is convenient to introduce the chord number operator $\hat{n}$ and its conjugate momentum, $p$:
\begin{equation}\label{eq:Normalized ops}
A=\rme^{\rmi p}\sqrt{\frac{1-q^{\hat{n}}}{1-q}},\quad A^\dagger=\sqrt{\frac{1-q^{\hat{n}}}{1-q}}\rme^{-\rmi p}~,
\end{equation}
which obey the relations
\begin{equation}\label{eq:chord numb op}
\begin{aligned}
    &\qty[\hat{n},~\rme^{\rmi p}]_1=\rme^{\rmi p}~,\quad\qty[\hat{n},~\rme^{-\rmi p}]_1=-\rme^{-\rmi p}~,\quad\hat{n}\ket{n}=n\ket{n}~.
    \end{aligned}
\end{equation}
The transfer matrix takes the form
\begin{equation}\label{eq:H DDSSYK momnetum2}
\begin{aligned}
T=&\frac{J}{\sqrt{\lambda(1-q)}}\qty(\rme^{\rmi p}\sqrt{1-q^{\hat{n}}}+\sqrt{1-q^{\hat{n}}}\rme^{-\rmi p})~.
\end{aligned}
\end{equation}
Furthermore, there is a special basis $\ket{\theta}$ where $T$ is diagonal, which is related to the eigenvalues of the Hamiltonian (\ref{eq:DDSSYK Hamiltonian}), $E(\theta)$, given as \cite{Berkooz:2018jqr,Berkooz:2018qkz}
\begin{equation}
    T\ket{\theta}=-E(\theta)\ket{\theta},\quad E(\theta)=-\frac{2J\cos(\theta)}{\sqrt{\lambda(1-q)}}
\end{equation}
with $\theta\in[0,~\pi]$. This angular basis is related to the chord number basis $\qty{\ket{n}}$ through q-Hermite polynomials:
\begin{equation}\label{eq: proj E0 theta}
    \bra{\theta}\ket{n}=\frac{H_n(\cos\theta|q)}{\sqrt{(q;~q)_n}}~,
\end{equation}
where $(a;~q)_n$ is the q-Pochhammer symbol:
\begin{equation}\label{eq:q-Pochhammer}
    (a;~q)_n\equiv\prod_{k=0}^{n-1}(1-aq^k)~,\quad (a_0,\dots, a_N;q)_n=\prod_{i=1}^N(a_i;~q)~,
\end{equation}
and $H_n(x|q)$ is the q-Hermite polynomial, which can be expressed as
\begin{equation}\label{eq:H_n def}
    H_n(\cos\theta|q)=\sum_{k=0}^n\begin{bmatrix}
        n\\
        k
    \end{bmatrix}_q\rme^{\rmi(n-2k)\theta}~,\quad
    \begin{bmatrix}
        n\\
        k
    \end{bmatrix}_{q}\equiv\frac{(q;~q)_{n}}{(q;~q)_{n-k}(q;~q)_{k}}~.
\end{equation}
The $\ket{\theta}$ basis is normalized such that
\begin{align}
\bra{\theta}\ket{\theta_0}&=\frac{2\pi}{\mu(\theta)}\delta(\theta-\theta_0)~,\quad\mu(\theta)=(q,~\rme^{\pm 2 \rmi \theta};q)_\infty\ ,\quad\label{eq:norm theta}\\
    \mathbb{1}&=\int_0^\pi\frac{\rmd\theta}{2\pi}\mu(\theta)\ket{\theta}\bra{\theta}~,\label{eq:identity theta}
\end{align}
where we introduced the notation $g(\pm x\pm y)=g(x+y)g(-x+y)g(x-y)g(x-y)$.

So far, our review has been focused on the counting rules for the Hamiltonian moments (\ref{eq:chord tech}). We present the results once matter operators (also called matter chords) are included, which we consider to have the form:
\begin{align}\label{eq:matter ops DSSYK}
\mathcal{O}_\Delta(t)&=\rmi^{\frac{p'}{2}}\sum_{i_1,\dots~,i_{p'}}K_{i_1\dots i_{p'}}\psi_{i_1}(t)\dots\psi_{i_{p'}}(t)~.
    \end{align}
    Here $K_{i_1\dots i_{p'}}$ are Gaussian random couplings, independent of $J_{i_1\dots i_{p}}$, and $\Delta\equiv p'/p$. Now, one has to account for the $H$ and $\mathcal{O}$-nodes where we perform the Wick contractions, and average over random couplings $J_{i_1\dots i_{p}}$ and $K_{i_1\dots i_{p'}}$. It has been found in \cite{Berkooz:2018jqr,Berkooz:2018qkz} that even n-point correlation functions can be expressed in terms of a counting problem similar to (\ref{eq:chord tech}), which takes the form
\begin{equation}
    \expval{\tr(H^{k_1}\mathcal{O}_\Delta(t_1)\dots H^{k_n}\mathcal{O}_\Delta(t_n)H^{k_{n+1}})}_{J,~K}\propto\sum_{\rm chord~diagrams}q^{\#(H\cap H)}q^{\Delta\#(H\cap\mathcal{O})}q^{\Delta^2\#(\mathcal{O}\cap \mathcal{O})}~.
\end{equation}
To simplify the evaluations, we assume there are no intersections of the form $(\mathcal{O}\cap\mathcal{O})$, which can be justified e.g. by considering bulk-free fields in SdS$_3$ space dual to the matter operator $\mathcal{O}^{\rm phys}_\Delta$.

Matter correlators will be involved in our discussions about Krylov and query complexity in Sec. \ref{sec:Krylov} and \ref{sec:Query} respectively.

\subsubsection{The doubled Hilbert space}\label{ssec:HxH and MERA}
We now will introduce a doubled Hilbert space description for the chord number states $\qty{\ket{n}}$. This allows a connection between the DSSYK model with tensor networks, as recently argued by \cite{Okuyama:2024yya}. This description will be useful to study the spread complexity in the doubled DSSYK model in the following section.

Let us denote $X$ as an arbitrary operator acting on the Hilbert space $\mathcal{H}$ of one of the DSSYK models. We introduce its state representation in a doubled Hilbert space $\mathcal{H}\otimes\mathcal{H}$ in the chord number basis as
\begin{equation}
\begin{aligned}
\hat{X}&=\sum_{m,\,n}X_{nm}\ket{m}\bra{n}~,\\
    |{X})&=\sum_{m,~n}X_{nm}\ket{m,~n}~,\label{eq:Op as state Oku}
    \end{aligned}
\end{equation}
where $X_{nm}\equiv\bra{m}X\ket{n}$ $\ket{m,~n}\equiv\ket{m}\otimes\ket{n}$. Notice that the inner product between operators $({Y}|{X})$ is determined by the chord basis elements.

Given that the chord number basis is orthonormal, one can then represent the identity operator as
\begin{equation}
    |\mathbb{1})=\sum_{n=0}^\infty \ket{n,~n}=\mathcal{E}\ket{0,~0}
\end{equation}
where
\begin{equation}\label{eq:entangler}
    \mathcal{E}=\sum_n\frac{(A^\dagger\otimes A^\dagger)^n}{(q;~q)_{n}}~.
\end{equation}
Manifestly, $\mathcal{E}$ maximally entangles the vacuum state $\ket{0,~0}$. It was pointed out \cite{Okuyama:2024yya} that $\mathcal{E}$ is reminiscent of an entangler operator in a MERA network \cite{Vidal:2007hda,Vidal:2008zz}. 

\subsection{Notions of complexity}
We briefly review the definition of the concrete proposals that we will examine in the main text: spread complexity \cite{Balasubramanian:2022tpr}; Krylov complexity \cite{Parker:2018yvk}; query complexity \cite{Chen:2020nlj}; and Nielsen complexity \cite{Nielsen1}.

\subsubsection{Spread complexity}\label{ssec:spread}
Starting from the Schrödinger picture for a generic pure quantum system, we would like to construct an ordered, orthonormal basis of states $\qty{\ket{B_n}}$ that minimizes $\sum_n c_n\abs{\bra{\phi(t)}\ket{B_n}}^2$ where c$_n$ is an arbitrary monotonically increasing real sequence, and
\begin{equation}
     \ket{\phi(t)}=\rme^{-\rmi Ht}\ket{\phi_0}~.
\end{equation}
It was found in \cite{Balasubramanian:2022tpr} that the solution to this problem is the so-called Krylov basis, $\ket{K_n}$, defined through the Lanczos algorithm shown below
\begin{align}
    \ket{A_{n+1}}&\equiv (H - a_n)\ket{K_n} - b_n \ket{K_{n-1}}~, \label{eq:lanczos}\\
    \ket{K_n} &\equiv b_n^{-1}\ket{A_n}~.
\end{align}
Here $\ket{K_0}\equiv\ket{\phi_0}$ and
\begin{equation}
    a_{n} \equiv \bra{K_n} H \ket{K_n}, \qquad b_{n} \equiv(\braket{A_n})^{1/2}~,
\end{equation}
are called the Lanczos coefficients. Using this basis, $\ket{\phi(t)}$ can be expressed as
\begin{equation}\label{eq: phi (t) spread}
    \ket{\phi(t)} = \sum^{\mathcal{K}}_{n=0} \phi_n (t) \ket{K_n}~.
\end{equation}
Here $\mathcal{K}$ denotes the Krylov space dimension, which satisfies $\mathcal{K}\leq D_{\mathcal{H}}$, with $D_{\mathcal{H}}$ the Hilbert space dimension. The Hamiltonian in this basis becomes tridiagonal, and we can express a recursive relation between the time-dependent components in (\ref{eq: phi (t) spread}) as a Schrödinger equation:
\begin{equation} \label{eq:Schro}
    \rmi\partial_t \phi_n(t) = a_n \phi_n (t) + b_{n+1}\phi_{n+1}(t) + b_n \phi_{n-1}(t)~,
\end{equation}
with $\sum_n|\phi_n(t)|^2=1$. Spread complexity is then defined as
\begin{equation} \label{eq:S Complexity}
   \mathcal{C}_{\rm S}(t) \equiv \sum_n n |\phi_n(t)|^2 ~.
\end{equation}
Intuitively, $\mathcal{C}_{\rm S}$ measures the average position in a one-dimensional chain generated by the Krylov basis, where each step along the chain represents an increasingly chaotic state since they roughly behave as $\ket{K_n} \approx H^n \ket{\phi_0}$.

Importantly for us, and as noticed in \cite{Lin:2022rbf,Rabinovici:2023yex}, the Krylov basis of the DSSYK model is given by the chord number of states, i.e. 
\begin{equation}\label{eq:Krylov basis DSSYK}
\ket{K_n}=\ket{n}~,\quad b_n=-J\sqrt{\frac{1-q^n}{\lambda(1-q)}}~,
\end{equation}
since the Hamiltonian (i.e. the transfer matrix (\ref{eq:Transfer matrix}) up to a sign) becomes tridiagonal in this basis. Therefore, spread complexity is related to the chord number operator through the relation
\begin{equation}\label{eq:spread chord number}
    \mathcal{C}_{\rm S}(t)=\bra{\phi(t)}\hat{n}\ket{\phi(t)}~.
\end{equation}
This allowed \cite{Rabinovici:2023yex} to identify the spread complexity of the $\ket{\phi(0)}=\ket{0}$ state with a wormhole length in the JT gravity dual description to the DSSYK model.

\subsubsection{Krylov complexity}\label{ssec:def Krylov}
One can also define a notion of complexity in terms of an ordered Krylov basis for operators in generic quantum systems. Starting from the Heisenberg picture, one may express an $\hat{O}$ in terms of states in operator space from a complete basis of states $\qty{\ket{\chi_n}}$ as
\begin{equation}\label{eq:Op as state Krylov}
\begin{aligned}
  |O)&\equiv\sum_{m,\,n}O_{nm}\ket{\chi_m,~\chi_n}~,
\end{aligned}
\end{equation}
where $O_{nm}\equiv\bra{\chi_m}\hat{O}\ket{\chi_n}$. We will consider the Frobenius product\footnote{Other choices of inner products inherent related to finite temperature ensembles can be found in \cite{Parker:2018yvk,Barbon:2019wsy}.} for defining the inner product of these states as:
\begin{equation}\label{eq:inner prod}
    (X|Y)=\frac{1}{D_{\mathcal{H}}}\tr(X^\dagger Y)~,
\end{equation}
where $D_{\mathcal{H}}$ refers to the Hilbert space dimension. 

We can represent the evolution of the operator through the Heisenberg equation as
\begin{align}\label{eq: Heisenberg time evol}
\partial_t|O(t))&=\rmi\mathcal{L}|O(t))~,
\end{align}
where $\mathcal{L}$ is called the Liouvillian super-operator,
\begin{equation}
    \mathcal{L}=\qty[H,~\cdot~],\quad O(t)=\rme^{\rmi\mathcal{L}t}O~.
\end{equation}
We can then solve (\ref{eq: Heisenberg time evol}) in terms of a Krylov basis, $\qty{|O_n)}$,
\begin{equation}\label{eq:amplitudes}
\begin{aligned}
    |O (t))&=\sum_{n=0}^{\mathcal{K}-1} i^n \varphi_n(t) |O_n)~,\\
    \varphi_n(t)&=(O_n|\rme^{\rmi \mathcal{L}t}|O_n)~,\quad (O_m|O_n)=\delta_{mn}~.
\end{aligned}
\end{equation}
Moreover, assuming that $O(t)$ is a Hermitian operator, the correlation function is an even function in $t$ that can be expanded as a Taylor series as
\begin{equation}\label{eq:2pnt correlator Krylov}
    \varphi_0(t)=(O(t)|O(0))=\sum_nm_{2n}\frac{(-1)^{n}t^{2n}}{(2n)!}~,
\end{equation}
where $m_{2n}$ are referred to as the moments. The Lanczos coefficients $b_n$ can be then determined from the moments using an algorithm \cite{Parker:2018yvk,viswanath1994recursion,Bhattacharjee:2022ave}
\begin{align}\label{eq:Alt Lanczos}
    b_n=\sqrt{Q_{2n}^{(n)}}~,\quad Q_{2k}^{(m)}=\frac{Q_{2k}^{(m-1)}}{b_{m-1}^2}-\frac{Q_{2k-2}^{(m-2)}}{b_{m-2}^2}~,
\end{align}
where $Q_{2k}^{(0)}=m_{2k}$, and $Q_{2k}^{(-1)}=0$.

The other amplitudes can be determined through the Lanczos algorithm and the Heisenberg equation (\ref{eq: Heisenberg time evol}), leading to the recursion relation:
\begin{equation}\label{eq:sch eq K operator}
    \partial_t\varphi_n(t)=b_n\varphi_{n-1}(t)-b_{n+1}\varphi_{n+1}(t)~.
\end{equation}
Krylov-complexity is then defined as
    \begin{equation}\label{eq:Krylov complexity}
        {\mathcal{C}_{\rm K}}(t)\equiv\sum_{n=0}^{\mathcal{K}-1}n|\varphi_n(t)|^2\,.
    \end{equation}
The definition above was originally motivated \cite{Parker:2018yvk} to describe the size of the operator under Hamiltonian evolution, as it measures the mean width of a wavepacket in the Krylov space.

For our purposes, (\ref{eq:Alt Lanczos}) can be straightforwardly applied to study the operator growth in the different sides of the dS holographic correspondence, given that the correlation functions have been previously determined and matched \cite{Narovlansky:2023lfz,Verlinde:2024znh,Verlinde:2024zrh}. See \cite{Parker:2018yvk,Bhattacharjee:2022ave} for previous work on Krylov complexity in the DSSYK model. We come back to this point in Sec. \ref{sec:Krylov}.

\subsubsection{Query complexity}\label{ssec:query}
We would like a notion of state complexity for a CFT that can be naturally adapted to the CS formulation of 3D gravity \cite{Witten:1998qj,Witten:1988hc,Witten:2007kt} so that we can define complexity in the LdS$_2$ CFT. A promising proposal with these characteristics was developed in \cite{Chen:2020nlj} (for global AdS$_3$ gravity), called ''\textit{query complexity}", which is based on the same concept applied in quantum algorithms\footnote{The proposal shares similarities to the initial motivation for the CV conjecture \cite{Susskind:2014rva}. The volume of a codimension-one surface in a spacetime filled with a tensor network essentially counts the total number of tensors, while query complexity counts the number of tensor contractions in a tensor network computing the expectation value of the set of operators in the network. It follows that complexity is heuristically given by the size of the network.} (see a recent review in \cite{ambainis2018understanding}). Intuitively, {query complexity} for a CFT is defined as the number of times that a subroutine in an algorithm computing multipoint correlation functions of the CFTs must be performed. In this subsection, we will briefly review the original proposal in \cite{Chen:2020nlj}, while in Sec. \ref{sec:Query} we discuss how to evaluate it for the different sides of the SdS$_3$ space/LdS$_2$ CFT/doubled DSSYK model correspondence.

We start by defining a state $\rho$ which translates operators in the CFT to expectation values
\begin{equation}
    \rho:\quad \mathcal{O}\rightarrow\tr(\mathcal{O}\rho)~.
\end{equation}
In the holographic context, the location of the cutoff surface in AdS space will modify the domain of the map above.\footnote{We are referring to vacuum AdS space for the present discussion, but one should in principle account for heavy and light states when the proposal is generalized to other holographic CFTs, see comments on this \cite{Chen:2022fbg}.} Without the cutoff state complexity would be trivially infinite. Thus, we would like to implement $\rho$ as an algorithm that reads an input of local CFT operators, $\mathcal{O}(x_1)$, $\mathcal{O}(x_2)$, \dots $\mathcal{O}(x_n)$, where $x_i$ parametrizes the location on the cutoff region; and that it evaluates n-point correlation functions $\bra{0}\mathcal{O}(x_1)\mathcal{O}(x_2)\dots\mathcal{O}(x_n)\ket{0}$, with $\ket{0}$ being the ground state of the CFT. Moreover, if the cutoff surface is performed along a geodesic path, the map $\rho$ cannot take more than one input $\mathcal{O}$, otherwise, it would be outside the constant mode sector of the CFT cutoff.

To study a concrete way of implementing this algorithm, we employ topological gravity in the AdS$_3$/CFT$_2$ setting. The correlation functions above can be repeatedly evaluated through fusion rules in the CFT, corresponding to the junction of Wilson lines in the bulk. We will be using the SL$(2,~\mathbb{R})\times$SL$(2,~\mathbb{R})$ CS formulation of global AdS$_3$ space,
\begin{equation}\label{eq:CS AdS3}
    I=\frac{k}{4\pi}\qty(\int\tr(\mathcal{A}\rmd \mathcal{A}+\frac{2}{3}\mathcal{A}\wedge\mathcal{A}\wedge\mathcal{A})-\int\tr(\bar{\mathcal{A}}\rmd \bar{\mathcal{A}}+\frac{2}{3}\bar{\mathcal{A}}\wedge\bar{\mathcal{A}}\wedge\bar{\mathcal{A}}))~,
\end{equation}
where $k$ is the coupling constant; $(\mathcal{A},~\bar{\mathcal{A}})$ are 1-form gauge fields; and $\tr$ denotes contraction using the Killing forms of the algebra. 

We study Wilson lines of the form
\begin{equation} 
W_R(\gamma)=\tr_{R}\qty(P\exp(-\int_{\gamma}\mathcal{A})P\exp(-\int_{\gamma}\bar{\mathcal{A}}))
\end{equation}
where $R$ denotes a continuous series representation of SL$(2,~\mathbb{R})\times$SL$(2,~\mathbb{R})$, $P$ represents path-ordering along the curve $\gamma$. If the path is closed, one has a Wilson loop, which is trivial in global AdS$_3$, while if $\gamma$ is open, its endpoints of $\gamma$ need to end at the asymptotic boundary to define gauge invariant quantities.

Expectation values of the Wilson lines are evaluated using the representation theory of SL$(2,~\mathbb{R})\times $SL$(2,~\mathbb{R})$, where primary states are denoted by
\begin{equation}
    \ket{h,~\overline{h}}=\mathcal{O}_{h\overline{h}}(z,~\overline{z})\ket{0}~.
\end{equation} 
Now, consider global time slices of AdS$_3$ gravity. Since we are interested in an algorithm computing n-point correlation functions, we study how to combine Wilson lines. We will call a junction of Wilson lines when a pair (or higher number) of Wilson lines merge. It was proposed in \cite{Chen:2020nlj} to define the rule to junction ($\mathcal{J}$) Wilson lines purely in terms of CFT operators by mapping at least two primary states (or their descendants) $\ket{h_1,~\overline{h}_1}$ and $\ket{h_2,~\overline{h}_2}$ to a new one $\ket{h_3,~\overline{h}_3}$ as:
\begin{equation}\label{eq:fusions}
\begin{aligned}
    \mathcal{J}&\qty(\mathcal{O}_{h_1\overline{h}_1}(u,~\bar{u})\ket{0},~\mathcal{O}_{h_2\overline{h}_2}(v,~\bar{v})\ket{0})\\
    &=\int\rmd^2 w\sum_{h_3,~\overline{h}_3}c^{h_3\overline{h}_3}_{h_1\overline{h}_1h_2\overline{h}_2}(u,~\bar{u},~v,~\bar{v},~w,~\bar{w})\mathcal{O}_{h_3\overline{h}_3}(w,~\bar{w})\ket{0}
\end{aligned}
\end{equation}
where the functional dependence of the coefficients $c^{h_3\overline{h}_3}_{h_1\overline{h}_1h_2\overline{h}_2}(u,~\bar{u},~v,~\bar{v},~w,~\bar{w})$ are determined by the transformation rules of SL$(2,~\mathbb{R})\times $SL$(2,~\mathbb{R})$. Namely, the coefficients are invariant under a gauge transformation that affects all the Wilson lines simultaneously, and they need to transform covariantly when the gauge transformation acts only on a single one of the Wilson lines in the junction; similar to the operator product expansion (OPE) of local CFT operators. This definition of junction, together with the equations of motion of (\ref{eq:CS AdS3}) (called flatness conditions) guarantees that the Wilson line network is deformable under diffeomorphisms in the bulk \cite{Chen:2020nlj}; and thus that it computes multipoint correlation functions from the fusion algebra of the CFT (through OPE expansions) if the network is placed on the asymptotic boundary.

\begin{figure}
    \centering
    \includegraphics[width=0.5\textwidth]{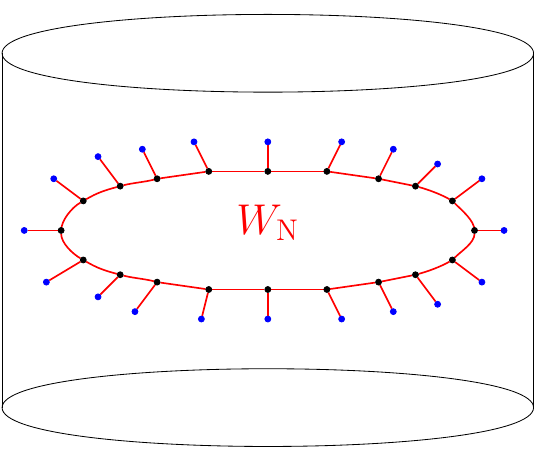}
    \caption{Wilson line network (labeled $W_N$) on a global time slice in pure AdS$_3$ space. The Wilson lines (red lines) have been junctioned together (black dots) according to the rule (\ref{eq:fusions}) and amputated (blue dots) along a cutoff surface in the bulk interior, which is not necessarily at a constant radial location.}
    \label{fig:Network_trunc}
\end{figure}

On the other hand, since maps $\rho$ are defined on a cutoff surface, one can introduce the concept of ``\emph{amputation}". This operation removes the ends of the Wilson lines that extend to the asymptotic boundary up to a given cutoff surface of the open Wilson lines, as illustrated in Fig. \ref{fig:Network_trunc}. In terms of the CFT, this operation represents a renormalization group (RG) flow in the sense that its input is the incoming representation of the local operators from a reference scale (e.g. close to the asymptotic boundary) and coarse grains them to the cutoff scale of the network, whose output will be a number (the n-point correlation function). As such, this operation will not be gauge invariant; instead, it will depend sensitively on the choice of cutoff. {As a remark, notice that if we input a trivial representation of $R$ on any of the open lines of the amputated network, this will reduce the order of the $n$-point correlation function to $n-1$.}

After defining the characteristics of the Wilson line network, one can now define query complexity in terms of the number of times that fusion rules in (\ref{eq:fusions}) are applied to compute a n-point correlation function, or equivalently, the number of junctions in the amputated network in Fig. \ref{fig:Network_trunc}. We can represent this relation as in (\ref{eq:Query n fusions}).

Given that the only differential invariants on the surface introduced in the Wilson line network in a static configuration are the {proper length of the induced curve, $\lambda$; its mean curvature $K$; and torsion $\mathcal{T}$,} \footnote{\label{fn:Powers K T}Arbitrary powers of these differential invariant quantities are in principle allowed, however, they will be ill-defined given that we consider junctions of Wilson lines to form the network, instead of smooth surfaces. Thus, they will not be considered.} it follows that the density of the state complexity will be given by
\begin{equation}\label{eq:complexity dlambda}
    \dv{\mathcal{C}_{\rm Q}}{\lambda}=c_1+c_2 K+c_3\mathcal{T}~,
\end{equation}
where $c_{i}\in\mathbb{R}$ ($i\in\qty{1,~2,~3}$) are constants. 

However, as we discussed at the beginning of the subsection, the map $\rho$ should take no more than one input operator $\mathcal{O}(x)$ if the network, formed by the Wilson lines, follow geodesic trajectories (for which $K=0$, $\mathcal{T}=0$); otherwise, {there would be more than single place where the representations in $R$ could originate from within a single cutoff surface, for which we associate no query complexity to this configuration}. This implies that $c_1=0$ in (\ref{eq:complexity dlambda}). After fixing this constant, one can then integrate (\ref{eq:complexity dlambda}), and use the Gauss-Bonnet theorem at a fixed global time slice:
\begin{equation}\label{eq:Query global time slice}
    \mathcal{C}_{\rm Q}=c_2\qty(-\int \mathcal{R}~\rmd V+2\pi)+c_3\int\rmd\lambda~\mathcal{T}~.
\end{equation}
Given that $\mathcal{R}=-\ell_{\rm AdS}^{-2}$ for pure AdS$_3$ space, then (\ref{eq:Query global time slice}) indicates a relation between query complexity with the CV proposal if one could fix $c_3=0$, although there is no a priori reason for it.

\subsubsection{Nielsen complexity}\label{app:Nielsen}
Nielsen operator complexity was introduced in \cite{Nielsen1} to provide lower and upper bounds on the computational complexity of quantum circuits. For recent reviews, the reader is referred to \cite{Chen:2021lnq,Chapman:2021jbh}; ours will be mostly based on \cite{Yang:2018nda,Yang:2018tpo,Yang:2019iav}.

\begin{figure}
    \centering
    \includegraphics[width=0.4\textwidth]{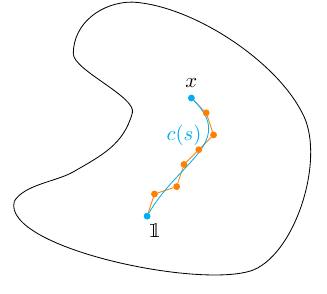}\hspace{0.5cm}\includegraphics[width=0.4\textwidth]{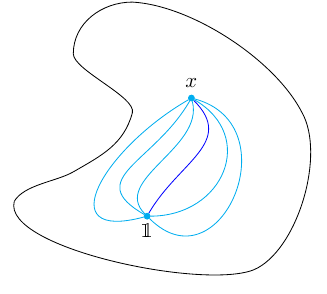}
    \caption{Nielsen's geometric approach to operator complexity. The group manifold of unitary operators (white blob) is approximated as a smooth region. \textit{Left}: A discrete set of elementary gates in a circuit (represented by orange dots) connecting the operators $\mathbb{1}$ and $x\in\;$SU(n) (cyan dots) is approximated through a continuous curve c($s$) (cyan). \textit{Right}: Nielsen operator complexity picks the minimal length geodesic (blue) among all (cyan) of those connecting $\mathbb{1}$ and $x$.}
    \label{fig:Nielsen approach}
\end{figure}

Consider the group manifold of unitary operators SU(n) acting on a finite-dimensional quantum mechanical system (e.g. the Majorana fermions in the DSSYK model). In Nielsen's geometric approach, the discrete nature of this manifold is approximated by a smooth one where continuous paths connect operators. The original motivation \cite{Nielsen1} for doing this is to provide an approximation to the total number of elementary gates of the form that are needed to reproduce an arbitrary unitary operator, $x\in\;$SU(n) to a given precision (relevant in optimization control of quantum circuits). The smooth geometric approximation becomes more accurate when the elementary gates have the form $\delta x=\rme^{-\rmi H\delta s}$, with $\delta s$ an infinitesimal parametrization (e.g. a small time step) of the gate, and $H$ is a generator of $U$ (such as the Hamiltonian). See Fig. \ref{fig:Nielsen approach} for an illustration.

We define \textit{Nielsen complexity} of an operator $x\in\;$SU(n), $\mathcal{C}_{\rm N}(x)$, as the \textit{minimal geodesic length} between the identify operator $\mathbb{1}$ and the operator $x$. This can be expressed as a map SU$(n)\rightarrow \mathbb{R}$, where one can impose certain axioms for it to describe a distance in the space of unitaries \cite{Yang:2018nda,Yang:2018tpo,Yang:2019iav}:
\begin{itemize}
    \item \textit{Non-negativity}:
    \begin{equation}
        \mathcal{C}_{\rm N}(x)\geq0~,~\forall x\in \text{SU(n)}~,\label{eq:non-negative}
    \end{equation}
    where $\mathcal{C}_{\rm N}(\mathbb{1})=0$.
    \item \textit{Triangle inequality}:
    \begin{equation}
        \mathcal{C}_{\rm N}(x)+\mathcal{C}_{\rm N}(y)\geq\mathcal{C}_{\rm N}(xy)~,~\forall x,~y\in \text{SU(n)}
    \end{equation}
    where we will consider the definition of operator product in (\ref{eq:inner prod}).
    \item \textit{Parallel decomposition}: Let $M[x]$ denote a matrix representation for $x\in$SU(n), then
    \begin{equation}
        \qty(\mathcal{C}_{\rm N}(M[x]\oplus M[y]))^Q=\qty(\mathcal{C}_{\rm N}(M[x]))^Q+\qty(\mathcal{C}_{\rm N}(M[y]))^Q~,
    \end{equation}
    where $Q\in\mathbb{Z}_+$.
    \item \textit{Smoothness}: Let $\delta x=\exp(\rmi H\delta s)$ represent the infinitesimal form of $x\in\;$SU(n), where $H$ is a traceless Hermitian operator and $\delta s\geq0$. We require
\begin{equation}\label{eq:Nielsen complexity}
    \mathcal{C}_{\rm N}(\delta x)=F(H)\delta s+\mathcal{O}(\delta s^2)~,
\end{equation}
where $F[H]$, called the cost function, is any analytic function.
\end{itemize}
Using these postulates for metric spaces, the continuous curve in the space of SU(n) operators can be represented as
\begin{equation}\label{eq:Most general NC}
    c(s)=P_{\xi}\exp\qty(-\rmi\int_0^sH_\xi(u)\rmd u)~,\quad\xi=\qty{\rm L,~R}~,
\end{equation}
where $s\in\mathbb{R}$ represents a parametrization of this curve, and $\xi$ determines the orientation of the path ordered integral (where R/L corresponds to building a quantum circuit from right to left; or left to right), such that under an infinitesimal displacement:
\begin{equation}
\begin{aligned}
    \delta c(s)&=-\rmi H_{\rm R}(s)c(s)\delta s=-\rmi c(s)H_{\rm L}(s)\delta s~.
\end{aligned}
\end{equation}
from which it follows that
\begin{equation}\label{eq:LR invariant H}
    H_{\rm R}(s)=c(s)H_{\rm L} c(s)^{-1}~.
\end{equation}
One can then define the length along the curve $c(s)$ starting from the identity $\mathbb{1}$ to operator $x\in\;$SU(n) in terms of (\ref{eq:Nielsen complexity}) as\footnote{We have used reparametrization invariance to set the limits $s\in[0,~1]$.}
\begin{equation}
    L_\xi[c]=\int\mathcal{C}[\delta x]=\int_0^1F(H_\xi(s))~\rmd s~.
\end{equation}
Nielsen operator complexity is defined as the minimum length of $c(s)$:
\begin{equation}\label{eq:Nielsen General}
    \mathcal{C}_{\rm N}^{(\xi)}(x)\equiv\min_{\qty{c(s):~c[0]=\mathbb{1},~c[1]=x}}L_\xi[c]~.
\end{equation}
Notice that since $H_{\rm R}\neq H_{\rm L}$ (\ref{eq:LR invariant H}), the Nielsen complexity will depend on the choice of orientation in the path integral (\ref{eq:Most general NC}). 

As we mentioned in the introduction, evaluating Nielsen complexity with (\ref{eq:Nielsen General}) is quite involved and often intractable. There is, however, a great simplification by demanding the following properties on the cost function:
\begin{itemize}
    \item \textit{Unitary invariance}: Let $x\in\;$SU(n),
    \begin{equation}\label{eq:f unitary}
    F(H_\xi)=F(x H_\xi x^\dagger)~.
    \end{equation}
    This implies that $F(H_{\rm L})=F(H_{\rm R})\equiv F(H)$ from (\ref{eq:LR invariant H}) given that $c(s)\in\;$SU(n). The resulting metric space is said to be bi-invariant, as (\ref{eq:Nielsen General}) is invariant under transformations $c(s)\rightarrow c(s)x$ and $c(s)\rightarrow x~ c(s)~\forall x\in\;$SU(n).
    \item \textit{Reversal invariance}: 
    \begin{equation}\label{eq:f time reverse}
        F(H)=F(-H)~.
    \end{equation}
    This property can be physically motivated when we identify $H$ as the generator of time translations, and we require that the map (\ref{eq:Nielsen General}) be time-reversal invariant.
\end{itemize}
It was shown in \cite{Yang:2018nda} (see also \cite{Yang:2018tpo,Yang:2019iav}) that (\ref{eq:non-negative}-\ref{eq:Nielsen complexity}) together with (\ref{eq:f unitary}, \ref{eq:f time reverse}) determines the form of the cost to be (up to a positive proportionality constant):
\begin{equation}\label{eq:Bi invariant cost}
    F(H(s))= \qty(\tr(H(s)H^\dagger(s))^{Q/2})^{1/Q}~.
\end{equation}
where the case $Q=2$ corresponds to a Riemannian metric on the SU(n) group manifold, and $Q\neq2$ to Finsler metrics (see e.g. \cite{bao2012introduction}). We will focus on the proposal (\ref{eq:Bi invariant cost}), and set $Q=2$ to make the minimization process much more tractable. 

We then study how to construct the unitary target operator of the form:
\begin{equation}\label{eq:unitaries}
    x(t)=\exp(-\rmi V)~,\quad V=H ~t+2\pi K_n~\mathbb{1}~,\quad K_n\in\mathbb{Z}~,
\end{equation}
where $H$ is the (traceless and Hermitian) Hamiltonian. Moreover, given that $V$ is the generator of SU(n) elements, we require $\tr(V)=0$; resulting in the constraint $\sum_{n=0}^\infty K_n=0$.

The corresponding bi-invariant Nielsen complexity, $\mathcal{C}_{\rm N}(x(t))=\mathcal{C}_{\rm N}(t)$, is then given by (\ref{eq:Nielsen General}) and (\ref{eq:Bi invariant cost}) with $Q=2$ (Riemannian case) as:
\begin{equation}\label{eq:biinvariant complexity}
\begin{aligned}
    \mathcal{C}_{\rm N}(t)&=\min_{\qty{K_n:~\sum_n K_n=0}}\sqrt{\tr(V V^\dagger)}~.
\end{aligned}
\end{equation}
The place of Nielsen complexity in the holographic dictionary is less understood in comparison with the others mentioned in this section. Nevertheless, its robust features, such as the growth of circuit complexity with system size, has motivated the different holographic complexity proposals mentioned in the introduction. We will study this proposal for the DSSYK model in Sec. \ref{sec:comments Nielsen}.

\section{Towards spread complexity in dS space}\label{sec:Spread}
In this section, we first study $\mathcal{C}_{\rm S}$ in the doubled Hilbert space description of the DSSYK model in terms of entanglers in a MERA network. Our arguments at this point are not restrained to dS holography, and they can be applied to the triple-scaled SYK/JT gravity correspondence. Afterward, we show that $\mathcal{C}_{\rm S}$ with $\ket{\mathbb{E}_0}$ as the reference state reproduces a geodesic distance in SdS$_3$ space from the holographic dictionary in \cite{Verlinde:2024znh}.

We start defining the operator
\begin{equation}\label{eq:N bb}
\begin{aligned}
    {\mathbb{N}}&\equiv \frac{1}{2}\qty(\hat{n}\otimes\mathbb{1}+\mathbb{1}\otimes\hat{n})~,\\
    |{\mathbb{N}})&=\sum_nn\ket{n,~n}~.
\end{aligned}
\end{equation}
Using the identification (\ref{eq:Krylov basis DSSYK}), we can then express the spread complexity of a time-evolved state $\phi(t)$ (\ref{eq:spread complexity}) in the doubled Hilbert space formalism as
\begin{equation}
    \mathcal{C}_{\rm S}(t)=\bra{\phi(t),~\phi^*(t)}{\mathbb{N}})~.
\end{equation}
Expanding the evolved state in its Krylov basis as $\ket{\phi(t)}=\sum_n\phi_n(t)\ket{n}$, and employing (\ref{eq:entangler}):
\begin{equation}\label{eq:spread as counting gates}
\begin{aligned}
    \mathcal{C}_{\rm S}(t)&=\bra{0,~0}\mathcal{O}_{\phi(t)}^\dagger~\mathbb{N}~\mathcal{E}\ket{0,~0}~,
\end{aligned}
\end{equation}
where we have defined
\begin{equation}\label{eq:general OP MERA}
    \mathcal{O}_{\phi(t)}\equiv\sum_{n,~m}\frac{\phi_n(t)\qty(A^\dagger)^n}{\sqrt{(q;~q)_n}}\otimes\frac{\phi_m(t)\qty(A^\dagger)^m}{\sqrt{(q;~q)_m}}
\end{equation}
It can be seen in (\ref{eq:spread as counting gates}) that spread complexity has a natural interpretation as a map from an operator ($\mathbb{N}~\mathcal{E}$) that counts entangled chord states in the state $\ket{\phi(t),~\phi^*(t)}$, which has been prepared from the vacuum through the map $\mathcal{O}_{\phi(t)}\ket{0,~0}$.

\subsubsection*{Interpretation in the doubled DSSYK model}\label{sssec:Application spread}
\begin{figure}[t!]
    \centering
    \includegraphics[width=0.3\textwidth]{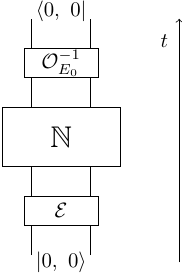}
    \caption{Illustration of spread complexity $\mathcal{C}_{\rm S}(t)$ (\ref{eq:spread complexity}) in $\mathcal{H}\otimes\mathcal{H}$. It maps the operator $\hat{\mathbb{N}}$ that counts the number of entangled chord states through $\mathcal{E}$ which are projected onto the state $\mathcal{O}_{E_0}\ket{0,~0}$ where $\mathcal{O}_{E_0}$ is defined in (\ref{eq:O E0}). See Sec. \ref{sec:conclusions} for comments on the DSSYK model and tensor networks.}
    \label{fig:spread MERA}
\end{figure}

We now specialize the previous discussion to the doubled DSSYK model \cite{Narovlansky:2023lfz,Verlinde:2024znh,Verlinde:2024zrh}. We take as reference state $\ket{\mathbb{E}_0}$ (\ref{eq:ref energy state}), which to the $\ket{\psi_{\rm dS}}$ in the bulk. We can express this state using the identity (\ref{eq:identity theta}) as:
\begin{equation}
    \ket{E_0}=\sum_n\frac{H_n(\cos\theta_0|q)}{\sqrt{(q;~q)_n}}\ket{n}~.
\end{equation}
(\ref{eq:spread as counting gates}) then simplifies to
\begin{equation}\label{eq:C S DDSSYK}
    \boxed{\mathcal{C}_{\rm S}(t)=\bra{0,~0}\mathcal{O}^{\dagger}_{E_0}\mathbb{N}~\mathcal{E}\ket{0,~0}~,}
\end{equation}
where
\begin{equation}\label{eq:O E0}
    \mathcal{O}_{E_0}=\sum_{n,~m}\frac{H_n(\cos\theta_0|q)\qty(A^\dagger)^n}{{(q;~q)_n}}\otimes\frac{H_m(\cos\theta_0|q)\qty(A^\dagger)^m}{{(q;~q)_m}}~.
\end{equation}
Notice that the time dependence has dropped out in (\ref{eq:C S DDSSYK}) due to $\ket{E_0}$ being an eigenstate of $T$. One can further simply (\ref{eq:O E0}) using the following identify (see e.g. \cite{Okuyama:2023byh}):
\begin{equation}
    \sum_n\frac{H_n(\cos\theta|q)t^n}{(q;~q)_n}=\frac{1}{(t~\rme^{\pm \rmi\theta};~q)_\infty}~,
\end{equation}
such that (\ref{eq:O E0}) becomes
\begin{equation}
    \mathcal{O}_{E(\theta)}=\frac{1}{(A^\dagger~\rme^{\pm \rmi\theta};~q)_\infty}\otimes \frac{1}{(A^\dagger~\rme^{\pm \rmi\theta};~q)_\infty}~.
\end{equation}
From the expressions above, we notice that $\mathcal{C}_{\rm S}$ manifestly counts the chord states that have been entangled through the gate $\mathcal{E}$, which are then projected to the maximal entropy state $\ket{\mathbb{E}_0}$, which is created from the vacuum $\ket{0,~0}$ by acting with $\mathcal{O}_{E_0}$. See Fig. \ref{fig:spread MERA} for an illustration.

\begin{figure}
    \centering
    \includegraphics[width=0.75\textwidth]{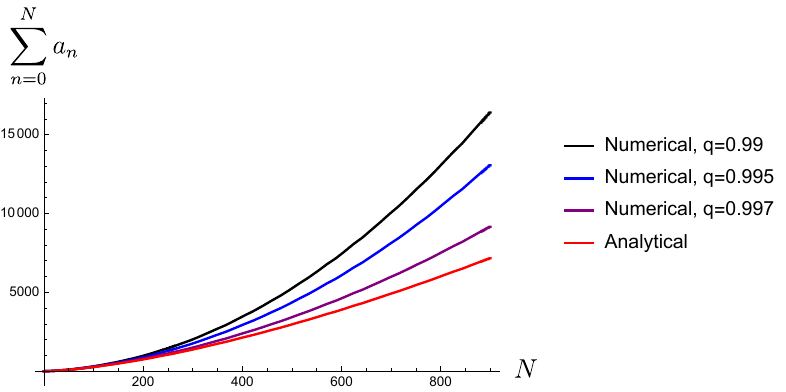}
    \caption{Evaluation of the series in (\ref{eq:towards CS exp}) with $q=0.99$ (black), $q=0.994$ (blue), $q=0.998$ (purple) and its analytic approximation in (\ref{eq:semiclassical spread}) (red), where, instead of the infinite summation, we have included $N$ as the upper limit of summation. Here, $a_n=n\frac{\qty(H_n(\cos\theta_0|q))^2}{\qty(q;~q)_n}$ in the former case; and $a_n=\frac{2^n\pi}{(n-1)!\Gamma\qty(\frac{1-n}{2})^2}$ in the latter. We observe that the analytic approximation lower bounds the numerical ones, and they diverge as we increase the upper bound $N$.}
    \label{fig:spread_approx}
\end{figure}

We now perform the evaluation of (\ref{eq:C S DDSSYK}) explicitly using (\ref{eq: proj E0 theta}):\footnote{We thank Nikolay Bobev for comments on this point.}
\begin{equation}\label{eq:towards CS exp}
    \mathcal{C}_{\rm S}=\sum_nn\frac{\qty(H_n(\cos\theta_0|q))^2}{\qty(q;~q)_n}~.
\end{equation}
In the $\lambda\rightarrow0$ regime, one can approximate \cite{Tang:2023ocr}
\begin{equation}
    H_n(x|q)\simeq\sqrt{\frac{\lambda}{2}}H_n\qty(x\sqrt{\frac{2}{\lambda}})~,
\end{equation}
where $H_n(x)$ is the Hermite polynomial of degree $n$. Moreover, $\lim_{\lambda\rightarrow0}(q;q)_n=\lambda^nn!$ from the definition (\ref{eq:q-Pochhammer}). Thus, in the semiclassical regime ($\lambda\rightarrow0$), and considering the maximal entropy state in (\ref{eq:towards CS exp}), $\theta_0=\pi/2$, one then recovers
\begin{equation}\label{eq:semiclassical spread}
    \mathcal{C}_{\rm S}\simeq\sum_{n=0}^\infty\frac{2^n\pi}{(n-1)!\Gamma\qty(\frac{1-n}{2})^2}~.
\end{equation}
This is a diverging series, given that $\lim_{n\rightarrow\infty}\frac{2^n\pi}{(n-1)!\Gamma\qty(\frac{1-n}{2})^2}\neq0$. We have confirmed the approximation bounds from below the (\ref{eq:towards CS exp}) when $q$ is close to $1$, as illustrated in Fig. \ref{fig:spread_approx}. 
This divergence implies that the chord number operator needs to be renormalized in the semiclassical limit. We will discuss this point in the following subsection, guided by the holographic dictionary.

We conclude this subsection with a few remarks:
\begin{itemize}
    \item \textbf{Microcanonical ensemble}: Notice that there is a straightforward extension, by switching from a canonical ensemble to a microcanonical one where we consider
\begin{equation}\label{eq:microcanonical state}
    \rho_{\rm dS}=\frac{1}{N}\sum_E\ket{E}\bra{E}~,
\end{equation}
centered around $E=E_0$ to evaluate the spread complexity \cite{Alishahiha:2022anw} as
\begin{equation}\label{eq:spread micro}
    \mathcal{C}_{\rm S}=\frac{1}{N}\sum_{E,~n}n\bra{E,~E}\ket{n,~n}=\frac{1}{N}\sum_{E}\bra{0,~0}\mathcal{O}_{E(\theta)}^{\dagger}~\mathbb{N}~\mathcal{E}\ket{0,~0}~.
\end{equation}
\item \textbf{Triple scaling limit}: In this regime, one might choose $\ket{n=0}$ as the reference state, representing the canonical ensemble thermofield doubled (TFD) state in the infinite temperature limit \cite{Rabinovici:2023yex}. In this case, (\ref{eq:general OP MERA}) adopts a simple expression in terms of Hamiltonian evolution (through the transfer matrix $T$) in each $\mathcal{H}$, which we denote:
\begin{equation}\label{eq:triple scaling op mera}
    \mathcal{O}_{T}(t)\equiv\rme^{\rmi t T}\otimes\rme^{-\rmi t T}~.
\end{equation}
\end{itemize}
We will now move on to study the dual interpretation of the spread complexity for the state $\ket{\mathbb{E_0}}$, and especially its time independence.

\subsection{Dual interpretations}
\begin{figure}[t!]
    \centering
    \includegraphics[width=0.5\textwidth]{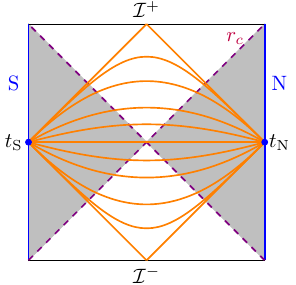}
    \caption{Geodesic curves (orange) joining antipodal static patch observers (blue lines S and N) in SdS$_3$ space, and probing the region outside the cosmological horizon ($r_c$, purple dashed lines). $\mathcal{C}_{\rm S}$ measures the time difference between the antipodal observers, which has been fixed to $t_{\rm N}=t_{\rm S}$ (blue dots) in the diagram.}
    \label{fig:dS geo}
\end{figure}

We would like to translate the above results using the bulk dictionary developed in \cite{Verlinde:2024znh}, where a bulk phase space variable $z$ was related to the chord number operator $\hat{n}$ though 
\begin{align}
    \rme^{-8\pi {G_N}(\hat{n}-1/2)/{\ell_{\rm dS}}}=-\rme^{-2z}~.\label{eq:classical phase}
\end{align}
Here $z$ is an operator whose expectation value on the $\ket{\psi_{\rm dS}}$ state corresponds to a (regularized) geodesic length in SdS$_3$ space measuring the static patch time difference between the antipodal observers (i.e. $\expval{z}_{\rm dS}=t_{\rm N}-t_{\rm S}$). Geometrically, $\expval{z}_{\rm dS}$ and the deficit angle in SdS$_3$ space determine the geodesic lengths connecting the antipodal observers, as shown in Fig. \ref{fig:dS geo}.

Using the dictionary, we take expectation values in $\ket{\psi_{\rm dS}}$ for the bulk side of (\ref{eq:classical phase}), and $\ket{E_0}$ on the chord number operator leads to:
\begin{equation}\label{eq:relation y and spread}
    \boxed{2\expval{z}_{\rm dS}=\frac{8\pi G_N}{\ell_{\rm dS}}\mathcal{C}_{\rm S}+\qty(\rmi\pi-\frac{8\pi G_N}{\ell_{\rm dS}})\bra{E_0}\ket{E_0}~.}
\end{equation}
The factor $\rmi\pi$ is related to the integration of the Wilson defining the holonomy variable which is involved in the identification of $z$ with a time difference \cite{Verlinde:2024znh}, and it can be shifted away. 

Next, we would like to interpret the relation (\ref{eq:norm theta}). We have that $\bra{E_0}\ket{E_0}\rightarrow\infty$, and as we have noticed in (\ref{eq:semiclassical spread}) $\mathcal{C}_{\rm S}$ also diverges in the $q\rightarrow1$ limit (i.e. when $\ket{E_0}$ corresponds to the pure dS state). We need to renormalize both of these terms in (\ref{eq:relation y and spread}). For instance, the dS space boost isometries give us the freedom to set $t_{\rm N}=t_{\rm S}$. This can be motivated on the doubled DSSYK side from the Hamiltonian constraint in physical states (\ref{eq:phy states}) which correspond to synchronizing the clocks (physical time) for the L/R system \cite{Narovlansky:2023lfz}. We will then normalize $\bra{E_0}\ket{E_0}_{q\rightarrow1}=\eval{\mathcal{C}_{\rm S}}_{q\rightarrow1}$ in (\ref{eq:relation y and spread}) such that
\begin{equation}
    \lim_{q\rightarrow1}\Re(\expval{z}_{\rm dS})=0~.\label{eq:redef E_0}
\end{equation}
We, therefore, conclude that the boost symmetries in SdS$_3$ space can be interpreted in terms of renormalization in the spread complexity of the $\ket{E_0^{\rm L},~E_0^{\rm R}}$ state in the DSSYK model. There are a few remarks about the above analysis:
\begin{itemize}
\item \textbf{Time independence}: The fact that the spread complexity for this state does not depend on the value of $t_{\rm N}$ or $t_{\rm S}$, might be interpreted in the bulk description with the total entropy perceived by the N and S pole observers being a time-independent constant (\ref{eq:GH entropy}). Given that $\mathcal{C}_{\rm S}$ is a constant, the precise bulk dictionary allowed us to set the time difference to vanish, which is related to the freedom in fixing the time difference between the observers by boost symmetry.

\item \textbf{Relation with entanglement in the doubled DSSYK model}: As we commented on in Sec. \ref{sssec:Application spread}, spread complexity counts the number of entangled chord states pairs in the maximal entropy state of the DSSYK model. Given that the spread complexity of the doubled DSSYK model is dual a geodesic length in the SdS$_3$ space (measuring a time difference), we find agreement with previous studies \cite{Lashkari:2013koa,Faulkner:2013ica,Swingle:2014uza,Faulkner:2017tkh,Haehl:2017sot,Agon:2020mvu,Agon:2021tia,VanRaamsdonk:2009ar,VanRaamsdonk:2010pw,Bianchi:2012ev,Maldacena:2013xja,Balasubramanian:2014sra} suggesting that entanglement builds spacetime, and in this case, spread complexity provides a measure of this relation.

\item \textbf{Connections with JT gravity}: In the triple scaled SYK model/ JT gravity correspondence, spread complexity is identified with a geodesic length between asymptotic AdS$_2$ boundaries \cite{Rabinovici:2023yex}, suggesting a relation with the CV conjecture \cite{Susskind:2014rva}. Our observations in the dS holographic context share some similarities. The spread complexity for the $\ket{E_0}$ state has a geodesic length interpretation, although in terms of a time-like coordinate difference between antipodal observers; and, the bulk has two space-like dimensions more than the theory where spread complexity is evaluated.

\item \textbf{Liouville-dS CFT}: As we mentioned in the introduction, the LdS$_2$ CFT, described by the field $\phi_\pm$, is located in a disk region $\Sigma$, whose boundary describes the time-like geodesic of a worldline observer. Since the N/S pole static patch time in SdS$_3$ space, $t$, is identified with the boundary time along $\partial\Sigma$, $\tau$ \cite{Verlinde:2024zrh}, $\mathcal{C}_{\rm S}$ can also be identified with a proper time difference between the L/R LdS$_2$ CFTs on the maximal entropy state, that is
\begin{equation}\label{eq:CS LdS}
    \tau_{\rm L}-\tau_{\rm R}\propto \mathcal{C}_{\rm S}~.
\end{equation}
Since the spread complexity counts the number of entangled modes in the doubled DSSYK model, this suggests there is entanglement between the pairs of CFTs.
\end{itemize}

\section{Towards Krylov complexity in dS space}\label{sec:Krylov}
This section investigates the evolution of Krylov complexity for physical operators $\mathcal{O}^{\rm phys}_\Delta$ (i.e. those obeying (\ref{eq:phy ops})) in the different sides of the correspondence, which are shown explicitly below:
\begin{itemize}
    \item Doubled DSSYK model:
    \begin{equation}
        \begin{aligned}\label{eq:O phys DDSSYK}
        &\mathcal{O}^{\rm phys}_\Delta(\tau)=\int\rmd t~O_{1-\Delta}^{\rm L}(t)O_{\Delta}^{\rm R}(\tau-t)~.
    \end{aligned}
    \end{equation}
    where $\mathcal{O}^{\rm L/R}_\Delta(\tau)$ are shown in (\ref{eq:matter ops DSSYK}). 
\item LdS$_2$ CFT:
\begin{equation}
\begin{aligned}\label{eq:O phys LdS}
        \mathcal{O}^{\rm phys}_\Delta(\tau)&=\int\rmd t~V_{1-\Delta}^-(t)~V_{\Delta}^+(\tau-t)~,\\
    V^\pm_\Delta&=\rme^{b_\pm\Delta \phi_{\pm}}~,
\end{aligned}
\end{equation}
where $V^\pm_\Delta(\tau)$ are boundary vertex operators, parametrized by the proper time $\tau$ in $\partial\Sigma$ (\ref{eq:action LdS}).
\item SdS$_3$ space: Scalar fields with conformal weight $\Delta$, 
\begin{equation}\label{eq:O phys SdS}
    \mathcal{O}^{\rm phys}_\Delta(x)=\phi_\Delta(x)~.
\end{equation}
\end{itemize}
We consider $|\mathcal{O}^{\rm phys}_\Delta)$ as the initial operator in the Lanczos algorithm to calculate their Krylov complexity. The first amplitude in (\ref{eq:2pnt correlator Krylov}) is determined by the 2-point correlation function:
\begin{equation}\label{eq:2pnt correlator Verlinde}
\begin{aligned}
    \varphi_0(\tau)&=\frac{(O^{\rm phys}_\Delta(0)|O^{\rm phys}_\Delta(\tau))}{(O^{\rm phys}_\Delta(0)|O^{\rm phys}_\Delta(0))}~.
\end{aligned}
\end{equation}
Meanwhile the case of SdS$_3$ space, we take $\tau=\tau(x_1,~x_2)$ as the proper time between the insertion of the fields $\phi_\Delta(x_1)$ and $\phi_\Delta(x_2)$ on time-like separated points $x_1$, $x_2$.

The correlation function of physical operators in (\ref{eq:2pnt correlator Verlinde}) has been computed for the state $\ket{E_0}$ and its holographic duals, and matched between them by \cite{Verlinde:2024znh,Verlinde:2024zrh}. Explicitly, $\varphi_0(\tau)$ in (\ref{eq:2pnt correlator Verlinde}) becomes:
\begin{equation}\label{eq:auto correlator dS}
\begin{aligned}
    \varphi_0(\tau)&=\mathcal{N}^{-1}\int_0^\pi\rmd\theta_1\frac{\mu(\theta_1)\rme^{-\rmi\tau\,E(\theta_1)}}{(q^{1-\Delta}\rme^{\pm\rmi\theta_0\pm\rmi\theta_1};~q)_\infty(q^{\Delta}\rme^{\pm\rmi\theta_0\pm\rmi\theta_1};~q)_\infty}~,\\
    \mathcal{N}&\equiv\int_0^\pi\rmd\theta_1\frac{\mu(\theta_1)}{(q^{1-\Delta}\rme^{\pm\rmi\theta_0\pm\rmi\theta_1};~q)_\infty(q^{\Delta}\rme^{\pm\rmi\theta_0\pm\rmi\theta_1};~q)_\infty}~,
\end{aligned}
\end{equation}
where $q\in[0,~1]$, and $\theta_0=\frac{\pi}{2}$ (corresponding to $E_0=0$ for the maximal entropy state). Given that we are considering Hermitian physical operators (\ref{eq:O phys DDSSYK}-\ref{eq:O phys SdS}), only the even moments in (\ref{eq:2pnt correlator Krylov}) will contribute to $\mathcal{C}_{\rm K}$. The moments, determined from (\ref{eq:auto correlator dS}), are
\begin{equation}
    m_{2n}=\mathcal{N}^{-1}\int_0^\pi\rmd\theta_1\frac{\mu(\theta_1)\rm(E(\theta_1))^{2n}}{(q^{1-\Delta}\rme^{\pm\rmi\theta_0\pm\rmi\theta_1};~q)_\infty(q^{\Delta}\rme^{\pm\rmi\theta_0\pm\rmi\theta_1};~q)_\infty}~.
\end{equation}
In principle, one can proceed to evaluate the Krylov complexity (\ref{eq:Krylov complexity}) exactly. To simplify the evaluation of the amplitudes $\varphi_n(\tau)$, we will work in $q\rightarrow1$ limit (i.e. $G_N\rightarrow0$),
\begin{align}
    \varphi_0(\tau)&=\frac{\sinh(\nu \tau)}{\nu\sinh(\tau)}~,\label{eq: correlator massive s q->1}\\
    m_{2n}&=\frac{2}{\nu}\sum_{k=0}^\infty\nu^{2n-2k+1}\frac{(1-2^{2k-1}){\rm B}_{2k}}{(2k)!(2n-2k+1)!}~,
\end{align}
where ${\rm B}_{n}$ are the Bernoulli numbers; $\tau$ has been rescaled by $\ell_{\rm dS}$ to make it dimensionless; and $\nu\equiv2\Delta-1$, which is related to the scalar particle's mass in SdS$_3$ space through $m^2\ell^2_{\rm dS}=4\Delta(1-\Delta)$, i.e.
\begin{equation}\label{eq:nu scalar field SdS3}
\nu=\sqrt{1-m^2\ell^2_{\rm dS}}    ~.
\end{equation}
Notice that $\nu\in[0,~1]$ when $m\ell_{\rm dS}\leq1$ and $\nu\in\rmi\mathbb{R}$ otherwise. Motivated by the duality with the doubled DSSYK model, where $\Delta\in[0,~1]$ (\ref{eq:O phys DDSSYK}), we will consider $m^2\ell^2_{\rm dS}\leq1$ in the evaluation. The Lanczos coefficients can be determined through the algorithm (\ref{eq:Alt Lanczos}), leading to\footnote{We thank Patrik Nandy for correspondence about this point.}
\begin{equation}\label{eq:Lanczos}
    b_n=n\sqrt{\frac{n^2-\nu^2}{4n^2-1}}~.
\end{equation}
Notice that the growth of $b_n$ is linear in $n$ for $n\gg1$, so that it satisfies Carleman's condition \cite{carleman1926fonctions}. The linear growth in the coefficients is generically found in chaotic systems \cite{Hashimoto:2023swv}, although it is also sensitive to the choice of the initial operator \cite{Espanol:2022cqr,Rabinovici:2022beu}.

We can now compute the amplitudes $\varphi_n(t)$ through the recursion relation (\ref{eq:sch eq K operator}). One can check that the amplitude for arbitrary $\nu$ and $n$ in the early and late time regime take the form:
\begin{equation}    
\varphi_n(\tau)=\frac{\tau^n\prod_{k=1}^n\sqrt{k^2-\nu^2}}{(2n-1)!!\sqrt{2n+1}}+\mathcal{O}(\tau^{n+2})~,\quad n\geq1~,
\end{equation}
\begin{equation}    
\varphi_n(\tau)=\rme^{(\nu-1)\tau}\frac{\sqrt{2n+1}}{\nu}\prod_{k=1}^n\frac{k-\nu}{\sqrt{k^2-\nu^2}}+\mathcal{O}(\rme^{-(1+\nu)\tau})~,\quad n\geq1~.
\end{equation}
Moreover, there is a particular non-trivial value, $\nu=1/2$, for which we find a closed form relation for the amplitudes $\varphi_n$ and the corresponding Krylov complexity (\ref{eq:Krylov complexity}) (see also \cite{Caputa:2021sib}):
\begin{align}
    \nu=1/2~:\quad {\varphi_n(\tau)}&=\sech(\frac{\tau}{2})\tanh^n\qty(\frac{\tau}{2})~,\quad b_n=n/2~,\\ {\mathcal{C}_{\rm K}(\tau)}&=\sinh^2\qty(\frac{\tau}{2})~.
\end{align}
Meanwhile, for $\nu\neq1/2$, we can still find the late time behavior of $\mathcal{C}_{\rm K}$ using a result shown in \cite{Parker:2018yvk}. Assuming smoothness of the Lanczos coefficients $b_n$ with $n$ for a local operator, it was shown that if $b_n=\frac{\lambda_{\rm K}}{2} n+\mathcal{O}(1)$ ($\lambda_{\rm K}\in\mathbb{R}$) for $n\gg1$, then $\mathcal{C}_{\rm K}(\tau)$ grows exponentially, with $\lambda_{\rm K}$ being the exponent. In our case, given that $b_n$ in (\ref{eq:Lanczos}) is smooth, and $\lambda_{\rm K}=1$; we conclude that
\begin{equation}\label{eq:late time CK}
    \boxed{\mathcal{C}_{\rm K}(\tau\gg1)\propto \rme^{\tau}~,\quad\forall\nu\in[0,~1)~.}
\end{equation}

\subsection{Dual interpretations}
The late-time exponential growth in $C_{\rm K}(\tau)$ was conjectured to be universally displayed by maximally chaotic systems in \cite{Parker:2018yvk}, so our results are consistent with the expectation that the DSSYK model is a maximally chaotic system \cite{Berkooz:2018jqr}, and with previous studies finding exponential growth of the Krylov complexity \cite{Parker:2018yvk,Bhattacharjee:2022ave}, albeit for the $\psi_i(\tau)$ operators in (\ref{eq:DDSSYK Hamiltonian}).\footnote{A first connection between Krylov complexity and dS holography appeared in \cite{Bhattacharjee:2022ave}. A ``cosmic time" scale appears in the exponent of $\mathcal{C}_{\rm K}$ which is equivalent to a rescaling $\tau\rightarrow p\tau$ in the correlation function of $\psi_i(\tau)$. In the double scaling limit (\ref{eq:double scaling}), the enhanced growth of $\mathcal{C}_{\rm K}$ was associated with hyperfast scrambling in the DSSYK model conjectured in \cite{Susskind:2021esx}.} The same holds for the other two members in the dS holographic proposal of \cite{Narovlansky:2023lfz,Verlinde:2024znh,Verlinde:2024zrh} since our calculations employ the known correlation functions of the physical operators in (\ref{eq:O phys DDSSYK} - \ref{eq:O phys SdS}). Moreover, according to the conjecture in \cite{Parker:2018yvk}, the exponent in (\ref{eq:late time CK}) corresponds to the maximal Lyapunov exponent measured by OTOCs \cite{Maldacena:2015waa,Murthy:2019fgs}: $2\pi/\beta$, where $\beta$ is the inverse temperature of the system. Our results are in \textbf{agreement} with the conjecture since the physical temperature identified in the correlator (\ref{eq: correlator massive s q->1}) corresponds to $\beta_{\rm dS}=2\pi$ \cite{Narovlansky:2023lfz}.

Lastly, we notice that $\mathcal{C}_{\rm K}=\,$const. (no operator spread) in the critical case $\nu=1$, given that the input correlator (\ref{eq: correlator massive s q->1}) corresponds to a scalar propagating on a time-like trajectory, or equivalently, no operator insertion in the doubled DSSYK and LdS$_2$ duals (\ref{eq:O phys DDSSYK}, \ref{eq:O phys LdS} respectively).

As a side comment, motivated by the recent discussions about spread and Krylov complexities for density matrices \cite{Caputa:2024vrn}, one might study the Krylov complexity of the density matrix $\rho_{\rm dS}$ in (\ref{eq:spread micro}) and compare with the features of the spread complexity encountered in Sec. \ref{sec:Spread}. Given that the evolution is controlled by the Liouville-von Neumann equation
\begin{equation}
    \partial_t\rho(t)=-\rmi\mathcal{L}\rho(t)~,
\end{equation}
with $|\rho(t=0))=|\rho_{\rm dS})$, it follows that the density matrix does not evolve in static patch time, i.e. $|\rho(t))=|\rho_{\rm dS})$, given that $\ket{E}$ are energy eigenstates. The Krylov complexity for $\rho_{\rm dS}$ is then trivial, in contrast to the spread complexity in (\ref{eq:spread micro}).

\section{Towards query complexity in dS space}\label{sec:Query}
In this section, we formulate an algorithm computing correlation functions for LdS$_2$ CFTs building on Sec. \ref{ssec:query}, and also guided by the diagrammatic structure of n-point correlation functions in the cylinder amplitude of the DSSYK model. We make contact with SdS$_3$ through its CS formulation, which is a SL$(2,~\mathbb{C})$ topological field theory, with $k\rightarrow\rmi\kappa$ and $\kappa\in\mathbb{R}$ in (\ref{eq:CS AdS3}).

In terms of the LdS$_2$ theory, it was argued in \cite{Verlinde:2024zrh} that the natural vacuum state dual to $\ket{\mathbb{E}_0}$ and $\ket{\psi_{\rm dS}}$, which we denote $\ket{s_0}$, corresponds to a FZZT brane with $\mu_{\rm B}=0$ in (\ref{eq:action LdS}) (equivalent to setting $E_0=0$ for the maximal entropy state). For higher energy states, let us consider the region $\Sigma$ where the LdS$_2$ CFT is defined (\ref{eq:action LdS}), and insert physical operators $\mathcal{O}_\Delta^{\rm phys}(\tau_1)$ and $\mathcal{O}_\Delta^{\rm phys}(\tau_0)$ on the vacuum state $\ket{s_0}$. Let us denote a segment $s_1\in\partial\Sigma$ where $\partial s_1=\qty{\tau_0,~\tau_1}$. Then, excited states can be represented as
\begin{equation}\label{eq: momentum eigenstates LdS}
\begin{aligned}
    \ket{s_1}=\hat{W}(\tau_0,~\tau_1)\ket{s_0}~,\quad\hat{W}(\tau_0,~\tau_1)\equiv\mathcal{O}_\Delta^{\rm phys}(\tau_1)\mathcal{O}_\Delta^{\rm phys}(\tau_0)~,
\end{aligned}
\end{equation}
where $\hat{W}$ is a CFT operator, whose expectation value on state $\ket{s_0}$ can be expressed in the bulk in terms of Wilson lines
\begin{equation}\label{eq:momentum eigenstates LdS}
    \bra{s_0}\hat{W}(\tau_0,~\tau_1)\ket{s_0}=\bra{\psi_{\rm dS}}\tr_{R_j}\qty(P\exp(-\int_{\gamma(\tau_0,\tau_1)} \mathcal{A})P\exp(-\int_{\gamma(\tau_0,\tau_1)} \bar{\mathcal{A}}))\ket{\psi_{\rm dS}}~,
\end{equation}
with $R_j$ a representation of $SL(2,~\mathbb{C})$; $j=1/2+\rmi~s_0$ is the spin; and ${\gamma}(\tau_0,~\tau_1)$ is a path between the insertion times ${\tau_0},~{\tau_1}$ along $\partial\Sigma$.

We will be considering the same definitions for the algorithm computing the arbitrary n-point correlation functions as we explained in Sec. \ref{ssec:query}. However, there will be key differences in its implementation with respect to the proposal in \cite{Chen:2020nlj}, which are intrinsically connected to the dual theories. We previously discussed that query complexity of a vacuum CFT trivializes if several operators are inserted on the exact same location in pure AdS space at a constant time slice. In contrast, for the LdS CFT case, if one were to remove $\partial\Sigma$ (corresponding to the worldline of the static patch observers in SdS$_3$ space), this potentially eliminates the degrees of freedom of the doubled DSSYK model. However, one can instead insert the physical operators of the LdS CFT at \textit{the same spatial location but at different proper times} $\tau_i$. 

\begin{figure}
    \centering
    \includegraphics[width=0.6\textwidth]{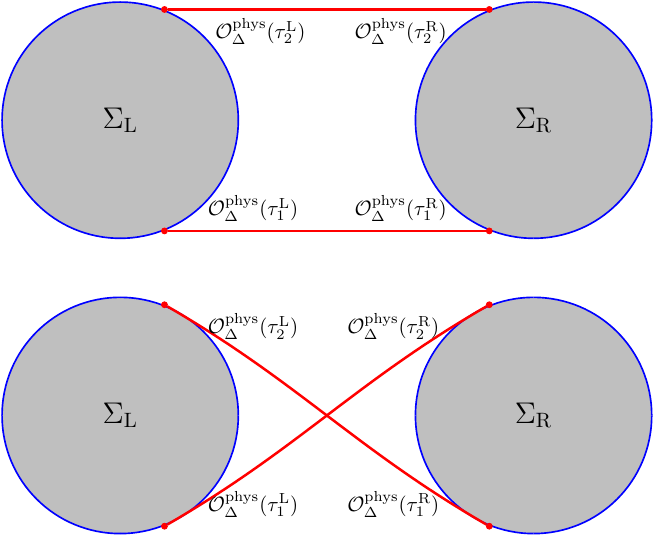}
    \caption{Pair of disks $\Sigma_{\rm L/R}$ where some operators $\mathcal{O}_{\Delta}^{\rm phys}(\tau^{\rm L/R}_i)$ are inserted (red dots) in $\partial\Sigma_{\rm L/R}$. We illustrate two out of all symmetric pairwise {contractions} between the matter operators on the L/R boundaries (red lines). \textit{Above}: $\mathcal{O}_{\Delta}^{\rm phys}(\tau_{1,~2}^{\rm R})$ and $\mathcal{O}_{\Delta}^{\rm phys}(\tau_{1,~2}^{\rm L})$ are contracted. \textit{Below}: $\mathcal{O}_{\Delta}^{\rm phys}(\tau_1^{\rm R/L})$ with $\mathcal{O}_{\Delta}^{\rm phys}(\tau_2^{\rm L/R})$.}
    \label{fig:LR_LdS}
\end{figure}

Moreover, counting the number of fusions for matter operators can be simplified substantially using cylinder amplitudes in the DSSYK model \cite{Okuyama:2023yat} (see Fig. \ref{fig:Correlators_DDSSYK}). Guided by the correspondence with the doubled DSSYK model, we choose to study the correlators where the matter fields are pairwise connected between the L/R LdS CFTs (corresponding to the N and S poles), as illustrated in Fig. \ref{fig:LR_LdS}.
Then, we will formulate the same type of CFT algorithm that we discussed in Sec. \ref{ssec:query}, in terms of the fusion algebra for the states $\qty{\ket{s}}$, but we consider operators from both the L and R side LdS CFTs. This means that the map (\ref{eq:fusions}) takes any number of incoming representations of the states (\ref{eq:momentum eigenstates LdS}) at an overlapping time (e.g. $\tau^{\rm L}_i$), and generates an additional one. For instance, in the case of two incoming matter chords and one outgoing:
\begin{equation}\label{eq:junction rule Liouville}
\begin{aligned}
    \mathcal{J}(W(\tau^{\rm L}_{i},~\tau^{\rm R}_j)\ket{s_0},~W(\tau^{\rm L}_{i},~\tau^{\rm R}_{k})\ket{s_0})=\int\rmd\tau^{\rm R}_l~c(\tau^{\rm L}_i,~\tau^{\rm R}_j,~\tau^{\rm R}_k,~\tau^{\rm R}_l)W(\tau^{\rm L}_{i},~\tau^{\rm R}_l)\ket{s_0}~,
\end{aligned}
\end{equation}
where $c(\tau^{\rm L}_i,~\tau^{\rm R}_j,~\tau^{\rm R}_k,~\tau^{\rm R}_l)$ corresponds to a conformal kinematical factor that obeys the deformability rule in Sec. \ref{ssec:query}, which is determined from OPE data of the representations in (\ref{eq:junction rule Liouville}). Then, correlation functions on a given multiplet representation of SL($2,~\mathbb{C}$) appear from contracting operators according to (\ref{eq:junction rule Liouville}) in the relevant multiplet representation. We will now adopt the definition for query complexity in the LdS CFT as the number of applications of fusion rules (\ref{eq:Query n fusions}) using (\ref{eq:junction rule Liouville}) iteratively, and study its dual interpretation.

\subsection{Dual interpretations}
\begin{figure}
    \centering
    \includegraphics[width=\textwidth]{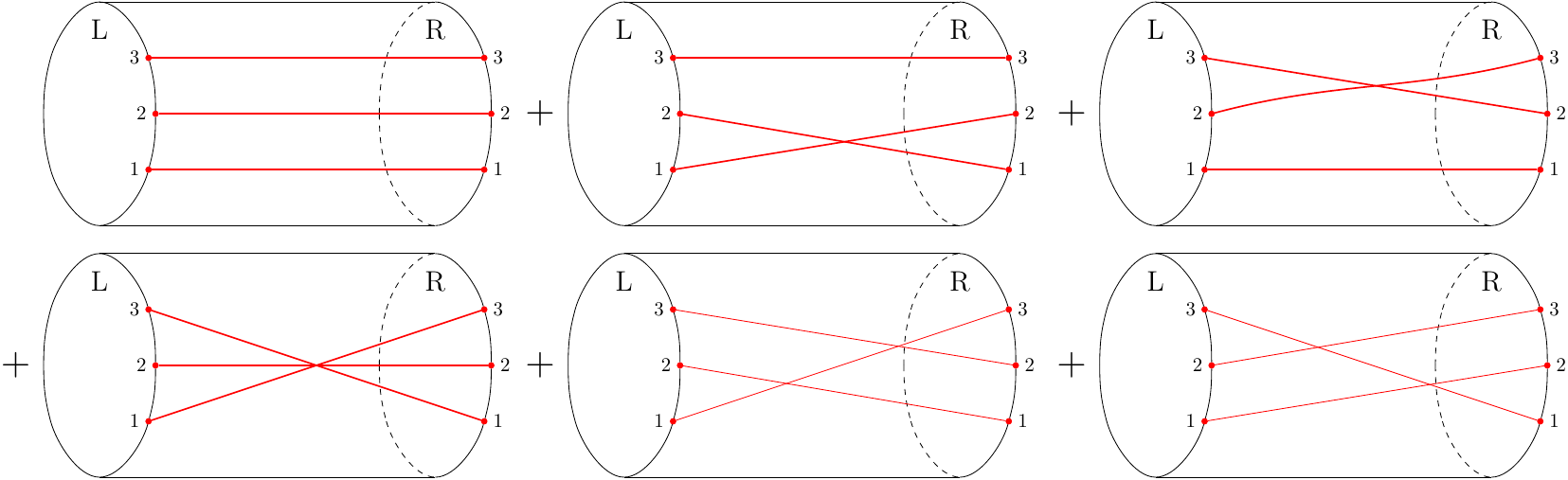}
    \caption{$S_{k=3}$ symmetric matter operator chords (red lines) in the cylinder amplitude of the doubled DSSYK model.}
    \label{fig:Correlators_DDSSYK}
\end{figure}

Our motivation for constructing the algorithm computing correlation functions between the L/R LdS CFTs is the wormhole amplitude describing the matter operator chords extending between two disks, which in our context corresponds to the pair of DSSYK models (L/R). See Fig. \ref{fig:Correlators_DDSSYK} for a representation of this system.
Based on the two-matrix model formulation of the DSSYK model \cite{Jafferis:2022wez,Jafferis:2022uhu}, an arbitrary ($2k$)-point function in the L/R edges of the cylinder is given by \cite{Okuyama:2023yat}
\begin{align}\label{eq:matter correlators query}
    \expval{\tr(\prod_{i=1}^k\rme^{\rmi t_i^{\rm L}T}\mathcal{O}^{\rm phys}_\Delta(t^{\rm L}_i))\tr(\prod_{i=1}^k\rme^{\rmi t_i^{\rm R}T}\mathcal{O}^{\rm phys}_\Delta(t^{\rm R}_i))}_{\rm cyl}=\sum_{\sigma\in S_k}\tr_{\mathcal{H}}\qty(\prod_{i=0}^k\rme^{\rmi\qty(t^{\rm L}_{i}+t^{\rm R}_{\sigma(i)})T}q^{\Delta \hat{n}})~.
\end{align}
Here $S_k$ represents the symmetric group, and $\mathcal{H}$ is the chord Hilbert space. 

If the dS holographic dictionary \cite{Verlinde:2024zrh} holds, the ($2k$)-point function (\ref{eq:matter correlators query}) corresponds to a ($2k$)-point function in the LdS$_2$ CFT, which can be computed using the rule (\ref{eq:junction rule Liouville}). It can be seen, for instance in Fig. \ref{fig:LR_LdS}, that to compute the $(2k)$-point correlator in the LdS$_2$ CFT one needs a total of ($2k$)-junctions between pairwise contractions of matter operators $\mathcal{O}_\Delta^{\rm phys}(\tau_i^{\rm L})$ and $\mathcal{O}_\Delta^{\rm phys}(\tau_j^{\rm R})$ in $\Sigma_{\rm L/R}$ respectively. The same can be seen in the cylinder diagram of the DSSYK model (Fig. \ref{fig:Correlators_DDSSYK}). 

If we sum over an $S_k$ orbit, one recovers an object (i.e. the $(2k)$-point correlator) that lives on the trivial representation of $S_k$.\footnote{This is similar to the s-wave reduction of the vacuum CFT in Sec. \ref{ssec:query}. We thank Bartek Czech for comments about this.} {Moreover, lower order-point correlation functions of the L/R LdS$_2$ CFTs can be computed by adding a trivial representation in the junction rule (\ref{eq:junction rule Liouville}). From the perspective of the DSSYK model, instead of summing an $S_k$ orbit for $k$ fixed to compute the ($2k$)-point function (\ref{eq:matter correlators query}), we have to evaluate over all the $(2k)$-point correlation functions for $k\leq k_{\rm max}$, such that lower than $k_{\rm max}$-point correlators are included in the DSSYK dual of query complexity.} Thus, the CFT definition in (\ref{eq:Query n fusions}) corresponds to a sum of matter insertions $k\leq k_{\rm max}$ in the doubled DSSYK model, resulting in:
\begin{equation}\label{eq:CQ DDSSYK}
    \boxed{\mathcal{C}_{\rm Q}= k_{\rm max}(k_{\rm max}+1)/2~,}
\end{equation}
where the number of insertions in the edges of the cylinder determines the number of $S_{k\leq k_{\rm max}}$ combinations of chord diagrams in (\ref{eq:matter correlators query}).

\begin{figure}
    \centering
    \includegraphics[width=0.7\textwidth]{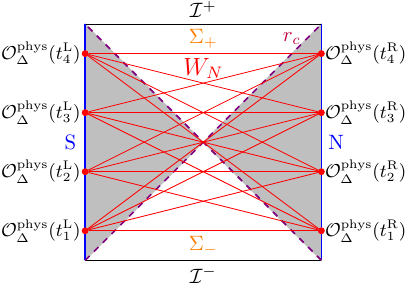}
    \caption{Wilson line network (labeled $W_N$, red lines) for a few operator insertions (red dots) along the static patch worldlines of SdS$_3$ space following the algorithm computing $S_{k\leq k_{\rm max}}$ symmetric ($2k$)-point correlation functions in the (\ref{eq:matter correlators query}), which is illustrated here for a fixed number, $k=4$. The future (past) surface where the network ends (starts) is labeled $\Sigma_+$ ($\Sigma_-$). Notice that there are $k$ number of junctions for every vertex.}
    \label{fig:SdS Wilson Network}
\end{figure}

Meanwhile, from the bulk perspective, query complexity counts the number of junctions of Wilson lines connecting the N to S patches, as shown in Fig. \ref{fig:SdS Wilson Network}.
In contrast to the case presented in Sec. \ref{ssec:query} there is no static time slice connecting both the N and S poles and, in principle, query complexity in this protocol needs to incorporate general local geometric invariants in the bulk spacetime of the form
\begin{equation}\label{eq:CAny}
    \boxed{\mathcal{C}_{\rm Q}=\int_{W_N}\rmd^{3}x~ \mathcal{F}[g_{\mu\nu}]+\sum_{\varepsilon=\pm}\int_{\Sigma_\varepsilon}\rmd^{2}\sigma~ \mathcal{G}_\varepsilon[g_{\mu\nu},~X_\varepsilon^\mu]~,}
\end{equation}
where $W_N$ corresponds to the Wilson line network manifold (see Fig. \ref{fig:SdS Wilson Network}); $\Sigma_{\pm}$ are codimension-one space-like slices on the future and past of $W_N$ which represent cutoff surfaces of the network near $\mathcal{I}^\pm$; $\sigma_i$ ($i=1,~2$) are coordinates on the $\Sigma_{\pm}$; while $\mathcal{F}[g_{\mu\nu}]$ is a scalar functional of the 3D bulk curvature invariants involving the metric $g_{\mu\nu}$; and similarly for $\mathcal{G}_{\pm}[g_{\mu\nu},~X_\pm^\mu]$, which are local invariant functionals constructed from the bulk metric and the embedding functions $X_\pm^\mu(\sigma_i)$ of $\Sigma_\pm$ respectively (e.g. the extrinsic curvature and torsion encountered in Sec. \ref{ssec:query}). {(\ref{eq:CAny}) is part of the family of codimension-zero CAny conjectures \cite{Belin:2021bga,Belin:2022xmt} for SdS$_3$ space \cite{Jorstad:2022mls,Aguilar-Gutierrez:2023zqm,Aguilar-Gutierrez:2023pnn,Aguilar-Gutierrez:2024rka}. In this case, the complexity surfaces would be anchored to the N and S worldline observers. Similar to footnote \ref{fn:Powers K T}, we can further restrict the form of $\mathcal{F}[\dots]$ and $\mathcal{G}_{\pm}[\dots]$ by requiring $\mathcal{C}_{\rm Q}$ to be well-behaved under the constraint that the Wilson line network is not smooth at the location of the junctions (i.e. where the operators $\mathcal{O}(x_i)$ are located in Fig. \ref{fig:SdS Wilson Network}). This requirement would then allow us to have a concrete holographic dual of query complexity, which we leave for future work.}

\section{Nielsen complexity in the DSSYK model}\label{sec:comments Nielsen}
In this section, we investigate Nielsen's geometric approach \cite{Nielsen1,Nielsen2,Nielsen3} in the DSSYK model using the bi-invariant proposal in (\ref{eq:biinvariant complexity}). It reproduces the linear growth expected for the CAny holographic complexity proposals in planar black holes in AdS space\cite{Belin:2021bga,Belin:2022xmt}, and in asymptotically dS spacetimes \cite{Aguilar-Gutierrez:2023zqm,Aguilar-Gutierrez:2023pnn,Aguilar-Gutierrez:2024rka}.

We begin evaluating (\ref{eq:biinvariant complexity}). In the context of the DSSYK model, the most natural choice for the Hilbert space where unitaries (\ref{eq:unitaries}) can act, and to evaluate the traces, is over the auxiliary chord Hilbert space $\mathcal{H}$. In that case, we can use $H=-T$ and (\ref{eq:identity theta}), to express (\ref{eq:unitaries}) as
\begin{align}
  x(t)=\rme^{-\rmi V}~,\quad  V=\int_0^\pi\frac{\rmd\theta}{2\pi}\mu(\theta)\qty(tE(\theta)+2\pi K_n)\ket{\theta}\bra{\theta}~.\label{eq:Nielsen intermediate}
\end{align}
(\ref{eq:biinvariant complexity}) with (\ref{eq:Nielsen intermediate}) then transforms into
\begin{equation}\label{eq:complicated CN}
    \mathcal{C}_{\rm N}(t)=\min_{\qty{K_n:~\sum_{n=0}^\infty K_n=0}}\sqrt{\sum_{m=0}^\infty{\int_0^\pi\frac{\rmd\theta}{2\pi}\mu(\theta)\qty(tE(\theta)+2\pi K_n)^2\frac{\qty(H_m(\cos\theta|q))^2~}{(q;~q)_m}}}~.
\end{equation}
We can perform the minimization above noticing that since $T$ is traceless, then it follows that $K_n=0$. (\ref{eq:complicated CN}) becomes:
\begin{equation}\label{eq:CN after minimization}
\begin{aligned}
\mathcal{C}_{\rm N}(t)&=t\sqrt{\sum_{m=0}^\infty{\int_0^\pi\frac{\rmd\theta}{2\pi}\mu(\theta)E(\theta)^2\frac{\qty(H_m(\cos\theta|q))^2~}{(q;~q)_m}}}\\
    &=t\sqrt{\bra{0,~0}\mathcal{E}^{\dagger}(T\otimes T)\mathcal{E}\ket{0,~0}}~,
\end{aligned}
\end{equation}
where in the second line, we have used the symmetry of $T$ in the orthonormal chord basis (\ref{eq:Transfer matrix}), and we reintroduced the entangler operator (\ref{eq:entangler}) in the doubled Hilbert space. Thus, the particular definition of Nielsen complexity (\ref{eq:biinvariant complexity}) in the DSSYK model measures the vacuum expectation value of an operator entangling chords in a doubled Hilbert space (such as for the doubled DSSYK model), with a similar structure to spread complexity in (\ref{eq:spread as counting gates}).

In the context of dS holography, the choice of bi-invariant $\mathcal{C}_{\rm N}(t)$ would have a corresponding dual observable exhibiting a static patch time linear growth in SdS$_3$ space. However, we \textit{have not} identified such observable with the dS holographic dictionary. Nevertheless, we hope this is a first step towards developing this side of the dictionary. Moreover, we emphasize that the evolution in (\ref{eq:CN after minimization}) is reproduced by certain codimension-one CAny proposals in asymptotically dS spacetimes \cite{Aguilar-Gutierrez:2023zqm}. Notice also there is a lot of freedom in the definition of Nielsen complexity (\ref{eq:Nielsen General}), so in principle, there could be other proposals where the hyperfast growth (\ref{eq:hyperfast}) could be reproduced instead \cite{Jorstad:2022mls}.

\subsection{JT gravity regime}
Next, we would like to evaluate (\ref{eq:complicated CN}) in the semiclassical limit, where $\lambda\rightarrow0$ (i.e. $q\rightarrow1$). The evaluation is still quite involved, but there is an analytically tractable limit, where the dominating terms in the sum are $m\sim\mathcal{O}(1/\lambda)$. Under these considerations, we can approximate the integral in (\ref{eq:complicated CN}) as \cite{Okuyama:2023yat}
\begin{align}
   &\int_0^\pi\frac{\rmd\theta}{2\pi}\mu(\theta)\tilde{G}(E(\theta))\frac{\qty(H_m(\cos\theta|q))^2}{(q;~q)_m}\simeq\int_0^\pi\frac{\rmd\theta}{2\pi}\int_0^{2\pi}\frac{\rmd\Phi}{2\pi}\rme^{-\lambda \tilde{F}(\theta,~\Phi)}\tilde{G}(E(\theta))~,\label{eq:int DiLog}
\end{align}
where $\tilde{G}(E(\theta))=t^2E(\theta)^2$ in our case, while
\begin{equation}
    \tilde{F}(\theta,~\Phi)\equiv\rmi\lambda m\Phi+\text{Li}_2(\rme^{2\rmi\Phi})-2\text{Li}_2(\rme^{\rmi\Phi})+\sum_{\varepsilon=\pm}\qty[\text{Li}_2(\rme^{\varepsilon2\rmi\theta})-\text{Li}_2(\rme^{\rmi\Phi+\varepsilon2\rmi\theta})]~,
\end{equation}
and Li$_2(x)=\sum_{k=1}^\infty x^k/k^2$ is the dilogarithm function. 

We now evaluate (\ref{eq:int DiLog}) with a saddle point approximation, which satisfies the conditions:
\begin{equation}
    \eval{\partial_{\theta}\tilde{F}(\theta,~\Phi)}_{\theta=\theta_S,~\Phi=\Phi_S}=\eval{\partial_{\Phi}\tilde{F}(\theta,~\Phi)}_{\theta=\theta_S,~\Phi=\Phi_S}=0~,
\end{equation}
resulting in $\sin^2\theta_S=q^m$, $\Phi_S=0$. Thus, combining (\ref{eq:complicated CN}) with (\ref{eq:int DiLog}) in the saddle point solution ($\theta_S$, $\Phi_S$), one has:
\begin{equation}\label{eq:Nielsen JT limit t}
    \mathcal{C}_{\rm N}(t)\simeq t\sqrt{\frac{J^2}{\lambda}\sum_{m=0}^\infty\frac{1-q^m}{1-q}}~.
\end{equation}
Note that the infinite sum above is divergent for $t\neq0$. This is expected for complexity without a regulating surface \cite{Chen:2020nlj} in the bulk of an asymptotically AdS spacetime. On the DSSYK side, one can instead consider that the cutoff can be implemented by truncating the series to a finite number of terms. 

As a remark, while the definition of $\mathcal{C}_{\rm N}$ in (\ref{eq:biinvariant complexity}) generically reproduces a late time linear growth; as we have emphasized below (\ref{eq:complicated CN}), many other possible behaviors could be recovered by properly choosing the cost function. For instance, one could use a different symmetry principle with respect to (\ref{eq:f unitary}, \ref{eq:f time reverse}). Also notice that while (\ref{eq:complicated CN}) is valid for the doubled DSSYK model, the condition (\ref{eq:int DiLog}) leading to (\ref{eq:Nielsen JT limit t}) assumes that $\lambda\rightarrow0$ and $m\sim\mathcal{O}(1/\lambda)$. In the holographic context, this regime is appropriate for studying JT gravity (see e.g. \cite{Lin:2022rbf}) instead of dS space.

\section{Discussion}\label{sec:conclusions}
In \emph{summary}, we studied concrete notions of complexity in the context of the holographic correspondence between the doubled DSSYK model, LdS$_2$ CFT, and SdS$_3$ space. The main results are shown in Table \ref{tab:results}. First, we showed that the \textit{spread complexity} in the doubled DSSYK model can be expressed as the number of entangled chord states in its maximal entropy state. We interpreted boost symmetries fixing the time difference between antipodal observers in SdS$_3$ space in terms of a renormalization condition of the spread complexity in the maximal entropy state. This leads to a connection between entanglement, geometry, and complexity in dS holography. Second, we used the correlation functions in the doubled DSSYK model, the LdS$_2$ CFT, and SdS$_3$ space to calculate the respective \textit{Krylov complexity} on all sides of the correspondence, and we showed they display the exponential time growth expected for maximally chaotic systems, with the expected maximal Lyapunov exponent. Later, we introduced the concept of \textit{query complexity} for the LdS$_2$ CFT, which counts the number of steps in an algorithm computing multipoint correlators between antipodal static patches. We described query complexity in terms of matter chord diagrams in a cylinder geometry in the doubled DSSYK model, and a network of Wilson lines in SdS$_3$ space. Here, we recognized a connection with the CAny proposals \cite{Belin:2021bga,Belin:2022xmt} in dS space. The geometric invariant terms involved in query complexity can be further constrained by demanding regularity on the network. Finally, we evaluated the \textit{Nielsen operator complexity} of the DSSYK model for a specific proposal where linear time growth is recovered. Although we did not identify its precise dual observable in the other sides of the correspondence in this latter proposal, it shares the late time behavior of certain holographic complexity conjectures in asymptotically dS spacetimes \cite{Aguilar-Gutierrez:2023zqm,Aguilar-Gutierrez:2023tic,Aguilar-Gutierrez:2023pnn,Aguilar-Gutierrez:2024rka}. 

Based on all these approaches, the doubled DSSYK model \cite{Narovlansky:2023lfz,Verlinde:2024znh,Verlinde:2024zrh} is a promising arena to develop complexity, and perhaps other quantum information-theoretic notions, in low-dimensional dS space holography. While several elements of the correspondence need to be developed further, we hope this work provides a step forward. We conclude with some questions left for future work.

\subsubsection*{Gibbons-Hawking entropy and thermodynamics} 
As we noticed in Sec. \ref{sec:Spread}, the spread complexity for the maximal entropy state is time-independent, which we related to the boost isometries allowing us to fix the time-difference between antipodal observers in SdS$_3$ space. Moreover, this observable has a natural interpretation in terms of counting entangled chord states. Perhaps, this time-independence for the maximal entropy states is a microscopic manifestation of the Gibbons-Hawking entropy being constant in time from the perspective of a worldline static patch observer. It would be interesting to study how the thermodynamic properties of the bulk theory are encoded in the explicit microscopic models and whether they can be manifested in the complexity proposals. For instance, there is a conjectured relation between the temperature, entropy and holographic complexity of AdS black holes \cite{Brown:2015bva,Brown:2015lvg} based on the Lloyd bound \cite{lloyd2000ultimate} (recently also hinted for SdS$_{d+1\geq4}$ black holes in \cite{Aguilar-Gutierrez:2024rka}):
\begin{equation}
    \text{Late~times:}\quad\dv{\mathcal{C}}{t}\sim TS~.
\end{equation}
It would be interesting to learn if a similar type of relation can be found in the dS holographic approach for any of the complexity proposals in our work.

On the other hand, an important aspect to develop in the dS holographic correspondence is the thermodynamic stability of the solutions, given that it has been found that dS$_3$ space with Dirichlet time-like boundaries is thermodynamically unstable \cite{Svesko:2022txo,Banihashemi:2022jys,Banihashemi:2022htw,Anninos:2024wpy}.\footnote{We thank Damian Galante and Andrew Svesko for related discussions.} Since the quantization surface involved in the derivation of the LdS$_2$ CFT from SdS$_3$ space required Dirichlet boundary conditions \cite{Verlinde:2024zrh}, it might be useful to study its thermodynamic stability, and, possibly, to consider other choices of boundary conditions (such as conformal boundary conditions) for the quantization surface.

\subsection*{MERA networks}
We have found that the spread and Nielsen complexity can be expressed in terms of entangler/disentangler operators (see (\ref{eq:C S DDSSYK}, \ref{eq:CN after minimization}) respectively) acting on the doubled Hilbert space of the DSSYK model \cite{Okuyama:2024yya}, which is reminiscent of a MERA network. However, to take a step further and associate this type of tensor network to dS$_3$ space\footnote{See \cite{Beny:2011vh,Bao:2017qmt,Bao:2017iye,Niermann:2021wco,Cao:2023gkw} for previous approaches to tensor networks and quantum circuit models of dS space.}, one would need to include a universal gate set of both (dis)entanglers and isometries in a hierarchical order that determines the causal structure of the MERA network \cite{Vidal:2008zz,Beny:2011vh}. This is currently not present (at least not manifestly) in any of the proposals. It would be interesting to pursue this quantum circuit description of dS space (see related work in \cite{Beny:2011vh}) emerging from the doubled DSSYK model.\footnote{We thank Pratik Nandy for interesting comments about this issue.} Perhaps this can be more naturally studied within the Fubini-Study distance approach to holographic complexity in \cite{Erdmenger:2022lov}.

\subsection*{Late time linear, or hyperfast growth?}
We studied a very particular notion of Nielsen complexity for the DSSYK model, which we showed grows linearly in time and it can be described in terms of entangler operators (\ref{eq:CN after minimization}). Given the universal scaling of computational complexity with system size, which motivated the CAny conjectures \cite{Belin:2021bga,Belin:2022xmt}, it would be useful to learn if the characteristics that we have encountered for the pair of DSSYK models are also generic for more intricate Nielsen complexity proposals, in view of related recent studies for bipartite multiparticle quantum systems in \cite{Baiguera:2023bhm}. It could be beneficial to show if the evolution of some of these definitions in the DSSYK model can be matched with the hyperfast growth of certain holographic complexity proposals (\ref{eq:hyperfast}) in asymptotically dS spacetimes \cite{Susskind:2021esx,Chapman:2021eyy,Jorstad:2022mls,Galante:2022nhj,Auzzi:2023qbm,Anegawa:2023wrk,Anegawa:2023dad,Baiguera:2023tpt,Aguilar-Gutierrez:2024rka}.

\subsection*{Non-unitary dS holography} Throughout our discussion, we have been working with a unitary microscopic theory, the doubled DSSYK model, as motivated by the cosmological central dogma in static patch holography. However, in higher dimensions, dS space is known to undergo vacuum decay due to bubble nucleation (see e.g. \cite{Kashyap:2015lva}). This prompts us to introduce non-Hermitian terms in (\ref{eq:DDSSYK Hamiltonian}), such as multiple DSSYK Hamiltonians (\ref{eq:DDSSYK Hamiltonian}) with a different number of fermion interactions (see \cite{Anninos:2022qgy,Anninos:2020cwo} for the original proposal, motivated from thermodynamic considerations):
\begin{equation}\label{eq:non unitary}
    H=H^{(p)}_{\rm DSSYK}+\sum_i\lambda_i H^{(q_i)}_{\rm DSSYK}~,
\end{equation}
where the superscript ($p$, $q_i$) denotes the number of fermion interactions in (\ref{eq:DDSSYK Hamiltonian}), with $\lambda_i\in\mathbb{C}$, and $q_i\rightarrow\infty$ in the double scaling limit (while keeping $q_i<p$ for the non-unitary term to be relevant). A similar type of interpolated model has recently appeared in \cite{Berkooz:2024evs,Berkooz:2024ofm}. One might perform a similar analysis of the holographic dictionary of \cite{Narovlansky:2023lfz,Verlinde:2024znh,Verlinde:2024zrh} for theories that incorporate (\ref{eq:non unitary}). Moreover, non-unitary evolution is important for modeling measurement-induced dynamics. This has been recently studied in the holographic systems by \cite{Antonini:2023aza,Antonini:2022lmg}, and using Krylov complexity in \cite{Bhattacharya:2023yec}.\footnote{We thank Rathindra Nath Das for useful discussions on this point.} It could be interesting to incorporate these effects in dS holography and probe them with the complexity proposals of our work.

\section*{Acknowledgements}
I thank Andreas Blommaert, Arghya Chattopadhyay, Ohad Mamroud, Andrew Svesko, Thomas Van Riet, and Erik Verlinde for illuminating discussions; and Stefano Baiguera, Bartlomiej Czech, Rathindra Nath Das, Damián Galante, Pratik Nandy, Kazumi Okuyama, Andrew Rolph and Nicolo Zenoni for their comments and questions on an earlier version of the draft. The work of SEAG is partially supported by the FWO Research Project G0H9318N and the inter-university project iBOF/21/084.

\bibliographystyle{JHEP}
\bibliography{references.bib}

\providecommand{\href}[2]{#2}\begingroup\raggedright\begin{thebibliography}{100}

\bibitem{Narovlansky:2023lfz}
V.~Narovlansky and H.~Verlinde, \emph{{Double-scaled SYK and de Sitter
  Holography}},  \href{https://arxiv.org/abs/2310.16994}{{\ttfamily
  2310.16994}}.

\bibitem{Verlinde:2024znh}
H.~Verlinde, \emph{{Double-scaled SYK, Chords and de Sitter Gravity}},
  \href{https://arxiv.org/abs/2402.00635}{{\ttfamily 2402.00635}}.

\bibitem{Verlinde:2024zrh}
H.~Verlinde and M.~Zhang, \emph{{SYK Correlators from 2D Liouville-de Sitter
  Gravity}},  \href{https://arxiv.org/abs/2402.02584}{{\ttfamily 2402.02584}}.

\bibitem{Maldacena:1997re}
J.M.~Maldacena, \emph{{The Large N limit of superconformal field theories and
  supergravity}}, \href{https://doi.org/10.4310/ATMP.1998.v2.n2.a1}{\emph{Adv.
  Theor. Math. Phys.} {\bfseries 2} (1998) 231}
  [\href{https://arxiv.org/abs/hep-th/9711200}{{\ttfamily hep-th/9711200}}].

\bibitem{Gubser:1998bc}
S.S.~Gubser, I.R.~Klebanov and A.M.~Polyakov, \emph{{Gauge theory correlators
  from noncritical string theory}},
  \href{https://doi.org/10.1016/S0370-2693(98)00377-3}{\emph{Phys. Lett. B}
  {\bfseries 428} (1998) 105}
  [\href{https://arxiv.org/abs/hep-th/9802109}{{\ttfamily hep-th/9802109}}].

\bibitem{Witten:1998qj}
E.~Witten, \emph{{Anti-de Sitter space and holography}},
  \href{https://doi.org/10.4310/ATMP.1998.v2.n2.a2}{\emph{Adv. Theor. Math.
  Phys.} {\bfseries 2} (1998) 253}
  [\href{https://arxiv.org/abs/hep-th/9802150}{{\ttfamily hep-th/9802150}}].

\bibitem{Strominger:2001pn}
A.~Strominger, \emph{{The dS / CFT correspondence}},
  \href{https://doi.org/10.1088/1126-6708/2001/10/034}{\emph{JHEP} {\bfseries
  10} (2001) 034} [\href{https://arxiv.org/abs/hep-th/0106113}{{\ttfamily
  hep-th/0106113}}].

\bibitem{Witten:2001kn}
E.~Witten, \emph{{Quantum gravity in de Sitter space}},  in \emph{{Strings
  2001: International Conference}}, 6, 2001
  [\href{https://arxiv.org/abs/hep-th/0106109}{{\ttfamily hep-th/0106109}}].

\bibitem{Maldacena:2002vr}
J.M.~Maldacena, \emph{{Non-Gaussian features of primordial fluctuations in
  single field inflationary models}},
  \href{https://doi.org/10.1088/1126-6708/2003/05/013}{\emph{JHEP} {\bfseries
  05} (2003) 013} [\href{https://arxiv.org/abs/astro-ph/0210603}{{\ttfamily
  astro-ph/0210603}}].

\bibitem{Gibbons:1977mu}
G.W.~Gibbons and S.W.~Hawking, \emph{{Cosmological Event Horizons,
  Thermodynamics, and Particle Creation}},
  \href{https://doi.org/10.1103/PhysRevD.15.2738}{\emph{Phys. Rev. D}
  {\bfseries 15} (1977) 2738}.

\bibitem{Bousso:1999dw}
R.~Bousso, \emph{{The Holographic principle for general backgrounds}},
  \href{https://doi.org/10.1088/0264-9381/17/5/309}{\emph{Class. Quant. Grav.}
  {\bfseries 17} (2000) 997}
  [\href{https://arxiv.org/abs/hep-th/9911002}{{\ttfamily hep-th/9911002}}].

\bibitem{Bousso:2000nf}
R.~Bousso, \emph{{Positive vacuum energy and the N bound}},
  \href{https://doi.org/10.1088/1126-6708/2000/11/038}{\emph{JHEP} {\bfseries
  11} (2000) 038} [\href{https://arxiv.org/abs/hep-th/0010252}{{\ttfamily
  hep-th/0010252}}].

\bibitem{Banks:2006rx}
T.~Banks, B.~Fiol and A.~Morisse, \emph{{Towards a quantum theory of de Sitter
  space}}, \href{https://doi.org/10.1088/1126-6708/2006/12/004}{\emph{JHEP}
  {\bfseries 12} (2006) 004}
  [\href{https://arxiv.org/abs/hep-th/0609062}{{\ttfamily hep-th/0609062}}].

\bibitem{Anninos:2011af}
D.~Anninos, S.A.~Hartnoll and D.M.~Hofman, \emph{{Static Patch Solipsism:
  Conformal Symmetry of the de Sitter Worldline}},
  \href{https://doi.org/10.1088/0264-9381/29/7/075002}{\emph{Class. Quant.
  Grav.} {\bfseries 29} (2012) 075002}
  [\href{https://arxiv.org/abs/1109.4942}{{\ttfamily 1109.4942}}].

\bibitem{Banks:2018ypk}
T.~Banks and W.~Fischler, \emph{{The holographic spacetime model of
  cosmology}}, \href{https://doi.org/10.1142/S0218271818460057}{\emph{Int. J.
  Mod. Phys. D} {\bfseries 27} (2018) 1846005}
  [\href{https://arxiv.org/abs/1806.01749}{{\ttfamily 1806.01749}}].

\bibitem{Shaghoulian:2021cef}
E.~Shaghoulian, \emph{{The central dogma and cosmological horizons}},
  \href{https://doi.org/10.1007/JHEP01(2022)132}{\emph{JHEP} {\bfseries 01}
  (2022) 132} [\href{https://arxiv.org/abs/2110.13210}{{\ttfamily
  2110.13210}}].

\bibitem{Almheiri:2020cfm}
A.~Almheiri, T.~Hartman, J.~Maldacena, E.~Shaghoulian and A.~Tajdini,
  \emph{{The entropy of Hawking radiation}},
  \href{https://doi.org/10.1103/RevModPhys.93.035002}{\emph{Rev. Mod. Phys.}
  {\bfseries 93} (2021) 035002}
  [\href{https://arxiv.org/abs/2006.06872}{{\ttfamily 2006.06872}}].

\bibitem{Spradlin:2001pw}
M.~Spradlin, A.~Strominger and A.~Volovich, \emph{{Les Houches lectures on de
  Sitter space}},  in \emph{{Les Houches Summer School: Session 76: Euro Summer
  School on Unity of Fundamental Physics: Gravity, Gauge Theory and Strings}},
  pp.~423--453, 10, 2001
  [\href{https://arxiv.org/abs/hep-th/0110007}{{\ttfamily hep-th/0110007}}].

\bibitem{Bousso:2002fq}
R.~Bousso, \emph{{Adventures in de Sitter space}},  in \emph{{Workshop on
  Conference on the Future of Theoretical Physics and Cosmology in Honor of
  Steven Hawking's 60th Birthday}}, pp.~539--569, 5, 2002
  [\href{https://arxiv.org/abs/hep-th/0205177}{{\ttfamily hep-th/0205177}}].

\bibitem{Anninos:2012qw}
D.~Anninos, \emph{{De Sitter Musings}},
  \href{https://doi.org/10.1142/S0217751X1230013X}{\emph{Int. J. Mod. Phys. A}
  {\bfseries 27} (2012) 1230013}
  [\href{https://arxiv.org/abs/1205.3855}{{\ttfamily 1205.3855}}].

\bibitem{Galante:2023uyf}
D.A.~Galante, \emph{{Modave lectures on de Sitter space \& holography}},
  \href{https://doi.org/10.22323/1.435.0003}{\emph{PoS} {\bfseries Modave2022}
  (2023) 003} [\href{https://arxiv.org/abs/2306.10141}{{\ttfamily
  2306.10141}}].

\bibitem{Leuven:2018ejp}
S.~Leuven, E.~Verlinde and M.~Visser, \emph{{Towards non-AdS Holography via the
  Long String Phenomenon}},
  \href{https://doi.org/10.1007/JHEP06(2018)097}{\emph{JHEP} {\bfseries 06}
  (2018) 097} [\href{https://arxiv.org/abs/1801.02589}{{\ttfamily
  1801.02589}}].

\bibitem{Susskind:2021esx}
L.~Susskind, \emph{{Entanglement and Chaos in De Sitter Space Holography: An
  SYK Example}}, \href{https://doi.org/10.22128/jhap.2021.455.1005}{\emph{JHAP}
  {\bfseries 1} (2021) 1} [\href{https://arxiv.org/abs/2109.14104}{{\ttfamily
  2109.14104}}].

\bibitem{Sachdev_1993}
S.~Sachdev and J.~Ye, \emph{Gapless spin-fluid ground state in a random quantum
  heisenberg magnet},
  \href{https://doi.org/10.1103/physrevlett.70.3339}{\emph{Physical Review
  Letters} {\bfseries 70} (1993) 3339–3342}.

\bibitem{kitaevTalks}
A.~Kitaev, ``{Talks given at the Fundamental Physics Prize Symposium and KITP
  seminars}.''

\bibitem{Rosenhaus:2018dtp}
V.~Rosenhaus, \emph{{An introduction to the SYK model}},
  \href{https://doi.org/10.1088/1751-8121/ab2ce1}{\emph{J. Phys. A} {\bfseries
  52} (2019) 323001} [\href{https://arxiv.org/abs/1807.03334}{{\ttfamily
  1807.03334}}].

\bibitem{Sachdev:2024gas}
S.~Sachdev, \emph{{Quantum glasses, reparameterization invariance,
  Sachdev-Ye-Kitaev models}},
  \href{https://arxiv.org/abs/2402.17824}{{\ttfamily 2402.17824}}.

\bibitem{Susskind:2022bia}
L.~Susskind, \emph{{De Sitter Space, Double-Scaled SYK, and the Separation of
  Scales in the Semiclassical Limit}},
  \href{https://arxiv.org/abs/2209.09999}{{\ttfamily 2209.09999}}.

\bibitem{Rahman:2022jsf}
A.A.~Rahman, \emph{{dS JT Gravity and Double-Scaled SYK}},
  \href{https://arxiv.org/abs/2209.09997}{{\ttfamily 2209.09997}}.

\bibitem{HVtalks}
H.~Verlinde, \emph{{Talks given at the QGQC5 conference, UC Davis; the Franqui
  Symposium, Brussels; `Quantum Gravity on Southern Cone', Argentina; and `SYK
  models and Gauge Theory' workshop at Weizmann Institute}},  2019.

\bibitem{Gorbenko:2018oov}
V.~Gorbenko, E.~Silverstein and G.~Torroba, \emph{{dS/dS and $ T\overline{T}
  $}}, \href{https://doi.org/10.1007/JHEP03(2019)085}{\emph{JHEP} {\bfseries
  03} (2019) 085} [\href{https://arxiv.org/abs/1811.07965}{{\ttfamily
  1811.07965}}].

\bibitem{Lewkowycz:2019xse}
A.~Lewkowycz, J.~Liu, E.~Silverstein and G.~Torroba, \emph{{$ T\overline{T} $
  and EE, with implications for (A)dS subregion encodings}},
  \href{https://doi.org/10.1007/JHEP04(2020)152}{\emph{JHEP} {\bfseries 04}
  (2020) 152} [\href{https://arxiv.org/abs/1909.13808}{{\ttfamily
  1909.13808}}].

\bibitem{Shyam:2021ciy}
V.~Shyam, \emph{{$ \mathrm{T}\overline{\mathrm{T}} $ +
  \ensuremath{\Lambda}$_{2}$ deformed CFT on the stretched dS$_{3}$ horizon}},
  \href{https://doi.org/10.1007/JHEP04(2022)052}{\emph{JHEP} {\bfseries 04}
  (2022) 052} [\href{https://arxiv.org/abs/2106.10227}{{\ttfamily
  2106.10227}}].

\bibitem{Coleman:2021nor}
E.~Coleman, E.A.~Mazenc, V.~Shyam, E.~Silverstein, R.M.~Soni, G.~Torroba
  et~al., \emph{{De Sitter microstates from T$ \overline{T} $ +
  \ensuremath{\Lambda}$_{2}$ and the Hawking-Page transition}},
  \href{https://doi.org/10.1007/JHEP07(2022)140}{\emph{JHEP} {\bfseries 07}
  (2022) 140} [\href{https://arxiv.org/abs/2110.14670}{{\ttfamily
  2110.14670}}].

\bibitem{Batra:2024kjl}
G.~Batra, G.B.~De~Luca, E.~Silverstein, G.~Torroba and S.~Yang,
  \emph{{Bulk-local dS$_3$ holography: the Matter with $T\bar T+\Lambda_2$}},
  \href{https://arxiv.org/abs/2403.01040}{{\ttfamily 2403.01040}}.

\bibitem{Svesko:2022txo}
A.~Svesko, E.~Verheijden, E.P.~Verlinde and M.R.~Visser, \emph{{Quasi-local
  energy and microcanonical entropy in two-dimensional nearly de Sitter
  gravity}}, \href{https://doi.org/10.1007/JHEP08(2022)075}{\emph{JHEP}
  {\bfseries 08} (2022) 075}
  [\href{https://arxiv.org/abs/2203.00700}{{\ttfamily 2203.00700}}].

\bibitem{Banihashemi:2022jys}
B.~Banihashemi and T.~Jacobson, \emph{{Thermodynamic ensembles with
  cosmological horizons}},
  \href{https://doi.org/10.1007/JHEP07(2022)042}{\emph{JHEP} {\bfseries 07}
  (2022) 042} [\href{https://arxiv.org/abs/2204.05324}{{\ttfamily
  2204.05324}}].

\bibitem{Banihashemi:2022htw}
B.~Banihashemi, T.~Jacobson, A.~Svesko and M.~Visser, \emph{{The minus sign in
  the first law of de Sitter horizons}},
  \href{https://doi.org/10.1007/JHEP01(2023)054}{\emph{JHEP} {\bfseries 01}
  (2023) 054} [\href{https://arxiv.org/abs/2208.11706}{{\ttfamily
  2208.11706}}].

\bibitem{Anninos:2024wpy}
D.~Anninos, D.A.~Galante and C.~Maneerat, \emph{{Cosmological Observatories}},
  \href{https://arxiv.org/abs/2402.04305}{{\ttfamily 2402.04305}}.

\bibitem{Milekhin:2023bjv}
A.~Milekhin and J.~Xu, \emph{{Revisiting Brownian SYK and its possible
  relations to de Sitter}},  \href{https://arxiv.org/abs/2312.03623}{{\ttfamily
  2312.03623}}.

\bibitem{Xu:2024hoc}
J.~Xu, \emph{{Von Neumann Algebras in Double-Scaled SYK}},
  \href{https://arxiv.org/abs/2403.09021}{{\ttfamily 2403.09021}}.

\bibitem{Balasubramanian:2001nb}
V.~Balasubramanian, J.~de~Boer and D.~Minic, \emph{{Mass, entropy and
  holography in asymptotically de Sitter spaces}},
  \href{https://doi.org/10.1103/PhysRevD.65.123508}{\emph{Phys. Rev. D}
  {\bfseries 65} (2002) 123508}
  [\href{https://arxiv.org/abs/hep-th/0110108}{{\ttfamily hep-th/0110108}}].

\bibitem{Ghezelbash:2001vs}
A.M.~Ghezelbash and R.B.~Mann, \emph{{Action, mass and entropy of
  Schwarzschild-de Sitter black holes and the de Sitter / CFT correspondence}},
  \href{https://doi.org/10.1088/1126-6708/2002/01/005}{\emph{JHEP} {\bfseries
  01} (2002) 005} [\href{https://arxiv.org/abs/hep-th/0111217}{{\ttfamily
  hep-th/0111217}}].

\bibitem{Witten:1988hc}
E.~Witten, \emph{{(2+1)-Dimensional Gravity as an Exactly Soluble System}},
  \href{https://doi.org/10.1016/0550-3213(88)90143-5}{\emph{Nucl. Phys. B}
  {\bfseries 311} (1988) 46}.

\bibitem{Castro:2011xb}
A.~Castro, N.~Lashkari and A.~Maloney, \emph{{A de Sitter Farey Tail}},
  \href{https://doi.org/10.1103/PhysRevD.83.124027}{\emph{Phys. Rev. D}
  {\bfseries 83} (2011) 124027}
  [\href{https://arxiv.org/abs/1103.4620}{{\ttfamily 1103.4620}}].

\bibitem{Castro:2023dxp}
A.~Castro, I.~Coman, J.R.~Fliss and C.~Zukowski, \emph{{Keeping matter in the
  loop in dS$_{3}$ quantum gravity}},
  \href{https://doi.org/10.1007/JHEP07(2023)120}{\emph{JHEP} {\bfseries 07}
  (2023) 120} [\href{https://arxiv.org/abs/2302.12281}{{\ttfamily
  2302.12281}}].

\bibitem{Castro:2023bvo}
A.~Castro, I.~Coman, J.R.~Fliss and C.~Zukowski, \emph{{Coupling Fields to 3D
  Quantum Gravity via Chern-Simons Theory}},
  \href{https://doi.org/10.1103/PhysRevLett.131.171602}{\emph{Phys. Rev. Lett.}
  {\bfseries 131} (2023) 171602}
  [\href{https://arxiv.org/abs/2304.02668}{{\ttfamily 2304.02668}}].

\bibitem{Ohtsuki:2002ud}
T.~Ohtsuki, \emph{{Quantum invariants: A study of knots, 3-manifolds, and their
  sets}}, De Gruyter (2002).

\bibitem{Berkooz:2022mfk}
M.~Berkooz, M.~Isachenkov, M.~Isachenkov, P.~Narayan and V.~Narovlansky,
  \emph{{Quantum groups, non-commutative AdS$_{2}$, and chords in the
  double-scaled SYK model}},
  \href{https://doi.org/10.1007/JHEP08(2023)076}{\emph{JHEP} {\bfseries 08}
  (2023) 076} [\href{https://arxiv.org/abs/2212.13668}{{\ttfamily
  2212.13668}}].

\bibitem{Blommaert:2023opb}
A.~Blommaert, T.G.~Mertens and S.~Yao, \emph{{Dynamical actions and
  q-representation theory for double-scaled SYK}},
  \href{https://doi.org/10.1007/JHEP02(2024)067}{\emph{JHEP} {\bfseries 02}
  (2024) 067} [\href{https://arxiv.org/abs/2306.00941}{{\ttfamily
  2306.00941}}].

\bibitem{Lin:2023trc}
H.W.~Lin and D.~Stanford, \emph{{A symmetry algebra in double-scaled SYK}},
  \href{https://doi.org/10.21468/SciPostPhys.15.6.234}{\emph{SciPost Phys.}
  {\bfseries 15} (2023) 234}
  [\href{https://arxiv.org/abs/2307.15725}{{\ttfamily 2307.15725}}].

\bibitem{Blommaert:2023wad}
A.~Blommaert, T.G.~Mertens and S.~Yao, \emph{{The q-Schwarzian and Liouville
  gravity}},  \href{https://arxiv.org/abs/2312.00871}{{\ttfamily 2312.00871}}.

\bibitem{Almheiri:2024ayc}
A.~Almheiri and F.K.~Popov, \emph{{Holography on the Quantum Disk}},
  \href{https://arxiv.org/abs/2401.05575}{{\ttfamily 2401.05575}}.

\bibitem{Maldacena:2016hyu}
J.~Maldacena and D.~Stanford, \emph{{Remarks on the Sachdev-Ye-Kitaev model}},
  \href{https://doi.org/10.1103/PhysRevD.94.106002}{\emph{Phys. Rev. D}
  {\bfseries 94} (2016) 106002}
  [\href{https://arxiv.org/abs/1604.07818}{{\ttfamily 1604.07818}}].

\bibitem{Cotler:2016fpe}
J.S.~Cotler, G.~Gur-Ari, M.~Hanada, J.~Polchinski, P.~Saad, S.H.~Shenker
  et~al., \emph{{Black Holes and Random Matrices}},
  \href{https://doi.org/10.1007/JHEP05(2017)118}{\emph{JHEP} {\bfseries 05}
  (2017) 118} [\href{https://arxiv.org/abs/1611.04650}{{\ttfamily
  1611.04650}}].

\bibitem{Berkooz:2018qkz}
M.~Berkooz, P.~Narayan and J.~Simon, \emph{{Chord diagrams, exact correlators
  in spin glasses and black hole bulk reconstruction}},
  \href{https://doi.org/10.1007/JHEP08(2018)192}{\emph{JHEP} {\bfseries 08}
  (2018) 192} [\href{https://arxiv.org/abs/1806.04380}{{\ttfamily
  1806.04380}}].

\bibitem{Berkooz:2018jqr}
M.~Berkooz, M.~Isachenkov, V.~Narovlansky and G.~Torrents, \emph{{Towards a
  full solution of the large N double-scaled SYK model}},
  \href{https://doi.org/10.1007/JHEP03(2019)079}{\emph{JHEP} {\bfseries 03}
  (2019) 079} [\href{https://arxiv.org/abs/1811.02584}{{\ttfamily
  1811.02584}}].

\bibitem{Lin:2022rbf}
H.W.~Lin, \emph{{The bulk Hilbert space of double scaled SYK}},
  \href{https://doi.org/10.1007/JHEP11(2022)060}{\emph{JHEP} {\bfseries 11}
  (2022) 060} [\href{https://arxiv.org/abs/2208.07032}{{\ttfamily
  2208.07032}}].

\bibitem{Chandrasekaran:2022cip}
V.~Chandrasekaran, R.~Longo, G.~Penington and E.~Witten, \emph{{An algebra of
  observables for de Sitter space}},
  \href{https://doi.org/10.1007/JHEP02(2023)082}{\emph{JHEP} {\bfseries 02}
  (2023) 082} [\href{https://arxiv.org/abs/2206.10780}{{\ttfamily
  2206.10780}}].

\bibitem{Jensen:2023yxy}
K.~Jensen, J.~Sorce and A.J.~Speranza, \emph{{Generalized entropy for general
  subregions in quantum gravity}},
  \href{https://doi.org/10.1007/JHEP12(2023)020}{\emph{JHEP} {\bfseries 12}
  (2023) 020} [\href{https://arxiv.org/abs/2306.01837}{{\ttfamily
  2306.01837}}].

\bibitem{Kudler-Flam:2023qfl}
J.~Kudler-Flam, S.~Leutheusser and G.~Satishchandran, \emph{{Generalized Black
  Hole Entropy is von Neumann Entropy}},
  \href{https://arxiv.org/abs/2309.15897}{{\ttfamily 2309.15897}}.

\bibitem{Witten:2023qsv}
E.~Witten, \emph{{Algebras, Regions, and Observers}},
  \href{https://arxiv.org/abs/2303.02837}{{\ttfamily 2303.02837}}.

\bibitem{Witten:2023xze}
E.~Witten, \emph{{A Background Independent Algebra in Quantum Gravity}},
  \href{https://arxiv.org/abs/2308.03663}{{\ttfamily 2308.03663}}.

\bibitem{Seo:2022pqj}
M.-S.~Seo, \emph{{Von Neumann algebra description of inflationary cosmology}},
  \href{https://doi.org/10.1140/epjc/s10052-023-12202-6}{\emph{Eur. Phys. J. C}
  {\bfseries 83} (2023) 1003}
  [\href{https://arxiv.org/abs/2212.05637}{{\ttfamily 2212.05637}}].

\bibitem{Gomez:2023wrq}
C.~Gomez, \emph{{Entanglement, Observers and Cosmology: a view from von Neumann
  Algebras}},  \href{https://arxiv.org/abs/2302.14747}{{\ttfamily 2302.14747}}.

\bibitem{Gomez:2023upk}
C.~Gomez, \emph{{Clocks, Algebras and Cosmology}},
  \href{https://arxiv.org/abs/2304.11845}{{\ttfamily 2304.11845}}.

\bibitem{Gomez:2023tkr}
C.~Gomez, \emph{{Traces and Time: a de Sitter Black Hole correspondence}},
  \href{https://arxiv.org/abs/2307.01841}{{\ttfamily 2307.01841}}.

\bibitem{Gomez:2023jbg}
C.~Gomez, \emph{{On the algebraic meaning of quantum gravity for closed
  Universes}},  \href{https://arxiv.org/abs/2311.01952}{{\ttfamily
  2311.01952}}.

\bibitem{Aguilar-Gutierrez:2023odp}
S.E.~Aguilar-Gutierrez, E.~Bahiru and R.~Esp\'\i{}ndola, \emph{{The
  centaur-algebra of observables}},
  \href{https://doi.org/10.1007/JHEP03(2024)008}{\emph{JHEP} {\bfseries 03}
  (2024) 008} [\href{https://arxiv.org/abs/2307.04233}{{\ttfamily
  2307.04233}}].

\bibitem{Basteiro:2024cuh}
P.~Basteiro, G.~Di~Giulio, J.~Erdmenger and Z.-Y.~Xian, \emph{{Entanglement in
  interacting Majorana chains and transitions of von Neumann algebras}},
  \href{https://arxiv.org/abs/2401.04764}{{\ttfamily 2401.04764}}.

\bibitem{Papadodimas:2013jku}
K.~Papadodimas and S.~Raju, \emph{{State-Dependent Bulk-Boundary Maps and Black
  Hole Complementarity}},
  \href{https://doi.org/10.1103/PhysRevD.89.086010}{\emph{Phys. Rev. D}
  {\bfseries 89} (2014) 086010}
  [\href{https://arxiv.org/abs/1310.6335}{{\ttfamily 1310.6335}}].

\bibitem{Jefferson:2018ksk}
R.~Jefferson, \emph{{Comments on black hole interiors and modular inclusions}},
  \href{https://doi.org/10.21468/SciPostPhys.6.4.042}{\emph{SciPost Phys.}
  {\bfseries 6} (2019) 042} [\href{https://arxiv.org/abs/1811.08900}{{\ttfamily
  1811.08900}}].

\bibitem{Leutheusser:2021frk}
S.A.W.~Leutheusser, \emph{{Emergent Times in Holographic Duality}},
  {\emph{Phys. Rev. D} {\bfseries 108} (2023) 086020}
  [\href{https://arxiv.org/abs/2112.12156}{{\ttfamily 2112.12156}}].

\bibitem{Witten:2021unn}
E.~Witten, \emph{{Gravity and the crossed product}},
  \href{https://doi.org/10.1007/JHEP10(2022)008}{\emph{JHEP} {\bfseries 10}
  (2022) 008} [\href{https://arxiv.org/abs/2112.12828}{{\ttfamily
  2112.12828}}].

\bibitem{Witten:2018zxz}
E.~Witten, \emph{{APS Medal for Exceptional Achievement in Research: Invited
  article on entanglement properties of quantum field theory}},
  \href{https://doi.org/10.1103/RevModPhys.90.045003}{\emph{Rev. Mod. Phys.}
  {\bfseries 90} (2018) 045003}
  [\href{https://arxiv.org/abs/1803.04993}{{\ttfamily 1803.04993}}].

\bibitem{Witten:2021jzq}
E.~Witten, \emph{{Why Does Quantum Field Theory In Curved Spacetime Make Sense?
  And What Happens To The Algebra of Observables In The Thermodynamic Limit?}},
   \href{https://arxiv.org/abs/2112.11614}{{\ttfamily 2112.11614}}.

\bibitem{Sorce:2023fdx}
J.~Sorce, \emph{{Notes on the type classification of von Neumann algebras}},
  \href{https://arxiv.org/abs/2302.01958}{{\ttfamily 2302.01958}}.

\bibitem{Casini:2022rlv}
H.~Casini and M.~Huerta, \emph{{Lectures on entanglement in quantum field
  theory}}, \href{https://doi.org/10.22323/1.403.0002}{\emph{PoS} {\bfseries
  TASI2021} (2023) 002} [\href{https://arxiv.org/abs/2201.13310}{{\ttfamily
  2201.13310}}].

\bibitem{Maldacena:2016upp}
J.~Maldacena, D.~Stanford and Z.~Yang, \emph{{Conformal symmetry and its
  breaking in two dimensional Nearly Anti-de-Sitter space}},
  \href{https://doi.org/10.1093/ptep/ptw124}{\emph{PTEP} {\bfseries 2016}
  (2016) 12C104} [\href{https://arxiv.org/abs/1606.01857}{{\ttfamily
  1606.01857}}].

\bibitem{JACKIW1985343}
R.~Jackiw, \emph{Lower dimensional gravity},
  \href{https://doi.org/https://doi.org/10.1016/0550-3213(85)90448-1}{\emph{Nuclear
  Physics B} {\bfseries 252} (1985) 343}.

\bibitem{TEITELBOIM198341}
C.~Teitelboim, \emph{Gravitation and hamiltonian structure in two spacetime
  dimensions},
  \href{https://doi.org/https://doi.org/10.1016/0370-2693(83)90012-6}{\emph{Physics
  Letters B} {\bfseries 126} (1983) 41}.

\bibitem{Sachdev:2010um}
S.~Sachdev, \emph{{Holographic metals and the fractionalized Fermi liquid}},
  \href{https://doi.org/10.1103/PhysRevLett.105.151602}{\emph{Phys. Rev. Lett.}
  {\bfseries 105} (2010) 151602}
  [\href{https://arxiv.org/abs/1006.3794}{{\ttfamily 1006.3794}}].

\bibitem{Nakayama:2004vk}
Y.~Nakayama, \emph{{Liouville field theory: A Decade after the revolution}},
  \href{https://doi.org/10.1142/S0217751X04019500}{\emph{Int. J. Mod. Phys. A}
  {\bfseries 19} (2004) 2771}
  [\href{https://arxiv.org/abs/hep-th/0402009}{{\ttfamily hep-th/0402009}}].

\bibitem{Teschner:2001rv}
J.~Teschner, \emph{{Liouville theory revisited}},
  \href{https://doi.org/10.1088/0264-9381/18/23/201}{\emph{Class. Quant. Grav.}
  {\bfseries 18} (2001) R153}
  [\href{https://arxiv.org/abs/hep-th/0104158}{{\ttfamily hep-th/0104158}}].

\bibitem{Vargas:2017swx}
V.~Vargas, \emph{{Lecture notes on Liouville theory and the DOZZ formula}},
  \href{https://arxiv.org/abs/1712.00829}{{\ttfamily 1712.00829}}.

\bibitem{Polyakov:1981rd}
A.M.~Polyakov, \emph{{Quantum Geometry of Bosonic Strings}},
  \href{https://doi.org/10.1016/0370-2693(81)90743-7}{\emph{Phys. Lett. B}
  {\bfseries 103} (1981) 207}.

\bibitem{Fateev:2000ik}
V.~Fateev, A.B.~Zamolodchikov and A.B.~Zamolodchikov, \emph{{Boundary Liouville
  field theory. 1. Boundary state and boundary two point function}},
  \href{https://arxiv.org/abs/hep-th/0001012}{{\ttfamily hep-th/0001012}}.

\bibitem{Teschner:2000md}
J.~Teschner, \emph{{Remarks on Liouville theory with boundary}},
  \href{https://doi.org/10.22323/1.006.0041}{\emph{PoS} {\bfseries tmr2000}
  (2000) 041} [\href{https://arxiv.org/abs/hep-th/0009138}{{\ttfamily
  hep-th/0009138}}].

\bibitem{Chapman:2021jbh}
S.~Chapman and G.~Policastro, \emph{{Quantum computational complexity from
  quantum information to black holes and back}},
  \href{https://doi.org/10.1140/epjc/s10052-022-10037-1}{\emph{Eur. Phys. J. C}
  {\bfseries 82} (2022) 128}
  [\href{https://arxiv.org/abs/2110.14672}{{\ttfamily 2110.14672}}].

\bibitem{nielsen_2010}
M.A.~Nielsen and I.L.~Chuang, \emph{Quantum computation and quantum
  information}, Cambridge university press (2010).

\bibitem{Caputa:2017yrh}
P.~Caputa, N.~Kundu, M.~Miyaji, T.~Takayanagi and K.~Watanabe, \emph{{Liouville
  Action as Path-Integral Complexity: From Continuous Tensor Networks to
  AdS/CFT}}, \href{https://doi.org/10.1007/JHEP11(2017)097}{\emph{JHEP}
  {\bfseries 11} (2017) 097}
  [\href{https://arxiv.org/abs/1706.07056}{{\ttfamily 1706.07056}}].

\bibitem{Jefferson:2017sdb}
R.~Jefferson and R.C.~Myers, \emph{{Circuit complexity in quantum field
  theory}}, \href{https://doi.org/10.1007/JHEP10(2017)107}{\emph{JHEP}
  {\bfseries 10} (2017) 107}
  [\href{https://arxiv.org/abs/1707.08570}{{\ttfamily 1707.08570}}].

\bibitem{Chapman:2017rqy}
S.~Chapman, M.P.~Heller, H.~Marrochio and F.~Pastawski, \emph{{Toward a
  Definition of Complexity for Quantum Field Theory States}},
  \href{https://doi.org/10.1103/PhysRevLett.120.121602}{\emph{Phys. Rev. Lett.}
  {\bfseries 120} (2018) 121602}
  [\href{https://arxiv.org/abs/1707.08582}{{\ttfamily 1707.08582}}].

\bibitem{Bhattacharyya:2018wym}
A.~Bhattacharyya, P.~Caputa, S.R.~Das, N.~Kundu, M.~Miyaji and T.~Takayanagi,
  \emph{{Path-Integral Complexity for Perturbed CFTs}},
  \href{https://doi.org/10.1007/JHEP07(2018)086}{\emph{JHEP} {\bfseries 07}
  (2018) 086} [\href{https://arxiv.org/abs/1804.01999}{{\ttfamily
  1804.01999}}].

\bibitem{Chapman:2018hou}
S.~Chapman, J.~Eisert, L.~Hackl, M.P.~Heller, R.~Jefferson, H.~Marrochio
  et~al., \emph{{Complexity and entanglement for thermofield double states}},
  \href{https://doi.org/10.21468/SciPostPhys.6.3.034}{\emph{SciPost Phys.}
  {\bfseries 6} (2019) 034} [\href{https://arxiv.org/abs/1810.05151}{{\ttfamily
  1810.05151}}].

\bibitem{Camargo:2018eof}
H.A.~Camargo, P.~Caputa, D.~Das, M.P.~Heller and R.~Jefferson,
  \emph{{Complexity as a novel probe of quantum quenches: universal scalings
  and purifications}},
  \href{https://doi.org/10.1103/PhysRevLett.122.081601}{\emph{Phys. Rev. Lett.}
  {\bfseries 122} (2019) 081601}
  [\href{https://arxiv.org/abs/1807.07075}{{\ttfamily 1807.07075}}].

\bibitem{Ge:2019mjt}
D.~Ge and G.~Policastro, \emph{{Circuit Complexity and 2D Bosonisation}},
  \href{https://doi.org/10.1007/JHEP10(2019)276}{\emph{JHEP} {\bfseries 10}
  (2019) 276} [\href{https://arxiv.org/abs/1904.03003}{{\ttfamily
  1904.03003}}].

\bibitem{Brown:2019whu}
A.R.~Brown and L.~Susskind, \emph{{Complexity geometry of a single qubit}},
  \href{https://doi.org/10.1103/PhysRevD.100.046020}{\emph{Phys. Rev. D}
  {\bfseries 100} (2019) 046020}
  [\href{https://arxiv.org/abs/1903.12621}{{\ttfamily 1903.12621}}].

\bibitem{Balasubramanian:2019wgd}
V.~Balasubramanian, M.~Decross, A.~Kar and O.~Parrikar, \emph{{Quantum
  Complexity of Time Evolution with Chaotic Hamiltonians}},
  \href{https://doi.org/10.1007/JHEP01(2020)134}{\emph{JHEP} {\bfseries 01}
  (2020) 134} [\href{https://arxiv.org/abs/1905.05765}{{\ttfamily
  1905.05765}}].

\bibitem{Chapman:2019clq}
S.~Chapman and H.Z.~Chen, \emph{{Charged Complexity and the Thermofield Double
  State}}, \href{https://doi.org/10.1007/JHEP02(2021)187}{\emph{JHEP}
  {\bfseries 02} (2021) 187}
  [\href{https://arxiv.org/abs/1910.07508}{{\ttfamily 1910.07508}}].

\bibitem{Caputa:2018kdj}
P.~Caputa and J.M.~Magan, \emph{{Quantum Computation as Gravity}},
  \href{https://doi.org/10.1103/PhysRevLett.122.231302}{\emph{Phys. Rev. Lett.}
  {\bfseries 122} (2019) 231302}
  [\href{https://arxiv.org/abs/1807.04422}{{\ttfamily 1807.04422}}].

\bibitem{Auzzi:2020idm}
R.~Auzzi, S.~Baiguera, G.B.~De~Luca, A.~Legramandi, G.~Nardelli and N.~Zenoni,
  \emph{{Geometry of quantum complexity}},
  \href{https://doi.org/10.1103/PhysRevD.103.106021}{\emph{Phys. Rev. D}
  {\bfseries 103} (2021) 106021}
  [\href{https://arxiv.org/abs/2011.07601}{{\ttfamily 2011.07601}}].

\bibitem{Caginalp:2020tzw}
R.J.~Caginalp and S.~Leutheusser, \emph{{Complexity in One- and Two-Qubit
  Systems}},  \href{https://arxiv.org/abs/2010.15099}{{\ttfamily 2010.15099}}.

\bibitem{Flory:2020eot}
M.~Flory and M.P.~Heller, \emph{{Geometry of Complexity in Conformal Field
  Theory}}, \href{https://doi.org/10.1103/PhysRevResearch.2.043438}{\emph{Phys.
  Rev. Res.} {\bfseries 2} (2020) 043438}
  [\href{https://arxiv.org/abs/2005.02415}{{\ttfamily 2005.02415}}].

\bibitem{Flory:2020dja}
M.~Flory and M.P.~Heller, \emph{{Conformal field theory complexity from
  Euler-Arnold equations}},
  \href{https://doi.org/10.1007/JHEP12(2020)091}{\emph{JHEP} {\bfseries 12}
  (2020) 091} [\href{https://arxiv.org/abs/2007.11555}{{\ttfamily
  2007.11555}}].

\bibitem{Chagnet:2021uvi}
N.~Chagnet, S.~Chapman, J.~de~Boer and C.~Zukowski, \emph{{Complexity for
  Conformal Field Theories in General Dimensions}},
  \href{https://doi.org/10.1103/PhysRevLett.128.051601}{\emph{Phys. Rev. Lett.}
  {\bfseries 128} (2022) 051601}
  [\href{https://arxiv.org/abs/2103.06920}{{\ttfamily 2103.06920}}].

\bibitem{Basteiro:2021ene}
P.~Basteiro, J.~Erdmenger, P.~Fries, F.~Goth, I.~Matthaiakakis and R.~Meyer,
  \emph{{Quantum complexity as hydrodynamics}},
  \href{https://doi.org/10.1103/PhysRevD.106.065016}{\emph{Phys. Rev. D}
  {\bfseries 106} (2022) 065016}
  [\href{https://arxiv.org/abs/2109.01152}{{\ttfamily 2109.01152}}].

\bibitem{Brown:2021uov}
A.R.~Brown, M.H.~Freedman, H.W.~Lin and L.~Susskind, \emph{{Universality in
  long-distance geometry and quantum complexity}},
  \href{https://doi.org/10.1038/s41586-023-06460-3}{\emph{Nature} {\bfseries
  622} (2023) 58} [\href{https://arxiv.org/abs/2111.12700}{{\ttfamily
  2111.12700}}].

\bibitem{Balasubramanian:2021mxo}
V.~Balasubramanian, M.~DeCross, A.~Kar, Y.C.~Li and O.~Parrikar,
  \emph{{Complexity growth in integrable and chaotic models}},
  \href{https://doi.org/10.1007/JHEP07(2021)011}{\emph{JHEP} {\bfseries 07}
  (2021) 011} [\href{https://arxiv.org/abs/2101.02209}{{\ttfamily
  2101.02209}}].

\bibitem{Brown:2022phc}
A.R.~Brown, \emph{{Polynomial Equivalence of Complexity Geometries}},
  \href{https://arxiv.org/abs/2205.04485}{{\ttfamily 2205.04485}}.

\bibitem{Erdmenger:2022lov}
J.~Erdmenger, A.-L.~Weigel, M.~Gerbershagen and M.P.~Heller, \emph{{From
  complexity geometry to holographic spacetime}},
  \href{https://doi.org/10.1103/PhysRevD.108.106020}{\emph{Phys. Rev. D}
  {\bfseries 108} (2023) 106020}
  [\href{https://arxiv.org/abs/2212.00043}{{\ttfamily 2212.00043}}].

\bibitem{Baiguera:2023bhm}
S.~Baiguera, S.~Chapman, G.~Policastro and T.~Schwartzman, \emph{{The
  Complexity of Being Entangled}},
  \href{https://arxiv.org/abs/2311.04277}{{\ttfamily 2311.04277}}.

\bibitem{Ali:2018fcz}
T.~Ali, A.~Bhattacharyya, S.~Shajidul~Haque, E.H.~Kim and N.~Moynihan,
  \emph{{Time Evolution of Complexity: A Critique of Three Methods}},
  \href{https://doi.org/10.1007/JHEP04(2019)087}{\emph{JHEP} {\bfseries 04}
  (2019) 087} [\href{https://arxiv.org/abs/1810.02734}{{\ttfamily
  1810.02734}}].

\bibitem{Bhattacharyya:2018bbv}
A.~Bhattacharyya, A.~Shekar and A.~Sinha, \emph{{Circuit complexity in
  interacting QFTs and RG flows}},
  \href{https://doi.org/10.1007/JHEP10(2018)140}{\emph{JHEP} {\bfseries 10}
  (2018) 140} [\href{https://arxiv.org/abs/1808.03105}{{\ttfamily
  1808.03105}}].

\bibitem{Bhattacharyya:2019kvj}
A.~Bhattacharyya, P.~Nandy and A.~Sinha, \emph{{Renormalized Circuit
  Complexity}},
  \href{https://doi.org/10.1103/PhysRevLett.124.101602}{\emph{Phys. Rev. Lett.}
  {\bfseries 124} (2020) 101602}
  [\href{https://arxiv.org/abs/1907.08223}{{\ttfamily 1907.08223}}].

\bibitem{Ali:2019zcj}
T.~Ali, A.~Bhattacharyya, S.S.~Haque, E.H.~Kim, N.~Moynihan and J.~Murugan,
  \emph{{Chaos and Complexity in Quantum Mechanics}},
  \href{https://doi.org/10.1103/PhysRevD.101.026021}{\emph{Phys. Rev. D}
  {\bfseries 101} (2020) 026021}
  [\href{https://arxiv.org/abs/1905.13534}{{\ttfamily 1905.13534}}].

\bibitem{Bhattacharyya:2019txx}
A.~Bhattacharyya, W.~Chemissany, S.~Shajidul~Haque and B.~Yan, \emph{{Towards
  the web of quantum chaos diagnostics}},
  \href{https://doi.org/10.1140/epjc/s10052-022-10035-3}{\emph{Eur. Phys. J. C}
  {\bfseries 82} (2022) 87} [\href{https://arxiv.org/abs/1909.01894}{{\ttfamily
  1909.01894}}].

\bibitem{Bhattacharyya:2020iic}
A.~Bhattacharyya, S.S.~Haque and E.H.~Kim, \emph{{Complexity from the reduced
  density matrix: a new diagnostic for chaos}},
  \href{https://doi.org/10.1007/JHEP10(2021)028}{\emph{JHEP} {\bfseries 10}
  (2021) 028} [\href{https://arxiv.org/abs/2011.04705}{{\ttfamily
  2011.04705}}].

\bibitem{Bhattacharyya:2022ren}
A.~Bhattacharyya, G.~Katoch and S.R.~Roy, \emph{{Complexity of warped conformal
  field theory}},
  \href{https://doi.org/10.1140/epjc/s10052-023-11212-8}{\emph{Eur. Phys. J. C}
  {\bfseries 83} (2023) 33} [\href{https://arxiv.org/abs/2202.09350}{{\ttfamily
  2202.09350}}].

\bibitem{Bhattacharyya:2023sjr}
A.~Bhattacharyya and P.~Nandi, \emph{{Circuit complexity for Carrollian
  Conformal (BMS) field theories}},
  \href{https://doi.org/10.1007/JHEP07(2023)105}{\emph{JHEP} {\bfseries 07}
  (2023) 105} [\href{https://arxiv.org/abs/2301.12845}{{\ttfamily
  2301.12845}}].

\bibitem{Bhattacharyya:2020art}
A.~Bhattacharyya, W.~Chemissany, S.S.~Haque, J.~Murugan and B.~Yan, \emph{{The
  Multi-faceted Inverted Harmonic Oscillator: Chaos and Complexity}},
  \href{https://doi.org/10.21468/SciPostPhysCore.4.1.002}{\emph{SciPost Phys.
  Core} {\bfseries 4} (2021) 002}
  [\href{https://arxiv.org/abs/2007.01232}{{\ttfamily 2007.01232}}].

\bibitem{Bhattacharyya:2020rpy}
A.~Bhattacharyya, S.~Das, S.~Shajidul~Haque and B.~Underwood,
  \emph{{Cosmological Complexity}},
  \href{https://doi.org/10.1103/PhysRevD.101.106020}{\emph{Phys. Rev. D}
  {\bfseries 101} (2020) 106020}
  [\href{https://arxiv.org/abs/2001.08664}{{\ttfamily 2001.08664}}].

\bibitem{Bhattacharyya:2020kgu}
A.~Bhattacharyya, S.~Das, S.S.~Haque and B.~Underwood, \emph{{Rise of
  cosmological complexity: Saturation of growth and chaos}},
  \href{https://doi.org/10.1103/PhysRevResearch.2.033273}{\emph{Phys. Rev.
  Res.} {\bfseries 2} (2020) 033273}
  [\href{https://arxiv.org/abs/2005.10854}{{\ttfamily 2005.10854}}].

\bibitem{Bhattacharyya:2024duw}
A.~Bhattacharyya, S.~Brahma, S.S.~Haque, J.S.~Lund and A.~Paul, \emph{{The
  Early Universe as an Open Quantum System: Complexity and Decoherence}},
  \href{https://arxiv.org/abs/2401.12134}{{\ttfamily 2401.12134}}.

\bibitem{Susskind:2014moa}
L.~Susskind, \emph{{Entanglement is not enough}},
  \href{https://doi.org/10.1002/prop.201500095}{\emph{Fortsch. Phys.}
  {\bfseries 64} (2016) 49} [\href{https://arxiv.org/abs/1411.0690}{{\ttfamily
  1411.0690}}].

\bibitem{Susskind:2014rva}
L.~Susskind, \emph{{Computational Complexity and Black Hole Horizons}},
  \href{https://doi.org/10.1002/prop.201500092}{\emph{Fortsch. Phys.}
  {\bfseries 64} (2016) 24} [\href{https://arxiv.org/abs/1403.5695}{{\ttfamily
  1403.5695}}].

\bibitem{Stanford:2014jda}
D.~Stanford and L.~Susskind, \emph{{Complexity and Shock Wave Geometries}},
  \href{https://doi.org/10.1103/PhysRevD.90.126007}{\emph{Phys. Rev. D}
  {\bfseries 90} (2014) 126007}
  [\href{https://arxiv.org/abs/1406.2678}{{\ttfamily 1406.2678}}].

\bibitem{Brown:2015bva}
A.R.~Brown, D.A.~Roberts, L.~Susskind, B.~Swingle and Y.~Zhao,
  \emph{{Holographic Complexity Equals Bulk Action?}},
  \href{https://doi.org/10.1103/PhysRevLett.116.191301}{\emph{Phys. Rev. Lett.}
  {\bfseries 116} (2016) 191301}
  [\href{https://arxiv.org/abs/1509.07876}{{\ttfamily 1509.07876}}].

\bibitem{Brown:2015lvg}
A.R.~Brown, D.A.~Roberts, L.~Susskind, B.~Swingle and Y.~Zhao,
  \emph{{Complexity, action, and black holes}},
  \href{https://doi.org/10.1103/PhysRevD.93.086006}{\emph{Phys. Rev. D}
  {\bfseries 93} (2016) 086006}
  [\href{https://arxiv.org/abs/1512.04993}{{\ttfamily 1512.04993}}].

\bibitem{Couch:2016exn}
J.~Couch, W.~Fischler and P.H.~Nguyen, \emph{{Noether charge, black hole
  volume, and complexity}},
  \href{https://doi.org/10.1007/JHEP03(2017)119}{\emph{JHEP} {\bfseries 03}
  (2017) 119} [\href{https://arxiv.org/abs/1610.02038}{{\ttfamily
  1610.02038}}].

\bibitem{Belin:2021bga}
A.~Belin, R.C.~Myers, S.-M.~Ruan, G.~S\'arosi and A.J.~Speranza, \emph{{Does
  Complexity Equal Anything?}},
  \href{https://doi.org/10.1103/PhysRevLett.128.081602}{\emph{Phys. Rev. Lett.}
  {\bfseries 128} (2022) 081602}
  [\href{https://arxiv.org/abs/2111.02429}{{\ttfamily 2111.02429}}].

\bibitem{Belin:2022xmt}
A.~Belin, R.C.~Myers, S.-M.~Ruan, G.~S\'arosi and A.J.~Speranza,
  \emph{{Complexity equals anything II}},
  \href{https://doi.org/10.1007/JHEP01(2023)154}{\emph{JHEP} {\bfseries 01}
  (2023) 154} [\href{https://arxiv.org/abs/2210.09647}{{\ttfamily
  2210.09647}}].

\bibitem{Jorstad:2022mls}
E.~J\o{}rstad, R.C.~Myers and S.-M.~Ruan, \emph{{Holographic complexity in
  dS$_{d+1}$}}, \href{https://doi.org/10.1007/JHEP05(2022)119}{\emph{JHEP}
  {\bfseries 05} (2022) 119}
  [\href{https://arxiv.org/abs/2202.10684}{{\ttfamily 2202.10684}}].

\bibitem{Chapman:2021eyy}
S.~Chapman, D.A.~Galante and E.D.~Kramer, \emph{{Holographic complexity and de
  Sitter space}}, \href{https://doi.org/10.1007/JHEP02(2022)198}{\emph{JHEP}
  {\bfseries 02} (2022) 198}
  [\href{https://arxiv.org/abs/2110.05522}{{\ttfamily 2110.05522}}].

\bibitem{Auzzi:2023qbm}
R.~Auzzi, G.~Nardelli, G.P.~Ungureanu and N.~Zenoni, \emph{{Volume complexity
  of dS bubbles}},
  \href{https://doi.org/10.1103/PhysRevD.108.026006}{\emph{Phys. Rev. D}
  {\bfseries 108} (2023) 026006}
  [\href{https://arxiv.org/abs/2302.03584}{{\ttfamily 2302.03584}}].

\bibitem{Anegawa:2023wrk}
T.~Anegawa, N.~Iizuka, S.K.~Sake and N.~Zenoni, \emph{{Is action complexity
  better for de Sitter space in Jackiw-Teitelboim gravity?}},
  \href{https://doi.org/10.1007/JHEP06(2023)213}{\emph{JHEP} {\bfseries 06}
  (2023) 213} [\href{https://arxiv.org/abs/2303.05025}{{\ttfamily
  2303.05025}}].

\bibitem{Anegawa:2023dad}
T.~Anegawa and N.~Iizuka, \emph{{Shock waves and delay of hyperfast growth in
  de Sitter complexity}},
  \href{https://doi.org/10.1007/JHEP08(2023)115}{\emph{JHEP} {\bfseries 08}
  (2023) 115} [\href{https://arxiv.org/abs/2304.14620}{{\ttfamily
  2304.14620}}].

\bibitem{Baiguera:2023tpt}
S.~Baiguera, R.~Berman, S.~Chapman and R.C.~Myers, \emph{{The cosmological
  switchback effect}},
  \href{https://doi.org/10.1007/JHEP07(2023)162}{\emph{JHEP} {\bfseries 07}
  (2023) 162} [\href{https://arxiv.org/abs/2304.15008}{{\ttfamily
  2304.15008}}].

\bibitem{Aguilar-Gutierrez:2023zqm}
S.E.~Aguilar-Gutierrez, M.P.~Heller and S.~Van~der Schueren, \emph{{Complexity
  = Anything Can Grow Forever in de Sitter}},
  \href{https://arxiv.org/abs/2305.11280}{{\ttfamily 2305.11280}}.

\bibitem{Aguilar-Gutierrez:2023tic}
S.E.~Aguilar-Gutierrez, A.K.~Patra and J.F.~Pedraza, \emph{{Entangled universes
  in dS wedge holography}},
  \href{https://doi.org/10.1007/JHEP10(2023)156}{\emph{JHEP} {\bfseries 10}
  (2023) 156} [\href{https://arxiv.org/abs/2308.05666}{{\ttfamily
  2308.05666}}].

\bibitem{Aguilar-Gutierrez:2023pnn}
S.E.~Aguilar-Gutierrez, \emph{{C=Anything and the switchback effect in
  Schwarzschild-de Sitter space}},
  \href{https://arxiv.org/abs/2309.05848}{{\ttfamily 2309.05848}}.

\bibitem{Aguilar-Gutierrez:2024rka}
S.E.~Aguilar-Gutierrez, S.~Baiguera and N.~Zenoni, \emph{{Holographic
  complexity of the extended Schwarzschild-de Sitter space}},
  \href{https://arxiv.org/abs/2402.01357}{{\ttfamily 2402.01357}}.

\bibitem{Maldacena:2015waa}
J.~Maldacena, S.H.~Shenker and D.~Stanford, \emph{{A bound on chaos}},
  \href{https://doi.org/10.1007/JHEP08(2016)106}{\emph{JHEP} {\bfseries 08}
  (2016) 106} [\href{https://arxiv.org/abs/1503.01409}{{\ttfamily
  1503.01409}}].

\bibitem{Balasubramanian:2022tpr}
V.~Balasubramanian, P.~Caputa, J.M.~Magan and Q.~Wu, \emph{{Quantum chaos and
  the complexity of spread of states}},
  \href{https://doi.org/10.1103/PhysRevD.106.046007}{\emph{Phys. Rev. D}
  {\bfseries 106} (2022) 046007}
  [\href{https://arxiv.org/abs/2202.06957}{{\ttfamily 2202.06957}}].

\bibitem{Parker:2018yvk}
D.E.~Parker, X.~Cao, A.~Avdoshkin, T.~Scaffidi and E.~Altman, \emph{{A
  Universal Operator Growth Hypothesis}},
  \href{https://doi.org/10.1103/PhysRevX.9.041017}{\emph{Phys. Rev. X}
  {\bfseries 9} (2019) 041017}
  [\href{https://arxiv.org/abs/1812.08657}{{\ttfamily 1812.08657}}].

\bibitem{Nandy:2024htc}
P.~Nandy, A.S.~Matsoukas-Roubeas, P.~Mart\'\i{}nez-Azcona, A.~Dymarsky and
  A.~del Campo, \emph{{Quantum Dynamics in Krylov Space: Methods and
  Applications}},  \href{https://arxiv.org/abs/2405.09628}{{\ttfamily
  2405.09628}}.

\bibitem{Barbon:2019wsy}
J.L.F.~Barb\'on, E.~Rabinovici, R.~Shir and R.~Sinha, \emph{{On The Evolution
  Of Operator Complexity Beyond Scrambling}},
  \href{https://doi.org/10.1007/JHEP10(2019)264}{\emph{JHEP} {\bfseries 10}
  (2019) 264} [\href{https://arxiv.org/abs/1907.05393}{{\ttfamily
  1907.05393}}].

\bibitem{Caputa:2021sib}
P.~Caputa, J.M.~Magan and D.~Patramanis, \emph{{Geometry of Krylov
  complexity}},
  \href{https://doi.org/10.1103/PhysRevResearch.4.013041}{\emph{Phys. Rev.
  Res.} {\bfseries 4} (2022) 013041}
  [\href{https://arxiv.org/abs/2109.03824}{{\ttfamily 2109.03824}}].

\bibitem{Murthy:2019fgs}
C.~Murthy and M.~Srednicki, \emph{{Bounds on chaos from the eigenstate
  thermalization hypothesis}},
  \href{https://doi.org/10.1103/PhysRevLett.123.230606}{\emph{Phys. Rev. Lett.}
  {\bfseries 123} (2019) 230606}
  [\href{https://arxiv.org/abs/1906.10808}{{\ttfamily 1906.10808}}].

\bibitem{Rozenbaum:2016mmv}
E.B.~Rozenbaum, S.~Ganeshan and V.~Galitski, \emph{{Lyapunov Exponent and
  Out-of-Time-Ordered Correlator\textquoteright{}s Growth Rate in a Chaotic
  System}}, \href{https://doi.org/10.1103/PhysRevLett.118.086801}{\emph{Phys.
  Rev. Lett.} {\bfseries 118} (2017) 086801}
  [\href{https://arxiv.org/abs/1609.01707}{{\ttfamily 1609.01707}}].

\bibitem{Bhattacharjee:2022vlt}
B.~Bhattacharjee, X.~Cao, P.~Nandy and T.~Pathak, \emph{{Krylov complexity in
  saddle-dominated scrambling}},
  \href{https://doi.org/10.1007/JHEP05(2022)174}{\emph{JHEP} {\bfseries 05}
  (2022) 174} [\href{https://arxiv.org/abs/2203.03534}{{\ttfamily
  2203.03534}}].

\bibitem{Avdoshkin:2022xuw}
A.~Avdoshkin, A.~Dymarsky and M.~Smolkin, \emph{{Krylov complexity in quantum
  field theory, and beyond}},
  \href{https://arxiv.org/abs/2212.14429}{{\ttfamily 2212.14429}}.

\bibitem{Erdmenger:2023wjg}
J.~Erdmenger, S.-K.~Jian and Z.-Y.~Xian, \emph{{Universal chaotic dynamics from
  Krylov space}}, \href{https://doi.org/10.1007/JHEP08(2023)176}{\emph{JHEP}
  {\bfseries 08} (2023) 176}
  [\href{https://arxiv.org/abs/2303.12151}{{\ttfamily 2303.12151}}].

\bibitem{Bhattacharjee:2023dik}
B.~Bhattacharjee, \emph{{A Lanczos approach to the Adiabatic Gauge Potential}},
   \href{https://arxiv.org/abs/2302.07228}{{\ttfamily 2302.07228}}.

\bibitem{Chattopadhyay:2023fob}
A.~Chattopadhyay, A.~Mitra and H.J.R.~van Zyl, \emph{{Spread complexity as
  classical dilaton solutions}},
  \href{https://doi.org/10.1103/PhysRevD.108.025013}{\emph{Phys. Rev. D}
  {\bfseries 108} (2023) 025013}
  [\href{https://arxiv.org/abs/2302.10489}{{\ttfamily 2302.10489}}].

\bibitem{Camargo:2024deu}
H.A.~Camargo, K.-B.~Huh, V.~Jahnke, H.-S.~Jeong, K.-Y.~Kim and M.~Nishida,
  \emph{{Spread and Spectral Complexity in Quantum Spin Chains: from
  Integrability to Chaos}},  \href{https://arxiv.org/abs/2405.11254}{{\ttfamily
  2405.11254}}.

\bibitem{Aguilar-Gutierrez:2023nyk}
S.E.~Aguilar-Gutierrez and A.~Rolph, \emph{{Krylov complexity is not a measure
  of distance between states or operators}},
  \href{https://arxiv.org/abs/2311.04093}{{\ttfamily 2311.04093}}.

\bibitem{Caputa:2022eye}
P.~Caputa and S.~Liu, \emph{{Quantum complexity and topological phases of
  matter}}, \href{https://doi.org/10.1103/PhysRevB.106.195125}{\emph{Phys. Rev.
  B} {\bfseries 106} (2022) 195125}
  [\href{https://arxiv.org/abs/2205.05688}{{\ttfamily 2205.05688}}].

\bibitem{Afrasiar:2022efk}
M.~Afrasiar, J.~Kumar~Basak, B.~Dey, K.~Pal and K.~Pal, \emph{{Time evolution
  of spread complexity in quenched Lipkin\textendash{}Meshkov\textendash{}Glick
  model}}, \href{https://doi.org/10.1088/1742-5468/ad0032}{\emph{J. Stat.
  Mech.} {\bfseries 2310} (2023) 103101}
  [\href{https://arxiv.org/abs/2208.10520}{{\ttfamily 2208.10520}}].

\bibitem{Caputa:2022yju}
P.~Caputa, N.~Gupta, S.S.~Haque, S.~Liu, J.~Murugan and H.J.R.~Van~Zyl,
  \emph{{Spread complexity and topological transitions in the Kitaev chain}},
  \href{https://doi.org/10.1007/JHEP01(2023)120}{\emph{JHEP} {\bfseries 01}
  (2023) 120} [\href{https://arxiv.org/abs/2208.06311}{{\ttfamily
  2208.06311}}].

\bibitem{Pal:2023yik}
K.~Pal, K.~Pal, A.~Gill and T.~Sarkar, \emph{{Time evolution of spread
  complexity and statistics of work done in quantum quenches}},
  \href{https://doi.org/10.1103/PhysRevB.108.104311}{\emph{Phys. Rev. B}
  {\bfseries 108} (2023) 104311}
  [\href{https://arxiv.org/abs/2304.09636}{{\ttfamily 2304.09636}}].

\bibitem{Dymarsky:2019elm}
A.~Dymarsky and A.~Gorsky, \emph{{Quantum chaos as delocalization in Krylov
  space}}, \href{https://doi.org/10.1103/PhysRevB.102.085137}{\emph{Phys. Rev.
  B} {\bfseries 102} (2020) 085137}
  [\href{https://arxiv.org/abs/1912.12227}{{\ttfamily 1912.12227}}].

\bibitem{Dymarsky:2021bjq}
A.~Dymarsky and M.~Smolkin, \emph{{Krylov complexity in conformal field
  theory}}, \href{https://doi.org/10.1103/PhysRevD.104.L081702}{\emph{Phys.
  Rev. D} {\bfseries 104} (2021) L081702}
  [\href{https://arxiv.org/abs/2104.09514}{{\ttfamily 2104.09514}}].

\bibitem{Kundu:2023hbk}
A.~Kundu, V.~Malvimat and R.~Sinha, \emph{{State dependence of Krylov
  complexity in 2d CFTs}},
  \href{https://doi.org/10.1007/JHEP09(2023)011}{\emph{JHEP} {\bfseries 09}
  (2023) 011} [\href{https://arxiv.org/abs/2303.03426}{{\ttfamily
  2303.03426}}].

\bibitem{Bhattacharya:2022gbz}
A.~Bhattacharya, P.~Nandy, P.P.~Nath and H.~Sahu, \emph{{Operator growth and
  Krylov construction in dissipative open quantum systems}},
  \href{https://doi.org/10.1007/JHEP12(2022)081}{\emph{JHEP} {\bfseries 12}
  (2022) 081} [\href{https://arxiv.org/abs/2207.05347}{{\ttfamily
  2207.05347}}].

\bibitem{Mohan:2023btr}
V.~Mohan, \emph{{Krylov complexity of open quantum systems: from hard spheres
  to black holes}}, \href{https://doi.org/10.1007/JHEP11(2023)222}{\emph{JHEP}
  {\bfseries 11} (2023) 222}
  [\href{https://arxiv.org/abs/2308.10945}{{\ttfamily 2308.10945}}].

\bibitem{Yates:2021asz}
D.J.~Yates and A.~Mitra, \emph{{Strong and almost strong modes of Floquet spin
  chains in Krylov subspaces}},
  \href{https://doi.org/10.1103/PhysRevB.104.195121}{\emph{Phys. Rev. B}
  {\bfseries 104} (2021) 195121}
  [\href{https://arxiv.org/abs/2105.13246}{{\ttfamily 2105.13246}}].

\bibitem{Caputa:2021ori}
P.~Caputa and S.~Datta, \emph{{Operator growth in 2d CFT}},
  \href{https://doi.org/10.1007/JHEP12(2021)188}{\emph{JHEP} {\bfseries 12}
  (2021) 188} [\href{https://arxiv.org/abs/2110.10519}{{\ttfamily
  2110.10519}}].

\bibitem{Patramanis:2021lkx}
D.~Patramanis, \emph{{Probing the entanglement of operator growth}},
  \href{https://doi.org/10.1093/ptep/ptac081}{\emph{PTEP} {\bfseries 2022}
  (2022) 063A01} [\href{https://arxiv.org/abs/2111.03424}{{\ttfamily
  2111.03424}}].

\bibitem{Trigueros:2021rwj}
F.B.~Trigueros and C.-J.~Lin, \emph{{Krylov complexity of many-body
  localization: Operator localization in Krylov basis}},
  \href{https://doi.org/10.21468/SciPostPhys.13.2.037}{\emph{SciPost Phys.}
  {\bfseries 13} (2022) 037}
  [\href{https://arxiv.org/abs/2112.04722}{{\ttfamily 2112.04722}}].

\bibitem{Rabinovici:2020ryf}
E.~Rabinovici, A.~S\'anchez-Garrido, R.~Shir and J.~Sonner, \emph{{Operator
  complexity: a journey to the edge of Krylov space}},
  \href{https://doi.org/10.1007/JHEP06(2021)062}{\emph{JHEP} {\bfseries 06}
  (2021) 062} [\href{https://arxiv.org/abs/2009.01862}{{\ttfamily
  2009.01862}}].

\bibitem{Rabinovici:2021qqt}
E.~Rabinovici, A.~S\'anchez-Garrido, R.~Shir and J.~Sonner, \emph{{Krylov
  localization and suppression of complexity}},
  \href{https://doi.org/10.1007/JHEP03(2022)211}{\emph{JHEP} {\bfseries 03}
  (2022) 211} [\href{https://arxiv.org/abs/2112.12128}{{\ttfamily
  2112.12128}}].

\bibitem{Rabinovici:2022beu}
E.~Rabinovici, A.~S\'anchez-Garrido, R.~Shir and J.~Sonner, \emph{{Krylov
  complexity from integrability to chaos}},
  \href{https://doi.org/10.1007/JHEP07(2022)151}{\emph{JHEP} {\bfseries 07}
  (2022) 151} [\href{https://arxiv.org/abs/2207.07701}{{\ttfamily
  2207.07701}}].

\bibitem{Bhattacharjee:2022qjw}
B.~Bhattacharjee, S.~Sur and P.~Nandy, \emph{{Probing quantum scars and weak
  ergodicity breaking through quantum complexity}},
  \href{https://doi.org/10.1103/PhysRevB.106.205150}{\emph{Phys. Rev. B}
  {\bfseries 106} (2022) 205150}
  [\href{https://arxiv.org/abs/2208.05503}{{\ttfamily 2208.05503}}].

\bibitem{Bhattacharjee:2022ave}
B.~Bhattacharjee, P.~Nandy and T.~Pathak, \emph{{Krylov complexity in large q
  and double-scaled SYK model}},
  \href{https://doi.org/10.1007/JHEP08(2023)099}{\emph{JHEP} {\bfseries 08}
  (2023) 099} [\href{https://arxiv.org/abs/2210.02474}{{\ttfamily
  2210.02474}}].

\bibitem{Takahashi:2023nkt}
K.~Takahashi and A.~del Campo, \emph{{Shortcuts to Adiabaticity in Krylov
  Space}}, \href{https://doi.org/10.1103/PhysRevX.14.011032}{\emph{Phys. Rev.
  X} {\bfseries 14} (2024) 011032}
  [\href{https://arxiv.org/abs/2302.05460}{{\ttfamily 2302.05460}}].

\bibitem{Camargo:2022rnt}
H.A.~Camargo, V.~Jahnke, K.-Y.~Kim and M.~Nishida, \emph{{Krylov complexity in
  free and interacting scalar field theories with bounded power spectrum}},
  \href{https://doi.org/10.1007/JHEP05(2023)226}{\emph{JHEP} {\bfseries 05}
  (2023) 226} [\href{https://arxiv.org/abs/2212.14702}{{\ttfamily
  2212.14702}}].

\bibitem{Hashimoto:2023swv}
K.~Hashimoto, K.~Murata, N.~Tanahashi and R.~Watanabe, \emph{{Krylov complexity
  and chaos in quantum mechanics}},
  \href{https://doi.org/10.1007/JHEP11(2023)040}{\emph{JHEP} {\bfseries 11}
  (2023) 040} [\href{https://arxiv.org/abs/2305.16669}{{\ttfamily
  2305.16669}}].

\bibitem{Camargo:2023eev}
H.A.~Camargo, V.~Jahnke, H.-S.~Jeong, K.-Y.~Kim and M.~Nishida, \emph{{Spectral
  and Krylov complexity in billiard systems}},
  \href{https://doi.org/10.1103/PhysRevD.109.046017}{\emph{Phys. Rev. D}
  {\bfseries 109} (2024) 046017}
  [\href{https://arxiv.org/abs/2306.11632}{{\ttfamily 2306.11632}}].

\bibitem{Iizuka:2023pov}
N.~Iizuka and M.~Nishida, \emph{{Krylov complexity in the IP matrix model}},
  \href{https://doi.org/10.1007/JHEP11(2023)065}{\emph{JHEP} {\bfseries 11}
  (2023) 065} [\href{https://arxiv.org/abs/2306.04805}{{\ttfamily
  2306.04805}}].

\bibitem{Caputa:2023vyr}
P.~Caputa, J.M.~Magan, D.~Patramanis and E.~Tonni, \emph{{Krylov complexity of
  modular Hamiltonian evolution}},
  \href{https://arxiv.org/abs/2306.14732}{{\ttfamily 2306.14732}}.

\bibitem{Fan:2023ohh}
Z.-Y.~Fan, \emph{{Generalised Krylov complexity}},
  \href{https://arxiv.org/abs/2306.16118}{{\ttfamily 2306.16118}}.

\bibitem{Vasli:2023syq}
M.J.~Vasli, K.~Babaei~Velni, M.R.~Mohammadi~Mozaffar, A.~Mollabashi and
  M.~Alishahiha, \emph{{Krylov complexity in Lifshitz-type scalar field
  theories}}, \href{https://doi.org/10.1140/epjc/s10052-024-12609-9}{\emph{Eur.
  Phys. J. C} {\bfseries 84} (2024) 235}
  [\href{https://arxiv.org/abs/2307.08307}{{\ttfamily 2307.08307}}].

\bibitem{Gautam:2023bcm}
M.~Gautam, K.~Pal, K.~Pal, A.~Gill, N.~Jaiswal and T.~Sarkar, \emph{{Spread
  complexity evolution in quenched interacting quantum systems}},
  \href{https://doi.org/10.1103/PhysRevB.109.014312}{\emph{Phys. Rev. B}
  {\bfseries 109} (2024) 014312}
  [\href{https://arxiv.org/abs/2308.00636}{{\ttfamily 2308.00636}}].

\bibitem{Iizuka:2023fba}
N.~Iizuka and M.~Nishida, \emph{{Krylov complexity in the IP matrix model. Part
  II}}, \href{https://doi.org/10.1007/JHEP11(2023)096}{\emph{JHEP} {\bfseries
  11} (2023) 096} [\href{https://arxiv.org/abs/2308.07567}{{\ttfamily
  2308.07567}}].

\bibitem{Huh:2023jxt}
K.-B.~Huh, H.-S.~Jeong and J.F.~Pedraza, \emph{{Spread complexity in
  saddle-dominated scrambling}},
  \href{https://arxiv.org/abs/2312.12593}{{\ttfamily 2312.12593}}.

\bibitem{Anegawa:2024wov}
T.~Anegawa, N.~Iizuka and M.~Nishida, \emph{{Krylov complexity as an order
  parameter for deconfinement phase transitions at large N}},
  \href{https://doi.org/10.1007/JHEP04(2024)119}{\emph{JHEP} {\bfseries 04}
  (2024) 119} [\href{https://arxiv.org/abs/2401.04383}{{\ttfamily
  2401.04383}}].

\bibitem{Caputa:2024vrn}
P.~Caputa, H.-S.~Jeong, S.~Liu, J.F.~Pedraza and L.-C.~Qu, \emph{{Krylov
  complexity of density matrix operators}},
  \href{https://arxiv.org/abs/2402.09522}{{\ttfamily 2402.09522}}.

\bibitem{Chen:2024imd}
L.~Chen, B.~Mu, H.~Wang and P.~Zhang, \emph{{Dissecting Quantum Many-body Chaos
  in the Krylov Space}},  \href{https://arxiv.org/abs/2404.08207}{{\ttfamily
  2404.08207}}.

\bibitem{Caputa:2024xkp}
P.~Caputa and K.~Kutak, \emph{{Krylov complexity and gluon cascades in the high
  energy limit}},  \href{https://arxiv.org/abs/2404.07657}{{\ttfamily
  2404.07657}}.

\bibitem{Chattopadhyay:2024pdj}
A.~Chattopadhyay, V.~Malvimat and A.~Mitra, \emph{{Krylov complexity of
  deformed conformal field theories}},
  \href{https://arxiv.org/abs/2405.03630}{{\ttfamily 2405.03630}}.

\bibitem{Bhattacharya:2023yec}
A.~Bhattacharya, R.N.~Das, B.~Dey and J.~Erdmenger, \emph{{Spread complexity
  for measurement-induced non-unitary dynamics and Zeno effect}},
  \href{https://arxiv.org/abs/2312.11635}{{\ttfamily 2312.11635}}.

\bibitem{Bhattacharya:2024hto}
A.~Bhattacharya, R.N.~Das, B.~Dey and J.~Erdmenger, \emph{{Spread complexity
  and localization in $\mathcal{PT}$-symmetric systems}},
  \href{https://arxiv.org/abs/2406.03524}{{\ttfamily 2406.03524}}.

\bibitem{Basu:2024tgg}
R.~Basu, A.~Ganguly, S.~Nath and O.~Parrikar, \emph{{Complexity Growth and the
  Krylov-Wigner function}},  \href{https://arxiv.org/abs/2402.13694}{{\ttfamily
  2402.13694}}.

\bibitem{Tang:2023ocr}
H.~Tang, \emph{{Operator Krylov complexity in random matrix theory}},
  \href{https://arxiv.org/abs/2312.17416}{{\ttfamily 2312.17416}}.

\bibitem{Banerjee:2022ime}
A.~Banerjee, A.~Bhattacharyya, P.~Drashni and S.~Pawar, \emph{{From CFTs to
  theories with Bondi-Metzner-Sachs symmetries: Complexity and
  out-of-time-ordered correlators}},
  \href{https://doi.org/10.1103/PhysRevD.106.126022}{\emph{Phys. Rev. D}
  {\bfseries 106} (2022) 126022}
  [\href{https://arxiv.org/abs/2205.15338}{{\ttfamily 2205.15338}}].

\bibitem{Nandy:2023brt}
S.~Nandy, B.~Mukherjee, A.~Bhattacharyya and A.~Banerjee, \emph{{Quantum state
  complexity meets many-body scars}},
  \href{https://doi.org/10.1088/1361-648X/ad1a7b}{\emph{J. Phys. Condens.
  Matter} {\bfseries 36} (2024) 155601}
  [\href{https://arxiv.org/abs/2305.13322}{{\ttfamily 2305.13322}}].

\bibitem{Bhattacharyya:2023grv}
A.~Bhattacharyya, S.S.~Haque, G.~Jafari, J.~Murugan and D.~Rapotu,
  \emph{{Krylov complexity and spectral form factor for noisy random matrix
  models}}, \href{https://doi.org/10.1007/JHEP10(2023)157}{\emph{JHEP}
  {\bfseries 10} (2023) 157}
  [\href{https://arxiv.org/abs/2307.15495}{{\ttfamily 2307.15495}}].

\bibitem{Bhattacharyya:2023dhp}
A.~Bhattacharyya, D.~Ghosh and P.~Nandi, \emph{{Operator growth and Krylov
  complexity in Bose-Hubbard model}},
  \href{https://doi.org/10.1007/JHEP12(2023)112}{\emph{JHEP} {\bfseries 12}
  (2023) 112} [\href{https://arxiv.org/abs/2306.05542}{{\ttfamily
  2306.05542}}].

\bibitem{Alishahiha:2022anw}
M.~Alishahiha and S.~Banerjee, \emph{{A universal approach to Krylov state and
  operator complexities}},
  \href{https://doi.org/10.21468/SciPostPhys.15.3.080}{\emph{SciPost Phys.}
  {\bfseries 15} (2023) 080}
  [\href{https://arxiv.org/abs/2212.10583}{{\ttfamily 2212.10583}}].

\bibitem{Rabinovici:2023yex}
E.~Rabinovici, A.~S\'anchez-Garrido, R.~Shir and J.~Sonner, \emph{{A bulk
  manifestation of Krylov complexity}},
  \href{https://doi.org/10.1007/JHEP08(2023)213}{\emph{JHEP} {\bfseries 08}
  (2023) 213} [\href{https://arxiv.org/abs/2305.04355}{{\ttfamily
  2305.04355}}].

\bibitem{Chen:2020nlj}
B.~Chen, B.~Czech and Z.-z.~Wang, \emph{{Query complexity and cutoff dependence
  of the CFT2 ground state}},
  \href{https://doi.org/10.1103/PhysRevD.103.026015}{\emph{Phys. Rev. D}
  {\bfseries 103} (2021) 026015}
  [\href{https://arxiv.org/abs/2004.11377}{{\ttfamily 2004.11377}}].

\bibitem{Chen:2022fbg}
B.~Chen, B.~Czech, J.~de~Boer, L.~Lamprou and Z.-z.~Wang, \emph{{Boundary and
  bulk notions of transport in the AdS$_3$/CFT$_2$ correspondence}},
  \href{https://doi.org/10.1007/JHEP05(2023)102}{\emph{JHEP} {\bfseries 05}
  (2023) 102} [\href{https://arxiv.org/abs/2211.15684}{{\ttfamily
  2211.15684}}].

\bibitem{Nielsen1}
M.A.~Nielsen, \emph{A geometric approach to quantum circuit lower bounds},
  \href{https://arxiv.org/abs/quant-ph/0502070}{{\ttfamily quant-ph/0502070}}.

\bibitem{Nielsen2}
M.A.~Nielsen, M.R.~Dowling, M.~Gu and A.C.~Doherty, \emph{Quantum computation
  as geometry}, \href{https://doi.org/10.1126/science.1121541}{\emph{Science}
  {\bfseries 311} (2006) 1133}.

\bibitem{Nielsen3}
M.R.~Dowling and M.A.~Nielsen, \emph{The geometry of quantum computation},
  \href{https://arxiv.org/abs/quant-ph/0701004}{{\ttfamily quant-ph/0701004}}.

\bibitem{Craps:2022ese}
B.~Craps, M.~De~Clerck, O.~Evnin, P.~Hacker and M.~Pavlov, \emph{{Bounds on
  quantum evolution complexity via lattice cryptography}},
  \href{https://doi.org/10.21468/SciPostPhys.13.4.090}{\emph{SciPost Phys.}
  {\bfseries 13} (2022) 090}
  [\href{https://arxiv.org/abs/2202.13924}{{\ttfamily 2202.13924}}].

\bibitem{Craps:2023rur}
B.~Craps, M.~De~Clerck, O.~Evnin and P.~Hacker, \emph{{Integrability and
  complexity in quantum spin chains}},
  \href{https://doi.org/10.21468/SciPostPhys.16.2.041}{\emph{SciPost Phys.}
  {\bfseries 16} (2024) 041}
  [\href{https://arxiv.org/abs/2305.00037}{{\ttfamily 2305.00037}}].

\bibitem{Craps:2023ivc}
B.~Craps, O.~Evnin and G.~Pascuzzi, \emph{{A relation between Krylov and
  Nielsen complexity}},  \href{https://arxiv.org/abs/2311.18401}{{\ttfamily
  2311.18401}}.

\bibitem{Okuyama:2024yya}
K.~Okuyama, \emph{{Doubled Hilbert space in double-scaled SYK}},
  \href{https://arxiv.org/abs/2401.07403}{{\ttfamily 2401.07403}}.

\bibitem{Vidal:2007hda}
G.~Vidal, \emph{{Entanglement Renormalization}},
  \href{https://doi.org/10.1103/PhysRevLett.99.220405}{\emph{Phys. Rev. Lett.}
  {\bfseries 99} (2007) 220405}
  [\href{https://arxiv.org/abs/cond-mat/0512165}{{\ttfamily
  cond-mat/0512165}}].

\bibitem{Vidal:2008zz}
G.~Vidal, \emph{{Class of Quantum Many-Body States That Can Be Efficiently
  Simulated}},
  \href{https://doi.org/10.1103/PhysRevLett.101.110501}{\emph{Phys. Rev. Lett.}
  {\bfseries 101} (2008) 110501}
  [\href{https://arxiv.org/abs/quant-ph/0610099}{{\ttfamily
  quant-ph/0610099}}].

\bibitem{Maldacena:2013xja}
J.~Maldacena and L.~Susskind, \emph{{Cool horizons for entangled black holes}},
  \href{https://doi.org/10.1002/prop.201300020}{\emph{Fortsch. Phys.}
  {\bfseries 61} (2013) 781} [\href{https://arxiv.org/abs/1306.0533}{{\ttfamily
  1306.0533}}].

\bibitem{Erd_s_2014}
L.~Erdős and D.~Schröder, \emph{Phase transition in the density of states of
  quantum spin glasses},
  \href{https://doi.org/10.1007/s11040-014-9164-3}{\emph{Mathematical Physics,
  Analysis and Geometry} {\bfseries 17} (2014) 441–464}.

\bibitem{viswanath1994recursion}
V.~Viswanath and G.~M{\"u}ller, \emph{The recursion method: application to many
  body dynamics}, vol.~23, Springer Science \& Business Media (1994).

\bibitem{Witten:2007kt}
E.~Witten, \emph{{Three-Dimensional Gravity Revisited}},
  \href{https://arxiv.org/abs/0706.3359}{{\ttfamily 0706.3359}}.

\bibitem{ambainis2018understanding}
A.~Ambainis, \emph{Understanding quantum algorithms via query complexity},  in
  \emph{Proceedings of the International Congress of Mathematicians: Rio de
  Janeiro 2018}, pp.~3265--3285, World Scientific, 2018.

\bibitem{Chen:2021lnq}
B.~Chen, B.~Czech and Z.-z.~Wang, \emph{{Quantum information in holographic
  duality}}, \href{https://doi.org/10.1088/1361-6633/ac51b5}{\emph{Rept. Prog.
  Phys.} {\bfseries 85} (2022) 046001}
  [\href{https://arxiv.org/abs/2108.09188}{{\ttfamily 2108.09188}}].

\bibitem{Yang:2018nda}
R.-Q.~Yang, Y.-S.~An, C.~Niu, C.-Y.~Zhang and K.-Y.~Kim, \emph{{Principles and
  symmetries of complexity in quantum field theory}},
  \href{https://doi.org/10.1140/epjc/s10052-019-6600-3}{\emph{Eur. Phys. J. C}
  {\bfseries 79} (2019) 109}
  [\href{https://arxiv.org/abs/1803.01797}{{\ttfamily 1803.01797}}].

\bibitem{Yang:2018tpo}
R.-Q.~Yang, Y.-S.~An, C.~Niu, C.-Y.~Zhang and K.-Y.~Kim, \emph{{More on
  complexity of operators in quantum field theory}},
  \href{https://doi.org/10.1007/JHEP03(2019)161}{\emph{JHEP} {\bfseries 03}
  (2019) 161} [\href{https://arxiv.org/abs/1809.06678}{{\ttfamily
  1809.06678}}].

\bibitem{Yang:2019iav}
R.-Q.~Yang and K.-Y.~Kim, \emph{{Time evolution of the complexity in chaotic
  systems: a concrete example}},
  \href{https://doi.org/10.1007/JHEP05(2020)045}{\emph{JHEP} {\bfseries 05}
  (2020) 045} [\href{https://arxiv.org/abs/1906.02052}{{\ttfamily
  1906.02052}}].

\bibitem{bao2012introduction}
D.~Bao, S.-S.~Chern and Z.~Shen, \emph{An introduction to Riemann-Finsler
  geometry}, vol.~200, Springer Science \& Business Media (2012).

\bibitem{Okuyama:2023byh}
K.~Okuyama, \emph{{End of the world brane in double scaled SYK}},
  \href{https://doi.org/10.1007/JHEP08(2023)053}{\emph{JHEP} {\bfseries 08}
  (2023) 053} [\href{https://arxiv.org/abs/2305.12674}{{\ttfamily
  2305.12674}}].

\bibitem{Lashkari:2013koa}
N.~Lashkari, M.B.~McDermott and M.~Van~Raamsdonk, \emph{{Gravitational dynamics
  from entanglement 'thermodynamics'}},
  \href{https://doi.org/10.1007/JHEP04(2014)195}{\emph{JHEP} {\bfseries 04}
  (2014) 195} [\href{https://arxiv.org/abs/1308.3716}{{\ttfamily 1308.3716}}].

\bibitem{Faulkner:2013ica}
T.~Faulkner, M.~Guica, T.~Hartman, R.C.~Myers and M.~Van~Raamsdonk,
  \emph{{Gravitation from Entanglement in Holographic CFTs}},
  \href{https://doi.org/10.1007/JHEP03(2014)051}{\emph{JHEP} {\bfseries 03}
  (2014) 051} [\href{https://arxiv.org/abs/1312.7856}{{\ttfamily 1312.7856}}].

\bibitem{Swingle:2014uza}
B.~Swingle and M.~Van~Raamsdonk, \emph{{Universality of Gravity from
  Entanglement}},  \href{https://arxiv.org/abs/1405.2933}{{\ttfamily
  1405.2933}}.

\bibitem{Faulkner:2017tkh}
T.~Faulkner, F.M.~Haehl, E.~Hijano, O.~Parrikar, C.~Rabideau and
  M.~Van~Raamsdonk, \emph{{Nonlinear Gravity from Entanglement in Conformal
  Field Theories}}, \href{https://doi.org/10.1007/JHEP08(2017)057}{\emph{JHEP}
  {\bfseries 08} (2017) 057}
  [\href{https://arxiv.org/abs/1705.03026}{{\ttfamily 1705.03026}}].

\bibitem{Haehl:2017sot}
F.M.~Haehl, E.~Hijano, O.~Parrikar and C.~Rabideau, \emph{{Higher Curvature
  Gravity from Entanglement in Conformal Field Theories}},
  \href{https://doi.org/10.1103/PhysRevLett.120.201602}{\emph{Phys. Rev. Lett.}
  {\bfseries 120} (2018) 201602}
  [\href{https://arxiv.org/abs/1712.06620}{{\ttfamily 1712.06620}}].

\bibitem{Agon:2020mvu}
C.A.~Ag\'on, E.~C\'aceres and J.F.~Pedraza, \emph{{Bit threads,
  Einstein\textquoteright{}s equations and bulk locality}},
  \href{https://doi.org/10.1007/JHEP01(2021)193}{\emph{JHEP} {\bfseries 01}
  (2021) 193} [\href{https://arxiv.org/abs/2007.07907}{{\ttfamily
  2007.07907}}].

\bibitem{Agon:2021tia}
C.A.~Ag\'on and J.F.~Pedraza, \emph{{Quantum bit threads and holographic
  entanglement}}, \href{https://doi.org/10.1007/JHEP02(2022)180}{\emph{JHEP}
  {\bfseries 02} (2022) 180}
  [\href{https://arxiv.org/abs/2105.08063}{{\ttfamily 2105.08063}}].

\bibitem{VanRaamsdonk:2009ar}
M.~Van~Raamsdonk, \emph{{Comments on quantum gravity and entanglement}},
  \href{https://arxiv.org/abs/0907.2939}{{\ttfamily 0907.2939}}.

\bibitem{VanRaamsdonk:2010pw}
M.~Van~Raamsdonk, \emph{{Building up spacetime with quantum entanglement}},
  \href{https://doi.org/10.1142/S0218271810018529}{\emph{Gen. Rel. Grav.}
  {\bfseries 42} (2010) 2323}
  [\href{https://arxiv.org/abs/1005.3035}{{\ttfamily 1005.3035}}].

\bibitem{Bianchi:2012ev}
E.~Bianchi and R.C.~Myers, \emph{{On the Architecture of Spacetime Geometry}},
  \href{https://doi.org/10.1088/0264-9381/31/21/214002}{\emph{Class. Quant.
  Grav.} {\bfseries 31} (2014) 214002}
  [\href{https://arxiv.org/abs/1212.5183}{{\ttfamily 1212.5183}}].

\bibitem{Balasubramanian:2014sra}
V.~Balasubramanian, B.D.~Chowdhury, B.~Czech and J.~de~Boer, \emph{{Entwinement
  and the emergence of spacetime}},
  \href{https://doi.org/10.1007/JHEP01(2015)048}{\emph{JHEP} {\bfseries 01}
  (2015) 048} [\href{https://arxiv.org/abs/1406.5859}{{\ttfamily 1406.5859}}].

\bibitem{carleman1926fonctions}
T.~Carleman, \emph{Les Fonctions quasi analytiques: le{\c{c}}ons profess{\'e}es
  au College de France}, Gauthier-Villars (1926).

\bibitem{Espanol:2022cqr}
B.L.~Espa\~nol and D.A.~Wisniacki, \emph{{Assessing the saturation of Krylov
  complexity as a measure of chaos}},
  \href{https://doi.org/10.1103/PhysRevE.107.024217}{\emph{Phys. Rev. E}
  {\bfseries 107} (2023) 024217}
  [\href{https://arxiv.org/abs/2212.06619}{{\ttfamily 2212.06619}}].

\bibitem{Okuyama:2023yat}
K.~Okuyama, \emph{{Matter correlators through a wormhole in double-scaled
  SYK}}, \href{https://doi.org/10.1007/JHEP02(2024)147}{\emph{JHEP} {\bfseries
  02} (2024) 147} [\href{https://arxiv.org/abs/2312.00880}{{\ttfamily
  2312.00880}}].

\bibitem{Jafferis:2022wez}
D.L.~Jafferis, D.K.~Kolchmeyer, B.~Mukhametzhanov and J.~Sonner,
  \emph{{Jackiw-Teitelboim gravity with matter, generalized eigenstate
  thermalization hypothesis, and random matrices}},
  \href{https://doi.org/10.1103/PhysRevD.108.066015}{\emph{Phys. Rev. D}
  {\bfseries 108} (2023) 066015}
  [\href{https://arxiv.org/abs/2209.02131}{{\ttfamily 2209.02131}}].

\bibitem{Jafferis:2022uhu}
D.L.~Jafferis, D.K.~Kolchmeyer, B.~Mukhametzhanov and J.~Sonner, \emph{{Matrix
  Models for Eigenstate Thermalization}},
  \href{https://doi.org/10.1103/PhysRevX.13.031033}{\emph{Phys. Rev. X}
  {\bfseries 13} (2023) 031033}
  [\href{https://arxiv.org/abs/2209.02130}{{\ttfamily 2209.02130}}].

\bibitem{lloyd2000ultimate}
S.~Lloyd, \emph{Ultimate physical limits to computation}, {\emph{Nature}
  {\bfseries 406} (2000) 1047}.

\bibitem{Beny:2011vh}
C.~Beny, \emph{{Causal structure of the entanglement renormalization ansatz}},
  \href{https://doi.org/10.1088/1367-2630/15/2/023020}{\emph{New J. Phys.}
  {\bfseries 15} (2013) 023020}
  [\href{https://arxiv.org/abs/1110.4872}{{\ttfamily 1110.4872}}].

\bibitem{Bao:2017qmt}
N.~Bao, C.~Cao, S.M.~Carroll and A.~Chatwin-Davies, \emph{{De Sitter Space as a
  Tensor Network: Cosmic No-Hair, Complementarity, and Complexity}},
  \href{https://doi.org/10.1103/PhysRevD.96.123536}{\emph{Phys. Rev. D}
  {\bfseries 96} (2017) 123536}
  [\href{https://arxiv.org/abs/1709.03513}{{\ttfamily 1709.03513}}].

\bibitem{Bao:2017iye}
N.~Bao, C.~Cao, S.M.~Carroll and L.~McAllister, \emph{{Quantum Circuit
  Cosmology: The Expansion of the Universe Since the First Qubit}},
  \href{https://arxiv.org/abs/1702.06959}{{\ttfamily 1702.06959}}.

\bibitem{Niermann:2021wco}
L.~Niermann and T.J.~Osborne, \emph{{Holographic networks for (1+1)-dimensional
  de~Sitter space-time}},
  \href{https://doi.org/10.1103/PhysRevD.105.125009}{\emph{Phys. Rev. D}
  {\bfseries 105} (2022) 125009}
  [\href{https://arxiv.org/abs/2102.09223}{{\ttfamily 2102.09223}}].

\bibitem{Cao:2023gkw}
C.~Cao, W.~Chemissany, A.~Jahn and Z.~Zimbor\'as, \emph{{Approximate
  observables from non-isometric maps: de Sitter tensor networks with
  overlapping qubits}},  \href{https://arxiv.org/abs/2304.02673}{{\ttfamily
  2304.02673}}.

\bibitem{Galante:2022nhj}
D.~Galante, \emph{{Geodesics, complexity and holography in (A)dS$_2$}},
  \href{https://doi.org/10.22323/1.406.0359}{\emph{PoS} {\bfseries CORFU2021}
  (2022) 359}.

\bibitem{Kashyap:2015lva}
S.P.~Kashyap, S.~Mondal, A.~Sen and M.~Verma, \emph{{Surviving in a Metastable
  de Sitter Space-Time}},
  \href{https://doi.org/10.1007/JHEP09(2015)139}{\emph{JHEP} {\bfseries 09}
  (2015) 139} [\href{https://arxiv.org/abs/1506.00772}{{\ttfamily
  1506.00772}}].

\bibitem{Anninos:2022qgy}
D.~Anninos, D.A.~Galante and S.U.~Sheorey, \emph{{Renormalisation group flows
  of deformed SYK models}},
  \href{https://doi.org/10.1007/JHEP11(2023)197}{\emph{JHEP} {\bfseries 11}
  (2023) 197} [\href{https://arxiv.org/abs/2212.04944}{{\ttfamily
  2212.04944}}].

\bibitem{Anninos:2020cwo}
D.~Anninos and D.A.~Galante, \emph{{Constructing AdS$_{2}$ flow geometries}},
  \href{https://doi.org/10.1007/JHEP02(2021)045}{\emph{JHEP} {\bfseries 02}
  (2021) 045} [\href{https://arxiv.org/abs/2011.01944}{{\ttfamily
  2011.01944}}].

\bibitem{Berkooz:2024evs}
M.~Berkooz, N.~Brukner, Y.~Jia and O.~Mamroud, \emph{{From Chaos to
  Integrability in Double Scaled SYK}},
  \href{https://arxiv.org/abs/2403.01950}{{\ttfamily 2403.01950}}.

\bibitem{Berkooz:2024ofm}
M.~Berkooz, N.~Brukner, Y.~Jia and O.~Mamroud, \emph{{A Path Integral for Chord
  Diagrams and Chaotic-Integrable Transitions in Double Scaled SYK}},
  \href{https://arxiv.org/abs/2403.05980}{{\ttfamily 2403.05980}}.

\bibitem{Antonini:2023aza}
S.~Antonini, B.~Grado-White, S.-K.~Jian and B.~Swingle, \emph{{Holographic
  measurement in CFT thermofield doubles}},
  \href{https://doi.org/10.1007/JHEP07(2023)014}{\emph{JHEP} {\bfseries 07}
  (2023) 014} [\href{https://arxiv.org/abs/2304.06743}{{\ttfamily
  2304.06743}}].

\bibitem{Antonini:2022lmg}
S.~Antonini, B.~Grado-White, S.-K.~Jian and B.~Swingle, \emph{{Holographic
  measurement and quantum teleportation in the SYK thermofield double}},
  \href{https://doi.org/10.1007/JHEP02(2023)095}{\emph{JHEP} {\bfseries 02}
  (2023) 095} [\href{https://arxiv.org/abs/2211.07658}{{\ttfamily
  2211.07658}}].

\end{thebibliography}\endgroup
\end{document}